\documentclass[ba]{imsart}
\pubyear{2022}
\volume{TBA}
\issue{TBA}
\firstpage{1}
\lastpage{1}

\usepackage{amsthm}
\usepackage{amsmath}
\usepackage{natbib}
\usepackage[colorlinks,citecolor=blue,urlcolor=blue,filecolor=blue,backref=page]{hyperref}
\usepackage{graphicx}
\graphicspath{{./figures/}}

\usepackage{bbm}
\usepackage{bm}
\usepackage{amssymb}
\usepackage{amsfonts}
\usepackage{adjustbox}
\usepackage{caption}
\usepackage{subcaption}
\usepackage{multirow}
\usepackage{xcolor}
\usepackage{soul}

\theoremstyle{definition}
\newtheorem{definition}{Definition}
\DeclareMathOperator*{\argmax}{argmax}

\newcommand{\MPPI}{\mathrm{MPPI}} 
 
\startlocaldefs
\numberwithin{equation}{section}
\theoremstyle{plain}

\endlocaldefs

\begin{document}

\begin{frontmatter}
\title{Bayesian Image-on-Scalar Regression with a Spatial Global-Local Spike-and-Slab Prior}
\runtitle{\footnotesize Bayesian Image-on-Scalar Regression with a Spatial Global-Local Spike-and-Slab Prior}

\begin{aug}
\author{\fnms{Zijian} \snm{Zeng} \thanksref{addr1}\ead[label=e1]{zz57@rice.edu}},
\author{\fnms{Meng} \snm{Li} \thanksref{addr2}\ead[label=e2]{meng@rice.edu}}
\and
\author{\fnms{Marina} \snm{Vannucci} \thanksref{addr3}\ead[label=e3]{marina@rice.edu }%
}

\runauthor{Z. Zeng, M. Li and M. Vannucci}

\address[addr1]{Department of Statistcs, Rice University, Houston, TX.  
    \printead{e1} 
}
\address[addr2]{Department of Statistcs, Rice University, Houston, TX. 
    \printead{e2}
}
\address[addr3]{Department of Statistcs, Rice University, Houston, TX. 
    \printead{e3}
}


\end{aug}

\begin{abstract}
In this article, we propose a novel spatial global-local spike-and-slab selection prior for image-on-scalar regression. We consider a Bayesian hierarchical Gaussian process model for image smoothing, that uses a flexible Inverse-Wishart process prior to handle within-image dependency, and propose a general global-local spatial selection prior that broadly relates to a rich class of well-studied selection priors. Unlike existing constructions, we achieve simultaneous global (i.e., at covariate-level) and local (i.e., at pixel/voxel-level) selection by introducing {\it participation rate} parameters that measure the probability for the individual covariates to affect the observed images. This along with a hard-thresholding strategy leads to dependency between selections at the two levels, introduces extra sparsity at the local level, and allows the global selection to be informed by the local selection, all in a model-based manner. We design an efficient Gibbs sampler that allows inference for large image data. We show on simulated data that parameters are interpretable and lead to efficient selection. Finally, we demonstrate performance of the proposed model by using data from the Autism Brain Imaging Data Exchange (ABIDE) study \citep{ABIDE2014}. 
\end{abstract}



\end{frontmatter}

\section{Introduction}
\label{sec:intro}
With the explosive growth in the amount of image data collected for various medical research there comes an increasing interest in discovering the relation between the image data and potential covariates measured on the same set of subjects. Image-on-scalar regression models have drawn increasing attention for this purpose, see \cite{Worsley2004, Zhu2014, Xinyi2020, Zhang2020, Yu2021}, among others. These models present several challenges: spatial dependency in the image data can be highly complex and hard to model; image data can be composed by a large number of pixels/voxels and  lead to extremely large covariance matrices with heavy computational burden; covariates can have partial influence on the image responses, i.e., they can affect only a few pixels/voxels in the image, making it hard to distinguish such covariates from noisy ones.

Conventional approaches for image-on-scalar regression are based on mass univariate analysis (MUA), for example by running pixel/voxel-wise independent general linear models to generate maps for statistics of interest, and then applying methods for post-inference \citep{Worsley2004, David2011}. These methods are computationally efficient and have well-studied theoretical properties. However, they completely ignore the spatial dependency within the images and are generally not optimal in regards to statistical power \citep{Chumbley2009}. To address these shortcomings, recent approaches consider the image data as realizations of functions on a given domain and apply functional data analysis (FDA) methods
that use basis expansions and spatially-varying coefficients to account for dependency within and across images \citep{Zhu2014,Xinyi2020,Yu2021}. Joint uncertainty quantification for all model parameters, however, is difficult to achieve for these methods in the frequentist literature. Here, we consider a Bayesian hierarchical Gaussian process (GP) model for image smoothing that avoids assumptions on functional forms and that uses a flexible Inverse-Wishart process to handle within-image dependency. This modeling structure extends an approach proposed by \cite{Cox2016} for longitudinal data to the case of image data.

An important aspect in image-on-scalar regression is the selection and interpretation of influential covariates. Ideally, one may want a covariate to be influential for the whole image. In practice, however, the covariate can only partially affect the image, i.e., being influential only for a few pixels/voxels. We refer to the aspect of selecting whether a covariate is influential for the images as ``global", and to the aspect of selecting which pixels/voxels are affected by the covariate as ``local". In the Bayesian framework, a global selection prior was proposed by \cite{Reich2010} and allows coefficients to be non-zero constant, spatially-varying function, or zero constant. For local selection, a common way of selecting pixels/voxels uses a two-component mixture prior, which models the spatially-varying coefficient via a latent continuous process and a binary selection indicator process, see \cite{Smith2007, Scheel2013, Goldsmith2014, Fan2015,Choi2016}. Recently \cite{Kang2018} proposed a soft-threshold Gaussian process prior that does not make use of the indicator process but rather achieves local sparsity by thresholding. This idea can be traced back to the earlier research of \cite{Nakajima2013}, who used a hard-threshold prior for longitudinal data to introduce sparsity at each time point. Overall, none of these prior constructions achieve simultaneous global- and local-level selection of the coefficients. Finally, in the more general framework of function-on-scalar regression, in which we consider images as 2/3-dimensional functions, Bayesian approaches employ basis functions and functional principal component analysis to model the within-function dependency, see for example \cite{Goldsmith2016, Kowal2020}. Built on the basis functions domain, this framework can be computationally more efficient. However, the covariates' effects are assumed on the basis functions, instead of directly on the observed functions. As a consequence, selection relies on the choice of the basis functions, especially for high-dimensional functional responses, and a two-level selection becomes less intuitive, as the local level selection would require all basis functions to be set to $0$ at some specific pixels/voxels.   

We propose a spatial global-local spike-and-slab process prior for image-on-scalar regression that broadly relates to a rich class of well-studied local selection priors. We achieve simultaneous global and local selection by introducing {\it participation rate} parameters, that measure the probability for the individual covariates to affect the observed images, and employing hard thresholding. The proposed prior performs bi-level selection, allowing the global selection to be informed by the local selection. We design an efficient Gibbs sampler that allows inference for large image data and use simulated data to show that prior parameters are interpretable and lead to efficient selection.  We also demonstrate the performance of the proposed model with respect to MUA methods. Finally, we apply our method to data from the Autism Brain Imaging Data Exchange (ABIDE) study \citep{ABIDE2014}. Results show that modeling dependency in the data leads to more localized selection.

The rest of the paper is organized as follows. In section~\ref{sec:method}, we introduce the proposed method, the prior construction and the sampler procedure. In Section~\ref{sec:sim}, we conduct simulations and compare the proposed approach with widely used MUA methods. In Section~\ref{sec:app}, we apply the method to image data from the ABIDE study.

\section{Methods}
\label{sec:method}
\subsection{Bayesian Image-on-Scalar Regression}
Suppose $n$ images $Y_i(\cdot)$ are observed on a $K$-dimensional common domain $\bm{S} \subseteq \mathbb{R}^{K}$, each associated with a $q$-dimensional covariate vector $\bm{x}_i = (x_{i1}, \ldots, x_{iq})^T$, for $i = 1, \ldots, n$. We begin with a Bayesian hierarchical model for image responses and scalar covariates
\begin{align}
\label{BHM.eq.1}
    & Y_i( \bm{s} )  =  Z_i(\bm{s}) + \epsilon_{i,\bm{s}}, \quad \epsilon_{i,\bm{s}} \stackrel{i.i.d}{\sim}  N\left( 0, \sigma^2_\epsilon\right), \quad \bm{s}  \in \bm{S} \\
    \label{BHM.eq.2} 
    &  Z_i(\cdot) \sim  \mathcal{GP}\left( \mu_i(\cdot),  \Sigma\left(\cdot, \cdot\right)\right), \quad \mu_i(\cdot ) = \beta_0(\cdot) + \sum^q_{j=1} x_{ij} \beta_j(\cdot), \quad i = 1, \ldots n,
\end{align}
where the noise-free mean surface of $Y_i(\cdot)$, $Z_i(\cdot)$, is modeled by a Gaussian process (GP) with covariate-dependent, subject-specific mean $\mu_i(\cdot)$ and a common covariance surface $\Sigma(\cdot, \cdot)$ for all images, $\beta_0(\cdot)$ is the intercept coefficient image and $\{\beta_j(\cdot)\}^q_{j=1}$ are the coefficient images linking covariates $\bm{x}_i$ with $\mu_i(\cdot)$. Here we assume Gaussian errors $\epsilon_{i,\bm{s}} \sim N\left( 0, \sigma^2_\epsilon\right)$ independently across both location $\bm{s}$ and subject $i$, with $\sigma^2_\epsilon \sim \text{Inverse-Gamma}(a_\epsilon, b_\epsilon)$. Throughout this article, we assume $\bm{S}$ to be a compact set. 

Equations~\eqref{BHM.eq.1}-\eqref{BHM.eq.2} define a Bayesian hierarchical model for image data. This model has considerable virtues: it enables simultaneous smoothing of individual observations and borrowing of information across observations, while being flexible through Bayes nonparametrics and interpretable. It has been widely used to analyze functional data. For example, in the case of longitudinal data (i.e., $K = 1$), \cite{Cox2016} focus on a mean-covariance structure in the absence of covariates, with a common mean function, \cite{Liu2019} study conditional quantiles of $Y_i(\cdot)$ by altering the Gaussian error to asymmetric Laplace, and \cite{Shamshoian2020} consider a sparse Bayesian infinite factor model for $Z_i(\cdot)$. 

\subsection{Spatial Global-Local Spike-and-Slab Prior}
\label{sec:SGLSS}
We are interested in performing selection at both image and location (pixel/voxel) levels, while estimating the model coefficients and the covariance structure. We achieve this via a novel spatial global-local spike-and-slab (SGLSS) prior for the coefficient images $\left\{\beta_j(\cdot)\right\}^{q}_{j=1}$. Here, selection at the global level represents the covariate selection, as eliminating a covariate would zero out the entire coefficient image, while the local level refers to individual locations (pixels/voxels) in the coefficient image.
The proposed SGLSS prior is a three-component process consisting of a continuous process, a local-level discrete selection process, and a global-level indicator: 
\begin{equation}
    \begin{aligned}
    \beta_j(\cdot) & = \tilde{\beta}_j(\cdot) \times \tau_j(\cdot) \times I \left( \pi_j \ge d\right), \\
    & = \left[\tau_j(\cdot)I \left( \pi_j \ge d\right)\right] \tilde{\beta}_j(\cdot) + \left[ 1 - \tau_j(\cdot)I \left( \pi_j \ge d\right)\right] \delta_0,
    \end{aligned} \label{SGLSS.eq.1}
\end{equation}
with $\delta_0$ a point mass distribution at $0$, $\tilde{\beta}_j(\cdot)$ the continuous process, $\tau_j(\cdot)$ the discrete local selection process and $I( \pi_j \ge d)$ the global indicator. Construction \eqref{SGLSS.eq.1} is completed by choosing priors for $\tilde{\beta}_j(\cdot)$, $\tau_j(\cdot)$, and $\pi_j$. The high dimensionality of image data poses substantial challenges to computational efficiency, particularly for Markov Chain Monte Carlo methods. To achieve computational scalability, we adopt the following prior setting that exploits conjugacy:  
\begin{equation}
    \begin{aligned}
        \tilde{\beta}_j(\bm{s}) | \tau_j(\bm{s}) & \sim  \tau_j(\bm{s}) N\left( \mu_{0j}(\bm{s}), \sigma^2_{0j}(\bm{s})\right) + (1 - \tau_j(\bm{s}))\delta_0, \quad \bm{s}\in \bm{S}\\
        \tau_j( \bm{s}) | \pi_j & \sim  \text{Bernoulli}\left(\pi_j \right),  \\
        \pi_j  & \sim  \text{Beta}(a_\pi, b_\pi).
    \end{aligned}\label{SGLSS.eq.2}
\end{equation}
The SGLSS prior for $\beta_j(\cdot)$ with the specification of Equation~\eqref{SGLSS.eq.2} can be re-written into a classic spike-and-slab prior as
\begin{equation*} 
    \begin{aligned}
    \beta_j(\bm{s}) | \tau_j(\bm{s}), \pi_j & \sim  \left[\tau_j(\bm{s}) I\left( \pi_j \ge d \right)\right] N\left( \mu_{0j}(\bm{s}), \sigma^2_{0j}(\bm{s})\right) + \left[1 - \tau_j(\bm{s}) I\left( \pi_j \ge d \right)\right]\delta_0, \\
    \tau_j( \bm{s}) | \pi_j & \sim  \text{Bernoulli}\left(\pi_j \right),  \\
    \pi_j &  \sim  \text{Beta}(a_\pi, b_\pi).
    \end{aligned} 
\end{equation*}
As for the intercept image $\beta_0(\bm{s})$, we fix $\tau_0(\bm{s})I(\pi_0\ge d) = 1$ for all $\bm{s} \in \bm{S}$ since this term is typically always included in the model. The proposed SGLSS prior construction has rich connections with a wide range of existing priors, as we point out in the section below. 

The parameter $\pi_j$, which we call {\em `participation rate'}, has the interpretation that $\pi_j$ percent of the $j$th coefficient image is expected to be non-zero, and can also be interpreted as the probability that $x_j$ has an influence on the observed images. 
The parameter $d$ defines the threshold at which we include covariate $x_j$, i.e., if $d\times 100$ percent of its corresponding coefficient images are expected to be non-zero. Therefore, the threshold parameter $d$ controls selection at the ``global" level, leading to the exclusion of those covariates with low participation rates. Here, without a priori information, we use a common $d$ for all covariates. When a priori information is available, covariate-dependent parameters $d_j$'s can be specified. The parameter $\pi_j$, along with the hard-thresholding structure, introduces extra sparsity at local level, as it can be seen by calculating the expectation of the global-local selection indicator
\begin{align}
        \nonumber E\left[ I(\pi_j \ge d) \tau_j(\bm{s}) \right] & =  E_{\pi_j} \left[ E\left[ I(\pi_j \ge d) \tau_j(\bm{s}) \right] | \pi_j \right] = E_{\pi_j}\left[ I(\pi_j \ge d) \pi_j \right]\\
        \nonumber 
       & = \int^1_d \pi_j \frac{1}{B(a_{\pi}, b_{\pi})} \pi_j^{a_{\pi} -1 } (1-\pi_j)^{b_{\pi} -1 } d \pi_j \\
       & = \frac{a_\pi}{a_\pi+b_\pi} \left[ 1 - F_{\text{Beta}}(d)\right] =  E\left[\tau_j(\bm{s})\right] \left[ 1 - F_{\text{Beta}}(d)\right] , \label{eq:discount.factor}
\end{align}
where $F_{\text{Beta}}(\cdot)$ is the cumulative distribution function of $\text{Beta}(a_{\pi}+1, b_{\pi})$. Hence, the extra factor $\left[ 1 - F_{\text{Beta}}(d)\right] \le 1$ in  Equation~\eqref{eq:discount.factor}, which is strictly decreasing in $d$, introduces more sparsity in the coefficient images on average from a prior perspective. The participation rate parameter $\pi_j$ and the threshold parameter $d \in [0,1]$ establish a bridge between global and local level selection, favorably endowing existing local level selection prior with simultaneous two-level selection, allowing global selection to be {\em informed} by the selection at the local level. 

In practice, image data are typically observed at a grid of discretized locations. Given a vector of $p$ locations of interest $\vec{\bm{s}} = (\bm{s}_1, ..., \bm{s}_p)$, we use $Z_i(\vec{\bm{s}}) = (Z_i(\bm{s}_1), ..., Z_i(\bm{s}_p))$ to denote a $p$-by-$1$ vector of the process values at location $\bm{s} \in \vec{\bm{s}}$ and $\Sigma(\vec{\bm{s}},\vec{\bm{s}}) = \{\Sigma(\bm{s},\bm{s}'); \bm{s}, \bm{s}' \in \vec{\bm{s}} \}$ a $p$-by-$p$ matrix of the within-image covariance. We remark that the proposed method is applicable to discretized locations in a general $K$-dimensional domain, i.e., the vector $\vec{\bm{s}}$ is not limited to integers nor needs to be equidistant.  

\subsubsection{Relationship to existing literature} 

The proposed SGLSS process prior broadly relates to a wide range of existing priors that can be obtained for different choices of the parameter $d$ and prior choices for $\tilde{\beta}_j(\cdot)$ and $\tau_j(\cdot)$. In particular, for $d=1$ the use of continuous priors on $\pi_j$ leads to $I\left( \pi_j \ge 1\right) = 0$ with probability $1$, thus no covariate will be included in the model almost surely. This degenerate SGLSS process prior results in a high-dimensional extension of the mean-covariance smoothing model of \cite{Cox2016} for the following choice of prior
\begin{equation*}
    \beta_0(\cdot) = \tilde{\beta}_0(\cdot) \sim \mathcal{GP}\left( \mu_0(\cdot), \frac{1}{c}\Sigma( \cdot, \cdot) \right),
\end{equation*}
extending the mean-covariance smoothing structure from one-dimensional time-series to high-dimensional images. 

When $d = 0$ is specified, the three-component mixture prior in Equation~\eqref{SGLSS.eq.1} degenerates to a two-component mixture of the type
\begin{equation*}
    \beta_j(\cdot) = \tilde{\beta}_j(\cdot) \times \tau_j(\cdot) \times 1.
\end{equation*}
This degenerate construction naturally relates to the prior constructions used in scalar-on-image regression, and easily accommodates spatially dependent priors on $\tilde{\beta}_j(\cdot)$ and $\tau_j(\cdot)$ \citep{Smith2007,Scheel2013, Goldsmith2014,Fan2015,Choi2016}. We note, however, that incorporating spatially-correlated priors for the regression coefficients within our general global-local construction poses substantial computational challenges and a more careful interpretation of the prior parameters. See also the Conclusion section. Thresholding priors can also be accommodating. For example, by using an autoregressive process for $\tilde{\beta}_j(\cdot)$ and setting the hard threshold, $\tau_j(\cdot) = I( \tilde{\beta}_j(\cdot)>d_j)$, the prior in \cite{Nakajima2013} can be obtained, and by setting $\tilde{\beta}_j(\cdot) = \text{sgn}\left( z_j(\cdot)\right)\left(\left|z_j(\cdot)\right| - \lambda_j \right)$, with threshold $\tau_j(\cdot) = I\left( \left|z_j(\cdot) \right| > \lambda_j \right)$ and $z_j(\cdot) \sim \mathcal{GP}$, the prior in \cite{Kang2018} can be obtained. Another partially reproducible prior is the global level selection prior of \cite{Reich2010}, 
\begin{equation}
    \beta_j(\cdot) = \gamma_{1j} \left(\beta_{0} +  \gamma_{2j} z_j(\cdot)\right),
\end{equation}
that sets $\beta_j(\cdot)$ to a constant non-zero coefficient, a spatially varying process $( \beta_0 + z_j(\cdot))$ or a zero image. Since the SGLSS prior does not distinguish various types of included coefficient images, such as a non-zero constant versus a spatially varying process, it cannot fully recover the prior of \cite{Reich2010}. However, it can distinguish zero images, when $\pi_j = 0$ with all indicators $\tau_j(\vec{s}) = 0$, and full images, when $\pi_j = 1$ with all indicators $\tau_j(\vec{s}) = 1$.

We also mention the closely related bi-level selection priors for covariates with group structure, see \cite{Vannucci2011, Xu2015, Chen2016, Liquet2017}. These priors also deal with a two-level selection, but the global level is a group of covariates and the local level is the single covariate in the group. For example, \cite{Vannucci2011} proposed a linear regression model for identifying pathways, i.e, groups of genes, related to a particular phenotype, and a two-layer selection of pathways and genes. The two selection priors, however, are not linked and this leads to the necessity of constraining the prior set of possible configurations, to avoid selection of an empty group.
\cite{Chen2016} addressed this issue by adding an indicator correction procedure to do post-inference for group selection. \cite{Xu2015} and \cite{Liquet2017} followed the idea of Bayesian group lasso and conducted inference based on posterior median estimators of coefficients. Unlike these constructions, the proposed SGLSS prior construction naturally leads to  dependency between selections at the global and local levels, via the participation rate parameter, $\pi$, therefore preventing the selection of empty ``groups", i.e., covariates with no effects on the images. Furthermore, with the bi-level selection priors, a group is selected in the regression if at least one of its members has non-zero effect, while in the proposed SGLSS prior construction the selection at global level is based on the probability of a covariate to affect the image, as measured by the participation rate. 

\subsection{Inverse-Wishart Process prior}\label{sec:inverse.wishart}
For the covariance surface $\Sigma(\cdot, \cdot)$, pre-specified parametric kernels such as the Mat\'{e}rn or squared exponential kernel lack flexibility, and the possible misspecification may introduce considerable bias that hampers inference. We employ a flexible process, called the {\em Inverse-Wishart process} (IWP), to mitigate this concern. As a nonparametric generalization of the finite-dimensional Inverse-Wishart (IW) distribution, the IWP has been used in time series to capture time-varying {\em volatility} and {\em co-volatility} \citep{Philipov2006, Gouriroux2009, Wilson2011, Heaukulani2019}, and as a flexible prior for covariance kernels in functional data analysis \citep{Cox2016}. Existing literature, such as \cite{Dunson2016, Cox2016}, often uses a one-dimensional support; we instead define an IWP for a general $K$-dimensional index set $\bm{S}$ in the following sense. 

\begin{definition}\label{def.iwp}
An Inverse-Wishart process is a stochastic process $\Sigma = (\Sigma(\bm{s}, \bm{s}'): (\bm{s}, \bm{s}') \in \bm{S} \times \bm{S})$ indexed by $\bm{S} \times \bm{S}$ such that the random matrix $\Sigma(\vec{\bm{s}}, \vec{\bm{s}}) = ((\!\:\Sigma({\bm{s}}_i, {\bm{s}}_j)\!\:))_{i,j}$ possesses an Inverse-Wishart distribution for any $\vec{\bm{s}} = (\bm{s}_1, \ldots, \bm{s}_p)$ and $p \in \mathbb{N}$ with $\bm{s}_i \in \bm{S}$ for $i,j=1,\ldots,p$. 
\end{definition}

The matrix-valued $\Sigma(\vec{\bm{s}}, \vec{\bm{s}})$ in the definition are finite-dimensional marginals evaluated on $\vec{\bm{s}}$. An IW distribution is determined by two parameters: the degrees of freedom and a scale matrix that is symmetric and positive semi-definite. However, we shall follow the parameterization in \cite{Dawid1981}, denoted by $\text{IW}(\delta, \Psi(\vec{\bm{s}}, \vec{\bm{s}}))$ with $\delta$ a positive integer and scale matrix $\Psi(\vec{\bm{s}}, \vec{\bm{s}})$. This parameterization guarantees a crucial consistency property of IW after marginalization. Let $\Psi: \bm{S} \times \bm{S} \to \mathbb{R}$ be a symmetric and positive semi-definite mapping, i.e., the matrix $\Psi(\vec{\bm{s}}, \vec{\bm{s}}) = ((\!\:\Psi(\bm{s}_i, \bm{s}_j)\!\:))_{i, j=1, \ldots, p}$ is symmetric and positive semi-definite for any $\vec{\bm{s}} = (\bm{s}_1, \ldots, \bm{s}_p)$. By the Kolmogorov extension theorem, there exists an IWP for integer $\delta > 0$ and $\Psi$, which we denote by $\text{IWP}\left(\delta, \Psi(\cdot, \cdot) \right)$; see Lemma 2 in the Appendix of \cite{Dunson2016} for an elaborate proof when the index set is $\mathbb{N}$ and Proposition 1 in \cite{Cox2016} for a related discussion. We typically choose $\delta > 4$ to ensure marginals of an IWP have finite second moments.

We put this IWP prior on $\Sigma(\cdot,\cdot)$,
\begin{equation}\label{IWP.eq.1}
     \Sigma(\cdot, \cdot) \sim \text{IWP}\left(\delta, \Psi(\cdot, \cdot)\right) ,
\end{equation}
and choose the Mat\'{e}rn covariance function for $\Psi(\cdot,\cdot)$, 
\begin{equation}\label{eq:matern.kernel}
    \begin{aligned}
     \Psi(\bm{s}, \bm{s}') & =  \text{Mat\'{e}rn}\left( || \bm{s} - \bm{s}' ||_{l_2}; \sigma^2_s, \rho, \nu \right), \quad \bm{s},\bm{s}' \in {\bm{S}} \\
        & =  \frac{\sigma^2_s}{\Gamma(\nu) 2^{\nu-1}} \left( \sqrt{2\nu}\frac{ || \bm{s} - \bm{s}' ||_{l_2} }{\rho}\right)^{\nu} K_{\nu} \left(\sqrt{2 \nu} \frac{|| \bm{s} - \bm{s}' ||_{l_2}}{\rho}\right) ,
     \end{aligned}
\end{equation}
where $||\cdot||_{l_2}$ is the $l_2$ norm.
For a given vector of locations $\vec{\bm{s}}$, the prior leads to
\begin{equation}\label{IWP.eq.2}
     \Sigma\left(\vec{\bm{s}}, \vec{\bm{s}}\right)    \sim \text{IW}(\delta, \Psi(\vec{\bm{s}}, \vec{\bm{s}})). 
\end{equation}
In the applications reported below, we fix $\nu = 5/2$, following \cite{Cox2016}, to have the analytical forms for both the Mat\'{e}rn kernel and its gradients, facilitating computation for large covariance matrices in image data, and choose the other two hyperparameters $(\sigma^2_s, \rho)$ by minimizing the mean square error between an empirical covariance estimate, obtained as the MUA estimate, and the Mat\'{e}rn$(\sigma^2_s, \rho, 5/2)$ kernel. 

\subsection{Posterior Inference} \label{sec.settings}
We derive an efficient Gibbs sampler for the proposed hierarchical model with SGLSS prior. Posterior sampling proceeds in three main steps as follows, with detailed derivations provided in the Supplement. 
\begin{itemize}
    \item \textbf{Update the BHM parameters $\left\{Z_i(\vec{\bm{s}})\right\}^n_{i=1}$ and $\sigma^2_\epsilon$ conditional on $\{\beta_j(\vec{\bm{s}})\}^q_{j=1} $ and $\Sigma(\vec{\bm{s}},\vec{\bm{s}})$}: \\
    Evaluated on locations $\vec{\bm{s}}$, Equations~\eqref{BHM.eq.1} and~\eqref{BHM.eq.2} yield 
    \begin{equation*}
        \begin{aligned}
        & Y_i(\vec{\bm{s}}) | Z_i(\vec{\bm{s}}), \sigma^2_\epsilon  \sim \text{MVN}( Z_i(\vec{\bm{s}}), \sigma^2_\epsilon I_p), \quad \sigma^2_\epsilon  \sim \text{Inverse-Gamma}(a_\epsilon, b_\epsilon),\\
        & Z_i(\vec{\bm{s}}) | \{\beta_j(\vec{\bm{s}})\}_{j=1}^q, \Sigma( \vec{\bm{s}},\vec{\bm{s}})  \sim \text{MVN}( \mu_i(\vec{\bm{s}}), \Sigma(\vec{\bm{s}},\vec{\bm{s}})), \quad \mu_i(\vec{\bm{s}}) = \beta_0(\vec{\bm{s}}) + \sum^q_{j=1}x_{ij} \beta_j(\vec{\bm{s}}). \\
        \end{aligned}
    \end{equation*}
    In view of conjugacy, we sample 
    $$ Z_i(\vec{\bm{s}}) | Y_i(\vec{\bm{s}}), \{\beta_j(\vec{\bm{s}})\}_{j=1}^q, \Sigma(\vec{\bm{s}},\vec{\bm{s}}), \sigma^2_\epsilon  \sim  MVN\left( \mu_{Z_i}, V_{Z_i}\right), 
    $$
    with $V_{Z_i} =  \left( \sigma^{-2}_\epsilon I_p + \Sigma(\vec{\bm{s}},\vec{ \bm{s}})^{-1}\right)^{-1}$ and $\mu_{Z_i} =  V_{Z_i} \left( \sigma^{-2}_\epsilon Y_i(\vec{\bm{s}}) + \Sigma(\vec{\bm{s}},\vec{\bm{s}})^{-1} \mu_i(\vec{\bm{s}})\right)$ independently for $i = 1, \ldots, n$, and
    \begin{equation*}
        \begin{aligned}
        & \sigma^2_{\epsilon} | \{Y_i(\bm{\vec{s}})\}_{i=1}^n, \{Z_i(\bm{\vec{s}})\}_{i=1}^n  \sim  \\ 
        & {\hskip4em\relax}\text{Inverse-Gamma}\left( a_\epsilon + \frac{np}{2},  b_{\epsilon}+ \frac{1}{2}\sum^n_{i=1} \left( Y_i(\vec{\bm{s}}) - Z_i(\vec{\bm{s}})\right)^T\left(Y_i(\vec{\bm{s}}) - Z_i(\vec{\bm{s}})\right) \right).
        \end{aligned}
    \end{equation*}
    \item \textbf{Update the SGLSS prior parameters $\left\{\beta_j(\vec{\bm{s}}),  \tau_j(\vec{\bm{s}}),\pi_j\right\}^q_{j=0}$ conditional on $\left\{Z_i(\vec{\bm{s}})\right\}^n_{i=1}$ and $\Sigma(\vec{\bm{s}},\vec{\bm{s}})$}:  \\
    In this step, we first sample the indicators $\left\{\tau_j(\vec{\bm{s}})\right\}^q_{j=1}$ and update $\left\{\pi_j\right\}^q_{j=1}$ to obtain selection indicators at both global and local levels. This is achieved via a blocked Gibbs strategy as in \cite{Reich2010} and location-wise Gibbs updates similar to \cite{Brown1998} and \cite{Smith2007}, adapted to the SGLSS prior. 
    
    We use a blocked Gibbs sampler with respect to each feature $j \geq 1$. Denote $\tilde{Z}_{ij}(\vec{\bm{s}}) = Z_i(\vec{\bm{s}}) - \sum_{j' \ne j} x_{ij'} \beta_{j'}(\vec{\bm{s}})$. Equations~\eqref{BHM.eq.2} and~\eqref{SGLSS.eq.2} lead to a location-wise model,
    \begin{equation} \label{BGBF.eq.1}
        \begin{aligned}
         \tilde{Z}_{ij}(\bm{s}) | & \tilde{\beta}_j(\bm{s}), \tau_j(\bm{s}) = 1, \Sigma(\bm{s},\bm{s}) \sim N( x_{ij}  \tilde{\beta}_j(\bm{s}), \Sigma(\bm{s},\bm{s})), \\    
         & \tilde{\beta}_j(\bm{s}) | \tau_j(\bm{s}) = 1 \sim N( \mu_{0j}(\bm{s}), \sigma^2_{0j}(\bm{s})), \\
         \tilde{Z}_{ij}(\bm{s}) | & \tau_j(\bm{s}) = 0, \Sigma(\bm{s},\bm{s}) \sim N( 0, \Sigma(\bm{s},\bm{s})).
        \end{aligned}
    \end{equation}
    The location-wise Bayes factor can be obtained by integrating out $\tilde{\beta}_j(\bm{s})$,
    \begin{equation} \label{BGBF.eq.2}
        \begin{aligned}
            & \theta_j(\bm{s}) = \\
            & \frac{ \prod^n_{i=1} p\left(  \tilde{Z}_{ij}(\bm{s}) | \tau_j(\bm{s}) = 0,  \Sigma(\bm{s},\bm{s}),\pi_j\right)p(\tau_j(\bm{s}) = 0 | \pi_j)}{\left\{ \int \prod^n_{i=1} p\left(  \tilde{Z}_{ij}(\bm{s}) | \tilde{\beta}_j(\bm{s}) , \tau_j(\bm{s}) = 1, \Sigma(\bm{s},\bm{s}),\pi_j\right)  p(\tilde{\beta}_j(\bm{s})) d\tilde{\beta}_j(\bm{s}) \right\}p(\tau_j(\bm{s}) = 1 | \pi_j)} \\ \\
            = & \\
            & \frac{1-\pi_j}{ \pi_j \times \left( \sigma^2_{0j}(\bm{s})\right)^{-\frac{1}{2}}\exp\left\{ - \frac{1}{2} \left(\mu_{0j}^2(\bm{s}) /\sigma^{2}_{0j}(\bm{s})\right)\right\}   \times \left(  \tilde{\nu}_{j}(\bm{s})\right)^{\frac{1}{2}} \exp\left\{ \frac{1}{2} \tilde{m}^2_{j}(\bm{s})\tilde{\nu}_{j}(\bm{s}) \right\}  }, 
        \end{aligned}
    \end{equation}
    with 
    \begin{equation*}
    \begin{aligned} 
    \tilde{\nu}_{j}(\bm{s})  & =  \left[ \sum^n_{i=1}x^2_{ij}/\Sigma(\bm{s},\bm{s}) + 1/\sigma^{2}_{0j}(\bm{s}) \right]^{-1}, \\ \tilde{m}_{j}(\bm{s}) & =  \sum^n_{i=1} x_{ij}\tilde{Z}_{ij}(\bm{s}) / \Sigma(\bm{s},\bm{s}) + \mu_{0j}(\bm{s})/\sigma^{2}_{0j}(\bm{s}).
    \end{aligned}
    \end{equation*}
    This Bayes factor allows us to sample local selection indicators $\tau_j(\vec{\bm{s}})$ and participation rates $\pi_j$ from the conditional posterior distributions
    \begin{equation}\label{BGBF.eq.3}
        \begin{aligned} 
        & \tau_{j}(\bm{s}) | \{\bm{\beta}_{j'}(\vec{\bm{s}})\}_{j'\ne j}, \{Z_i(\vec{\bm{s}})\}^n_{i=1},\Sigma(\vec{\bm{s}},\vec{\bm{s}}),  \pi_j  \sim \text{Bernoulli}\left( \frac{1}{1 + \theta_{j}(\bm{s})}\right),\\
        & \pi_j | \tau_{j}(\vec{\bm{s}})  \sim \text{Beta}\left( a_{\pi_j} + \sum_{\bm{s} \in \vec{\bm{s}}} \tau_{j}(\bm{s}), b_{\pi_j} + p - \sum_{\bm{s} \in \vec{\bm{s}}} \tau_{j}(\bm{s})\right),
        \end{aligned}
    \end{equation}
    which also gives samples of global selection indicators $I\left( \pi_j \ge d \right)$. When $j = 0$, both $\tau_0(\vec{\bm{s}})$ and $\pi_j$ are fixed at $1$. 
    
    Conditional on selection indicators at the two levels, we sample the coefficient image $\beta_j(\vec{\bm{s}})$ for $j \geq 0$ as follows.
    If $\tau_j(\bm{s}) \times I\left(\pi_j \ge d \right) = 0$, we set $\beta_j(\bm{s})  = 0$; otherwise, using Equation (\ref{BGBF.eq.1}) we sample $\tilde{\beta}_j(\bm{s})$ from
        \begin{equation*}
            \tilde{\beta}_j(\bm{s}) | \{\bm{\beta}_{j'}(\vec{\bm{s}})\}_{j'\ne j}, \{Z_i(\vec{\bm{s}})\}^n_{i=1},\Sigma(\vec{\bm{s}},\vec{\bm{s}}) \sim N(\tilde{\nu}_{j}(\bm{s}) \tilde{m}_{j}(\bm{s}),\tilde{\nu}_{j}(\bm{s})),
        \end{equation*}
    and set $\beta_j(\bm{s}) = \tilde{\beta}_j(\bm{s}).$ 
    Note that the covariate $x_{i0} = 1$ when sampling the intercept $\beta_0(\vec{\bm{s}})$. The joint update of coefficient images and selection indicators avoids  reversible jump \citep{Savitsky2011}.
    
    \item \textbf{Update the IWP prior parameter $\Sigma(\vec{\bm{s}},\vec{\bm{s}})$ conditional on  $\left\{Z_i(\vec{\bm{s}})\right\}^n_{i=1}$ and $\{\beta_j(\vec{\bm{s}})\}^q_{j=0}$}: 
    
    Equations~\eqref{BHM.eq.2} and~\eqref{IWP.eq.2} lead to the conditional posterior distribution,
    \begin{equation*}
     \begin{aligned}
        &\Sigma(\vec{\bm{s}},\vec{\bm{s}}) |  \{Z_i(\vec{\bm{s}})\}^n_{i=1},\{\beta_j(\vec{\bm{s}})\}^q_{j=0}   \sim  \\
        &  {\hskip5em\relax} \text{IW}\left( n +  \delta, \sum^n_{i=1}\left(Z_i(\vec{\bm{s}}) - \mu_i(\vec{\bm{s}})\right)\left( Z_i(\vec{\bm{s}}) - \mu_i(\vec{\bm{s}}) \right)^T  + \Psi(\vec{\bm{s}},\vec{\bm{s}})  \right),
        \end{aligned}
    \end{equation*}
    where $\mu_i(\vec{\bm{s}}) = \beta_0(\vec{\bm{s}}) + \sum^q_{j=1}x_{ij} \beta_j(\vec{\bm{s}}).$
\end{itemize}

\vskip 2mm
\noindent At each iteration of the MCMC algorithm, for each covariate, the local level informs the selection at the global level, as the local selection indicator is sampled first and used to calculate the participation rate. The covariate is then selected if the participation rate is greater than $d$. Meanwhile, at the next iteration, the participation rate serves as the prior in the Binomial-Beta conjugate update of the local level indicators, providing feedback from the local level at the previous iteration. At convergence, global-level selection is done by calculating the marginal posterior probabilities of inclusion (MPPIs) of $I(\pi_j \ge d)$. Following \cite{Barbieri2004}, we use the median probability model and include the covariate if $\MPPI>0.5$, i.e., if more than half of the posterior samples give $I(\pi_j \ge d) = 1$. Similarly, local-level selection for covariate $j$ is determined by thresholding the MPPIs of $I(\pi_j \ge d)\tau_j(\bm{s})$ at 0.5. Commonly used values can be specified for the sparsity parameter $d$, such as $d =0.05$ or $d = 0.1$, to induce a desired level of sparsity. See results from the applications below and the supplementary material. Given the selected covariates and locations, we estimate the corresponding $\beta_j({\bm{s}})$ via posterior means obtained from the MCMC samples. We also estimate $ \left\{Z_i(\vec{\bm{s}})\right\}^n_{i=1}$, $\sigma^2_{\epsilon},$ and  $\Sigma(\vec{\bm{s}}, \vec{\bm{s}})$ via the posterior means.

We recommend setting the initial value of $\{\beta_j(\vec{\bm{s}})\}^q_{j=0}$ at $\hat{\bm{\beta}}_{\text{MUA}}(\vec{\bm{s}})$, and the initial value of $\Sigma(\vec{\bm{s}}, \vec{\bm{s}})$ at $\Psi(\vec{\bm{s}}, \vec{\bm{s}})$. Aside from the initial values, the sampler needs very little tuning. We provide a python implementation where we have optimized the sampling of $Z_i(\cdot)$ and $\Sigma(\cdot, \cdot)$ by utilizing various matrix decompositions to avoid redundant matrix inversions through equivalent formulations, and further building upon  pytorch, which allows automated efficient large matrix operations.  We remark here that the independent prior specified in Equation~(\ref{SGLSS.eq.2}) allows a parallel update for the $\left(\beta_j(\bm{s}), \tau_j(\bm{s}) \right)$ parameters at each local pixel/voxel $\bm{s} \in \bm{S}$ in each iteration of the MCMC, therefore reducing the computational cost from a multivariate Gaussian, roughly $O(|\bm{S}|^3)$ for the inverse of covariance matrix, to $|\bm{S}|$ univariate Gaussian.

\section{Simulation Study}
\label{sec:sim}
In this section we conduct simulations to assess the performances of the proposed method, which we call BHM, and perform comparisons with alternative approaches.

\begin{figure}[!hbt]
    \centering
    \begin{subfigure}[b]{0.24\textwidth}
        \centering
        \includegraphics[width=1\textwidth]{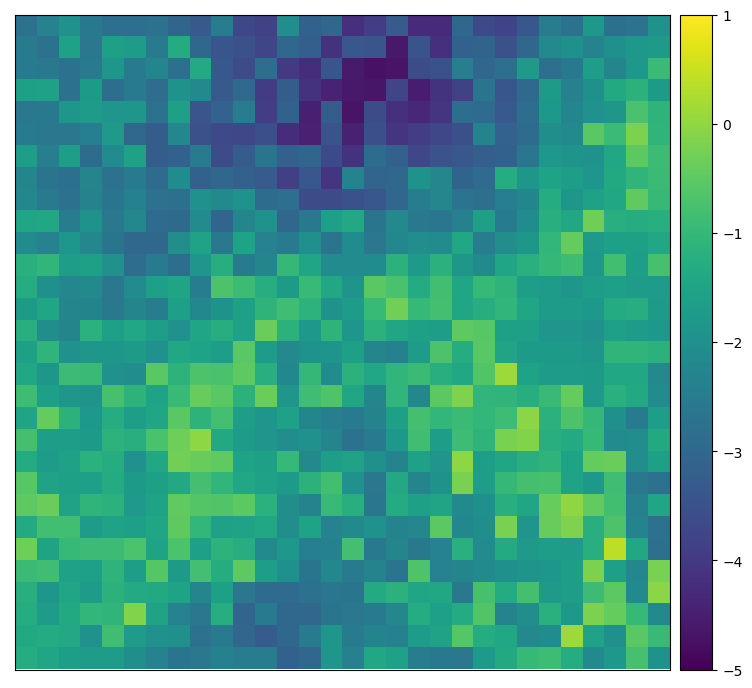}
    \end{subfigure}
    \begin{subfigure}[b]{0.75\textwidth}
        \centering
        \includegraphics[width=0.32\textwidth]{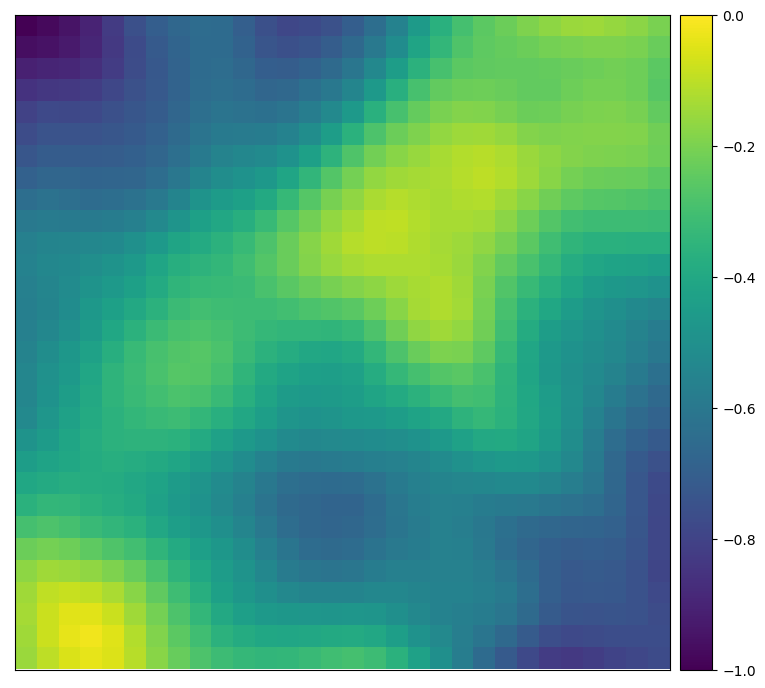}
        \includegraphics[width=0.32\textwidth]{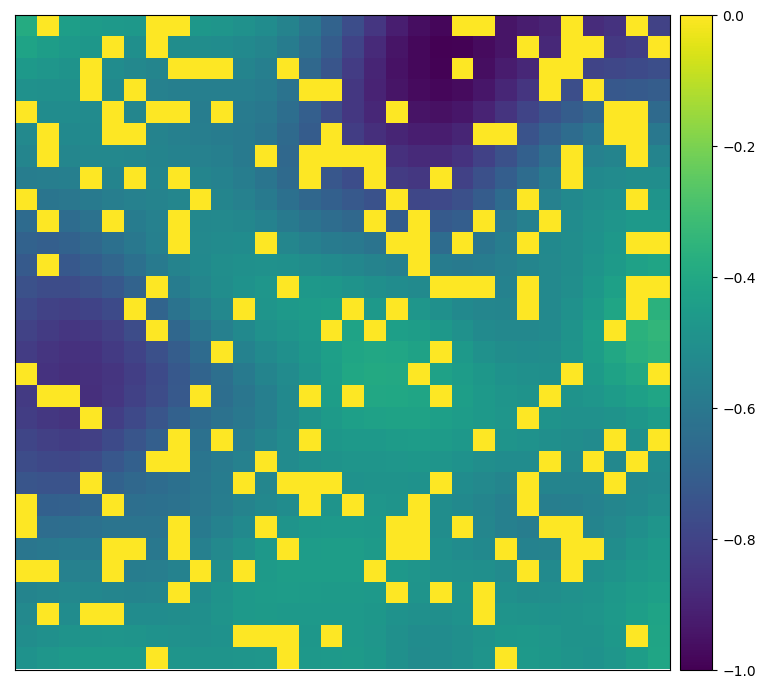}
        \includegraphics[width=0.32\textwidth]{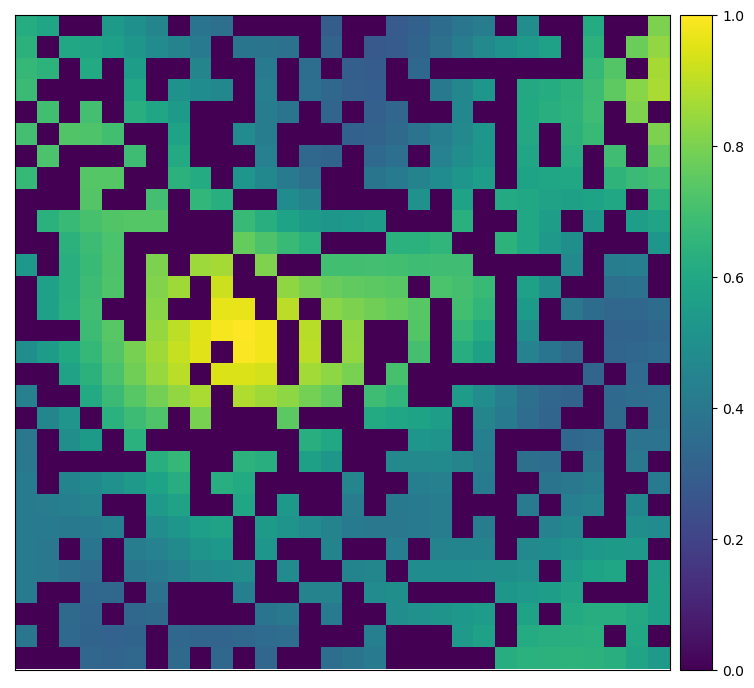}
    \end{subfigure}\\
    \begin{subfigure}[b]{0.24\textwidth}
        \centering
        \includegraphics[width=1\textwidth]{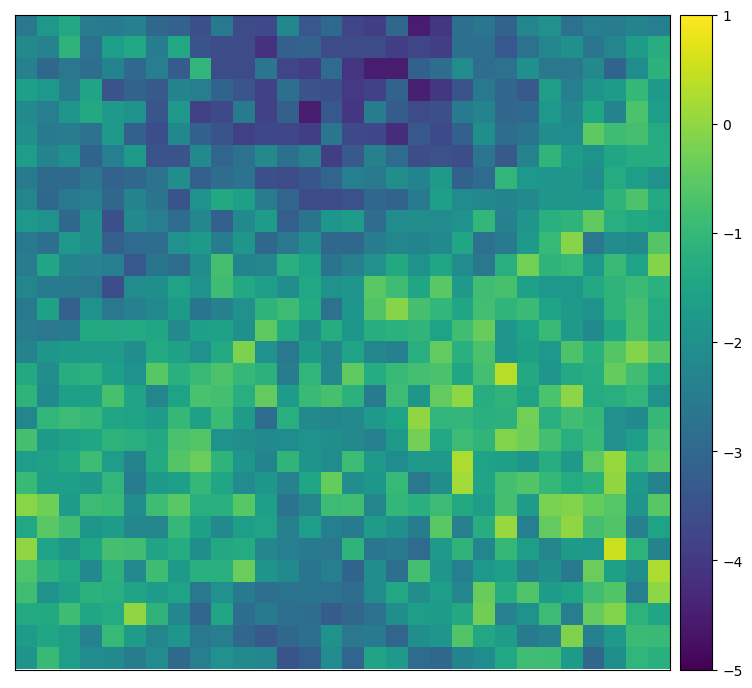}
    \end{subfigure}
    \begin{subfigure}[b]{0.75\textwidth}
        \centering
        \includegraphics[width=0.32\textwidth]{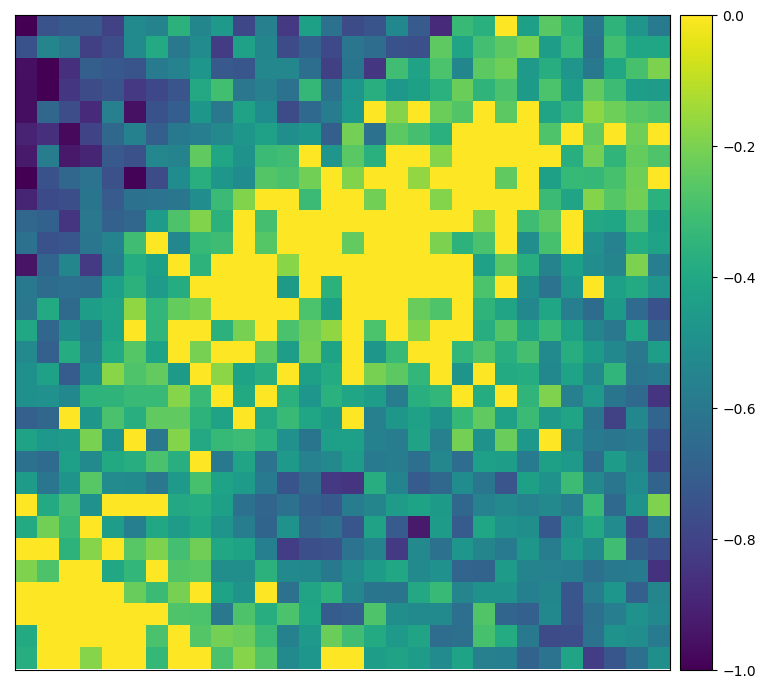}
        \includegraphics[width=0.32\textwidth]{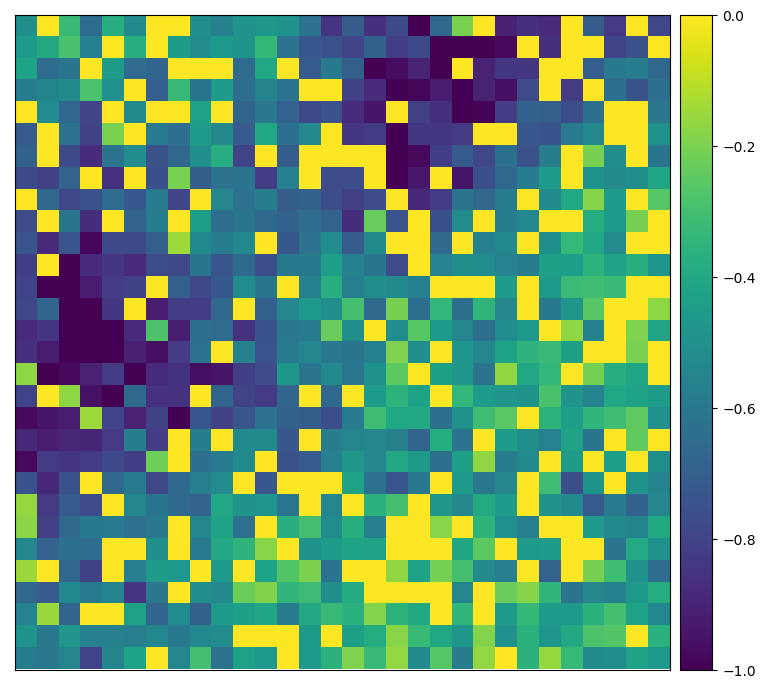}
        \includegraphics[width=0.32\textwidth]{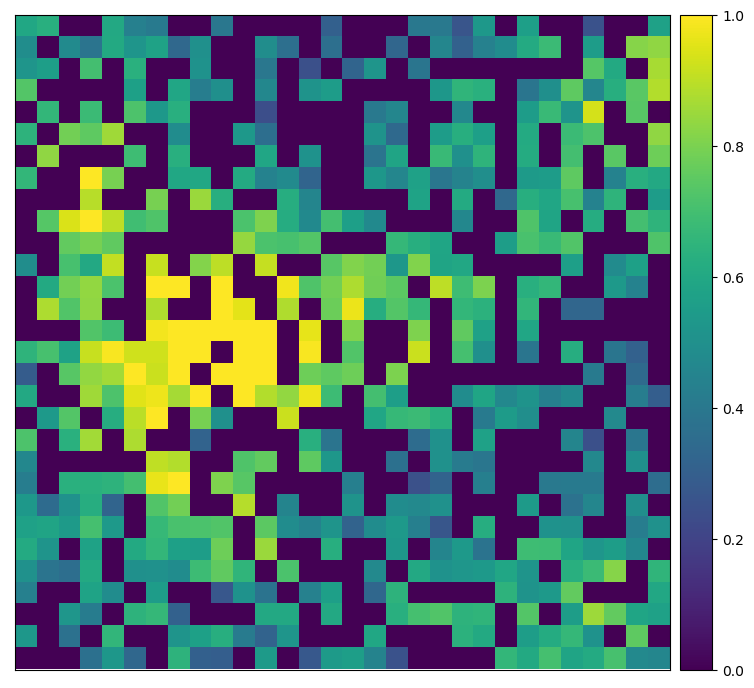}
    \end{subfigure}\\
        \begin{subfigure}[b]{0.24\textwidth}
        \centering
        \includegraphics[width=1\textwidth]{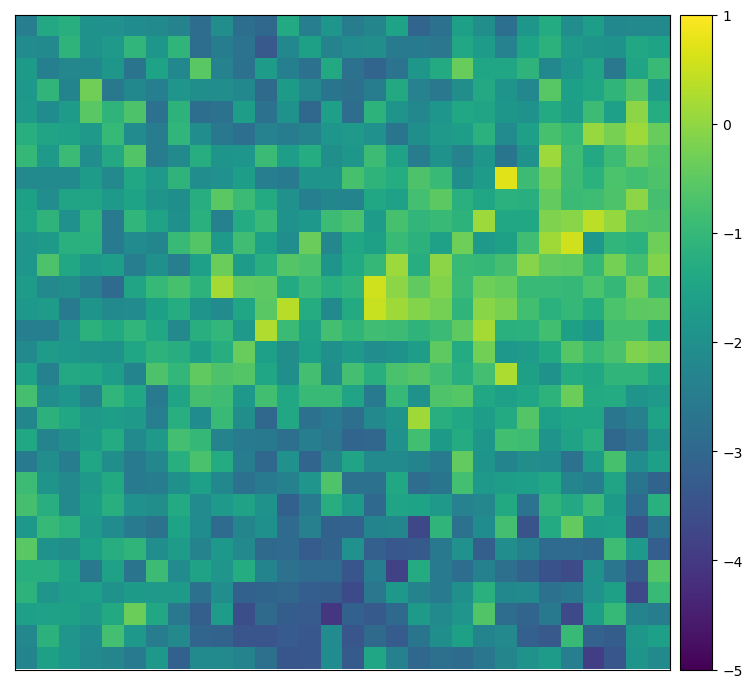}
    \caption*{  $Z(\vec{\bm{s}})$ }
    \end{subfigure}
    \begin{subfigure}[b]{0.75\textwidth}
        \centering
        \includegraphics[width=0.32\textwidth]{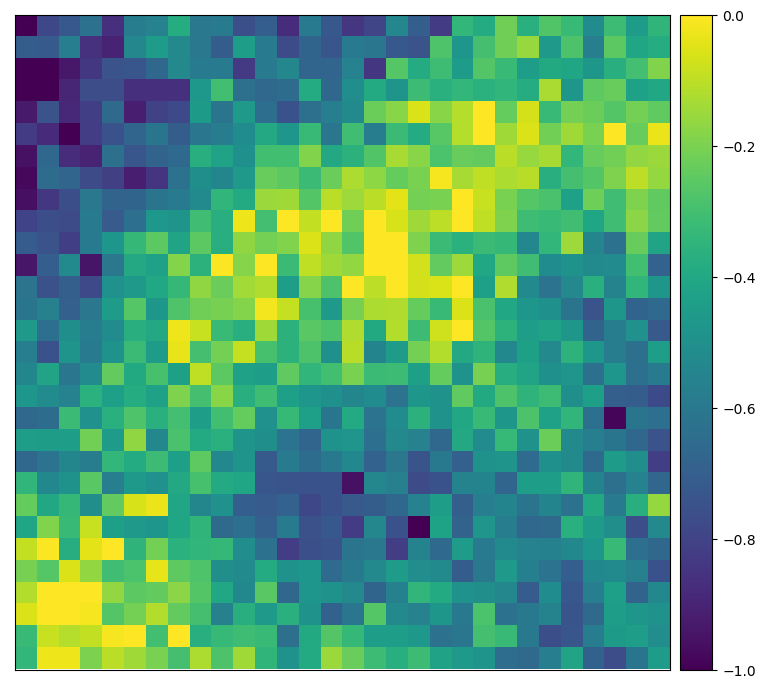}
        \includegraphics[width=0.32\textwidth]{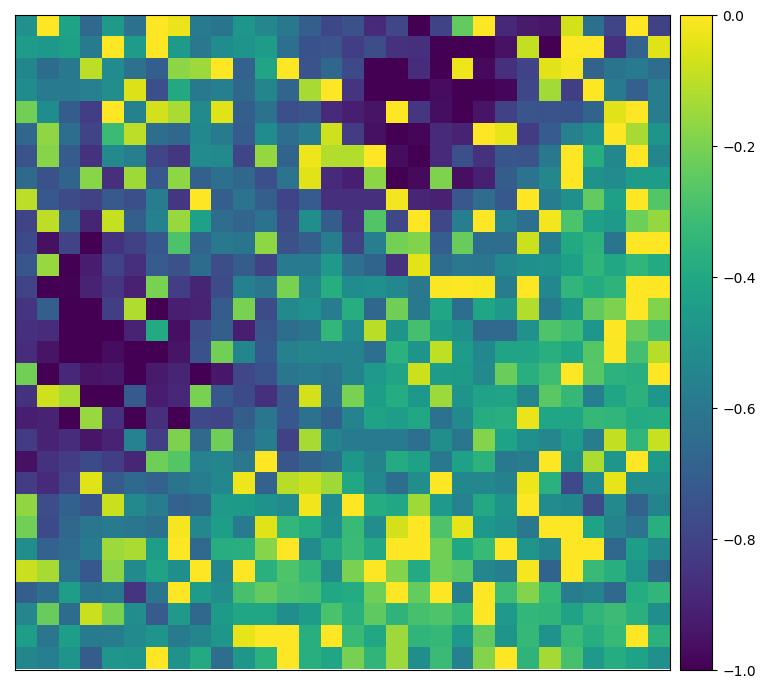}
        \includegraphics[width=0.32\textwidth]{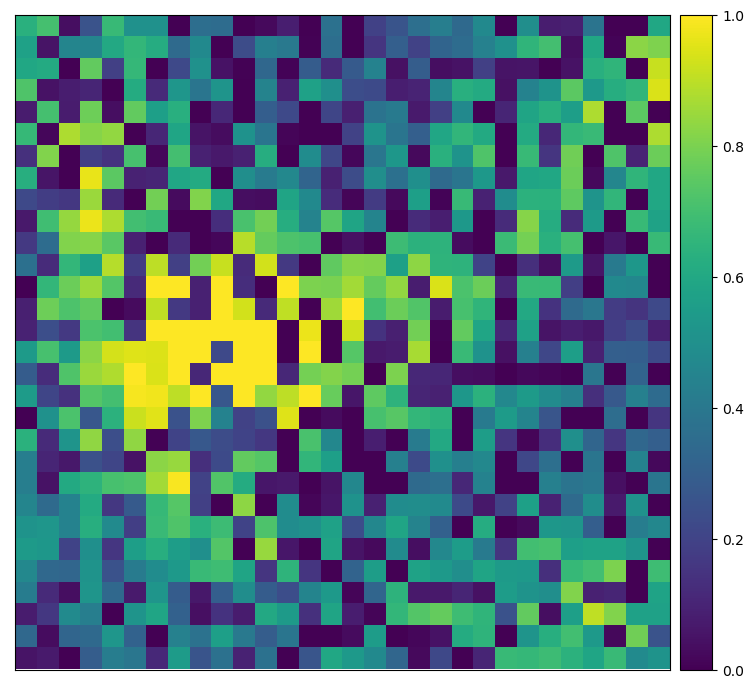}
    \caption*{ $\beta(\vec{\bm{s}})$ }
    \end{subfigure}
     \caption{First simulated scenario: Example images of $Z(\vec{\bm{s}})$ and $\beta(\vec{\bm{s}})$, where 1st row is generated data; 2nd row is the estimates of BHM; and 3rd row is the estimates of MUA.   }.
    \label{fig:sim.examples.1}
\end{figure}

\begin{figure}[!hbt]
    \centering
    \begin{subfigure}[b]{0.24\textwidth}
        \centering
        \includegraphics[width=1\textwidth]{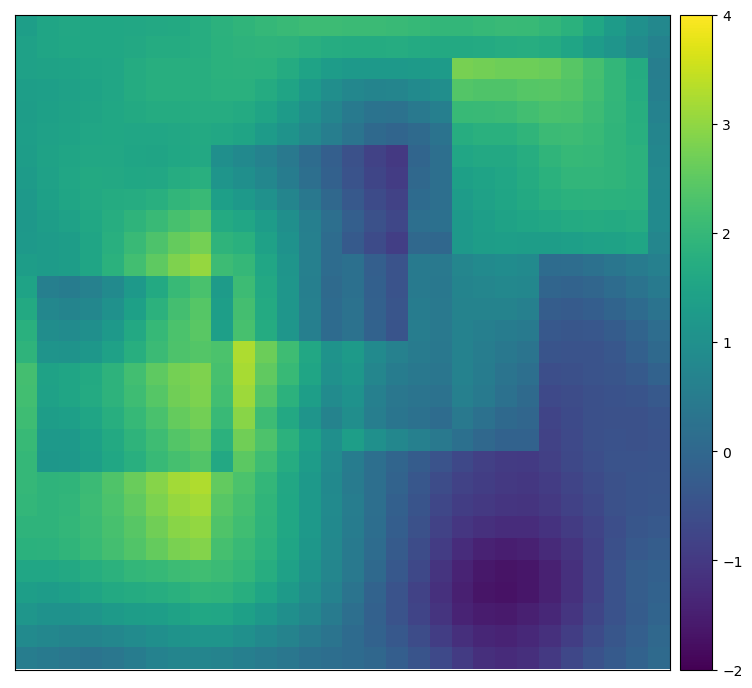}
    \end{subfigure}
    \begin{subfigure}[b]{0.75\textwidth}
        \centering
        \includegraphics[width=0.32\textwidth]{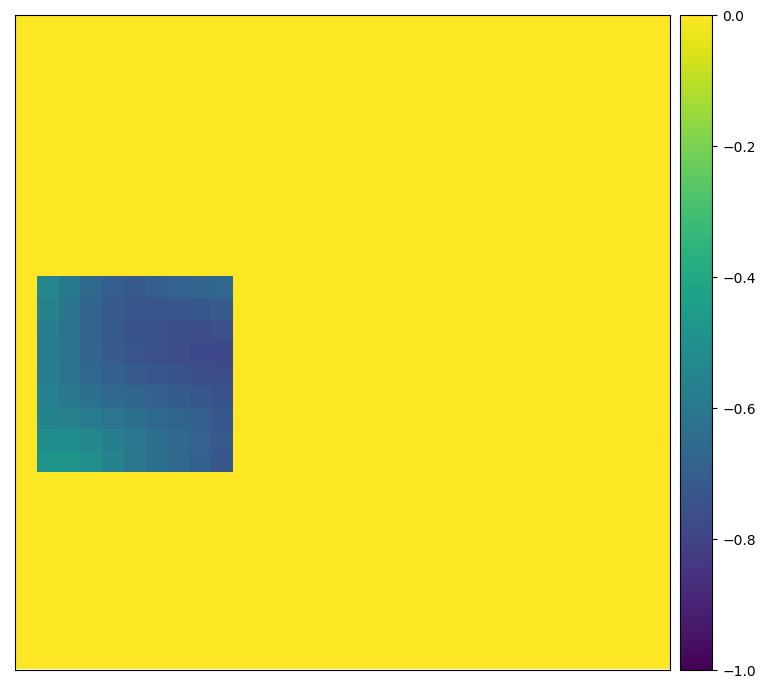}
        \includegraphics[width=0.32\textwidth]{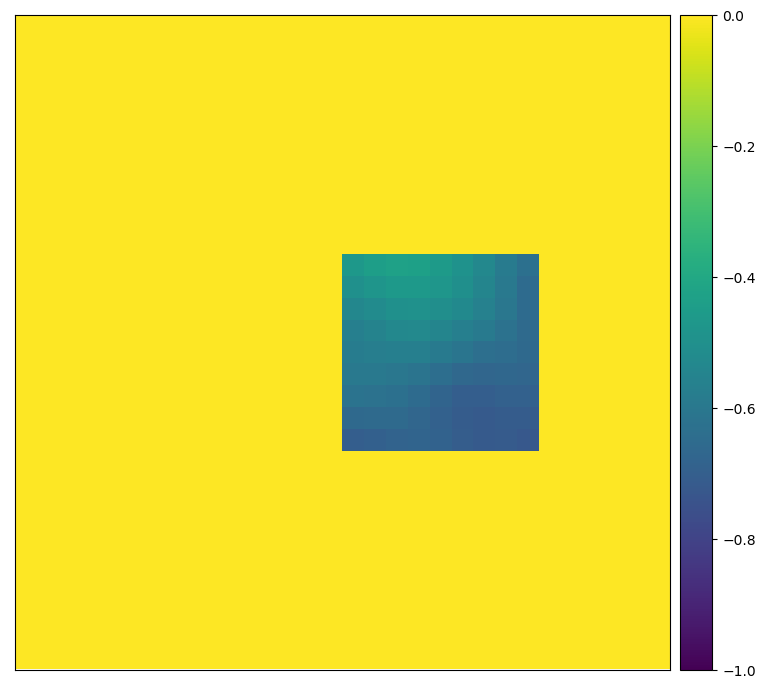}
        \includegraphics[width=0.32\textwidth]{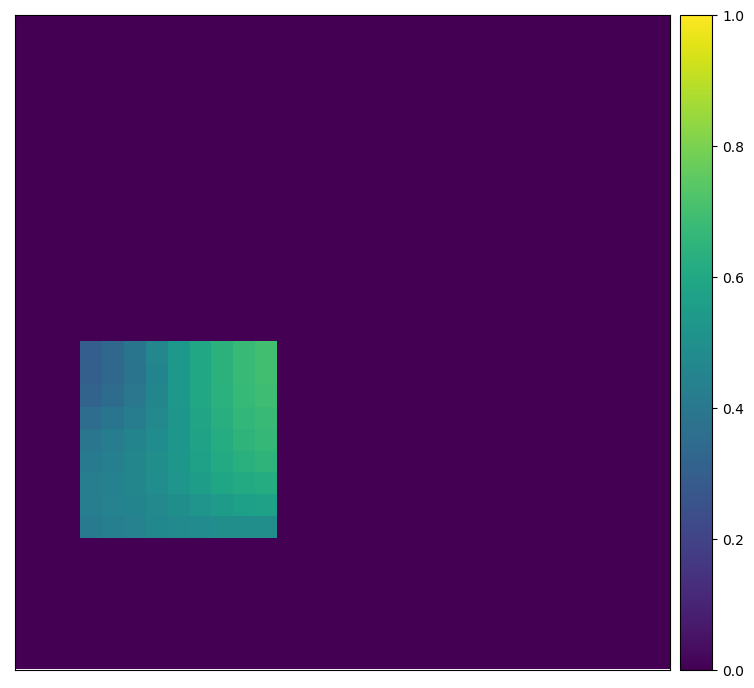}
    \end{subfigure}\\
    \begin{subfigure}[b]{0.24\textwidth}
        \centering
        \includegraphics[width=1\textwidth]{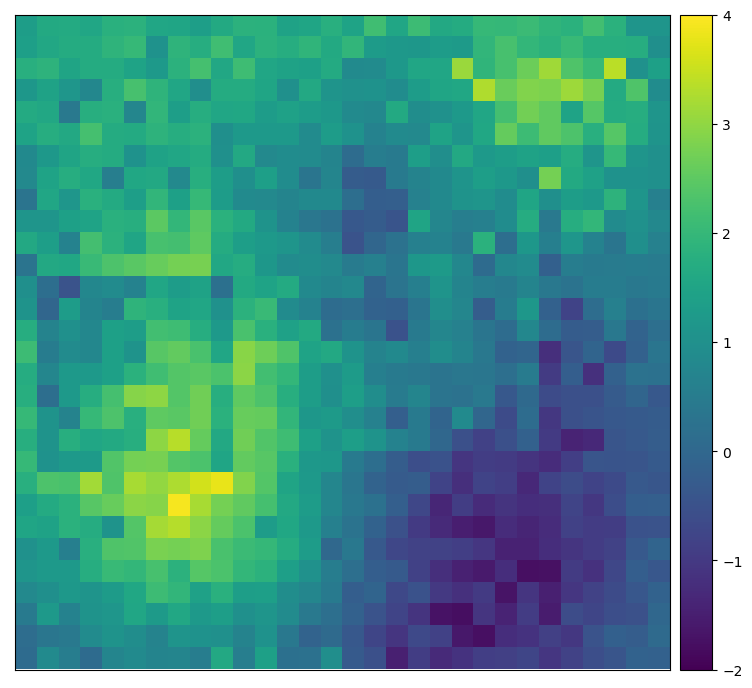}
    \end{subfigure}
    \begin{subfigure}[b]{0.75\textwidth}
        \centering
        \includegraphics[width=0.32\textwidth]{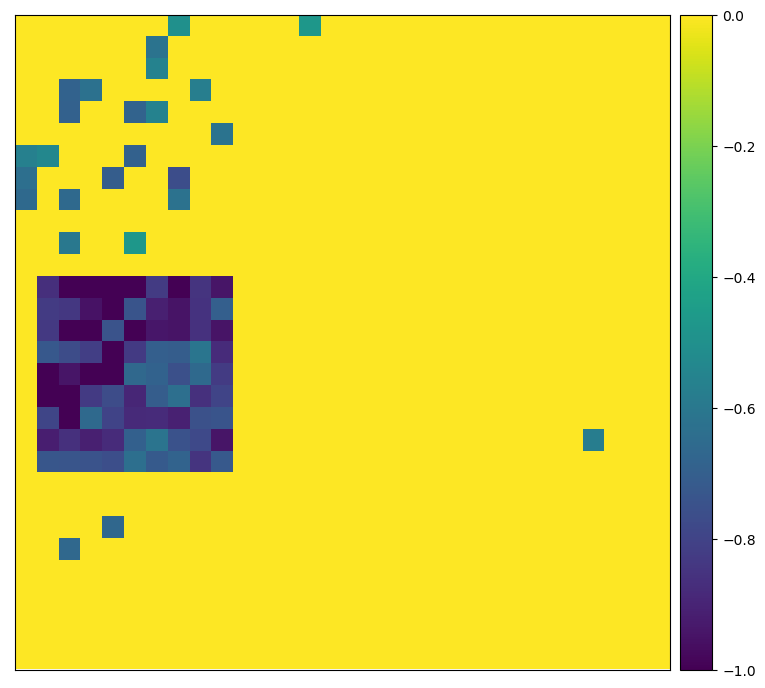}
        \includegraphics[width=0.32\textwidth]{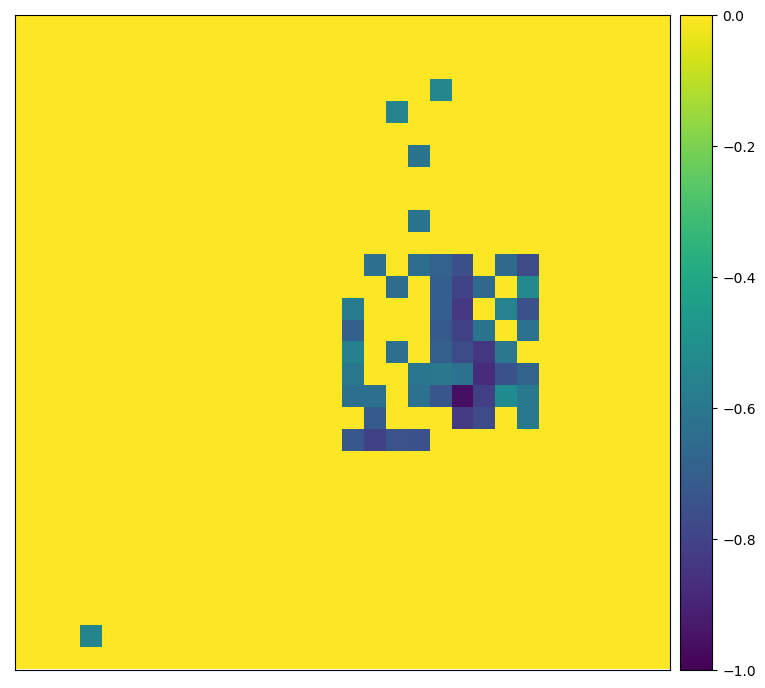}
        \includegraphics[width=0.32\textwidth]{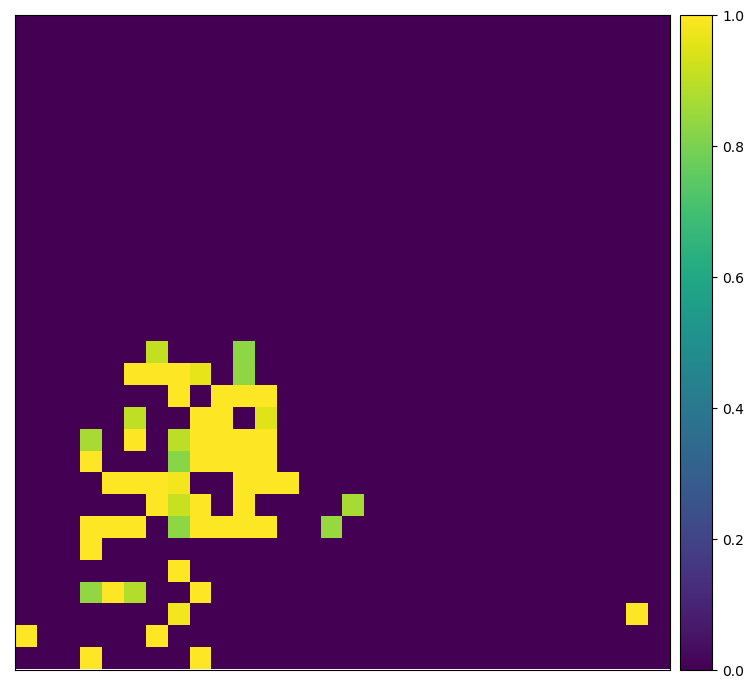}
    \end{subfigure}\\
        \begin{subfigure}[b]{0.24\textwidth}
        \centering
        \includegraphics[width=1\textwidth]{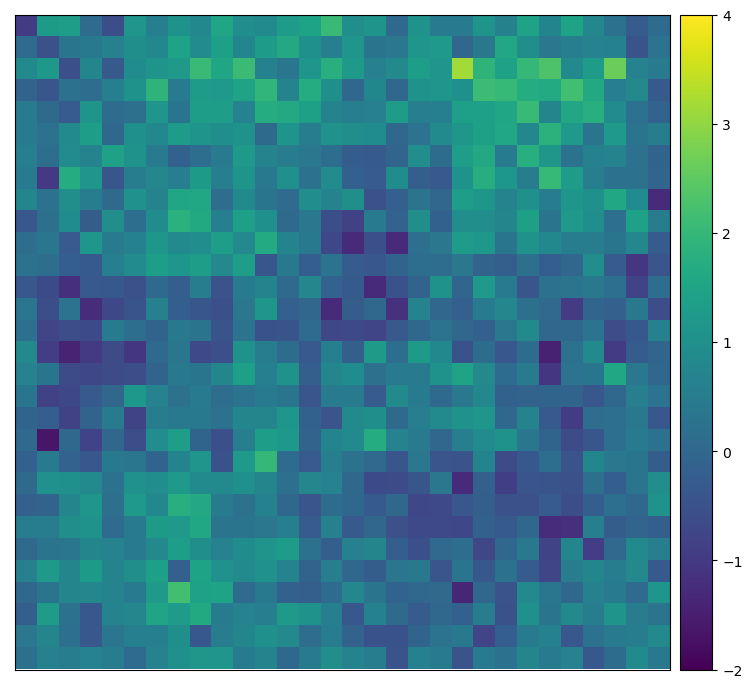}
    \caption*{  $Z(\vec{\bm{s}})$ }
    \end{subfigure}
    \begin{subfigure}[b]{0.75\textwidth}
        \centering
        \includegraphics[width=0.32\textwidth]{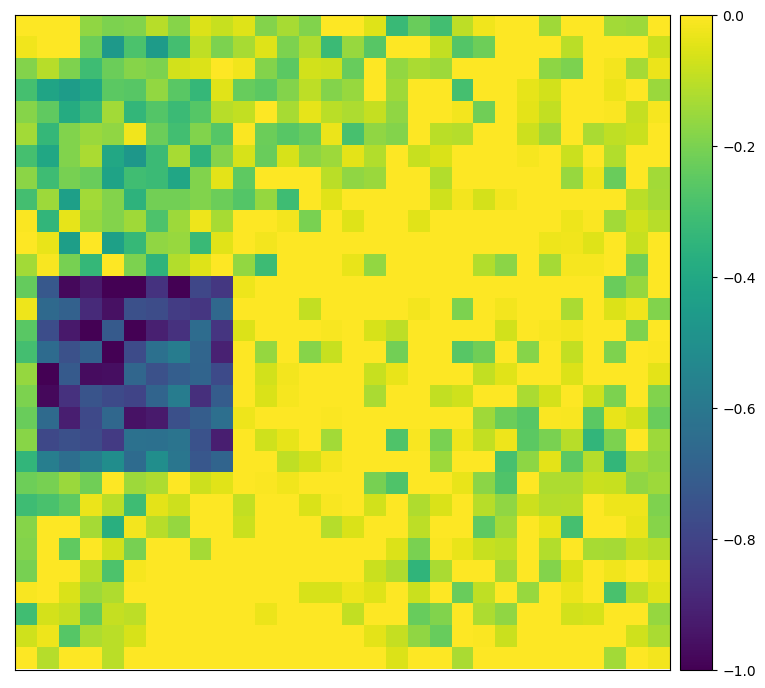}
        \includegraphics[width=0.32\textwidth]{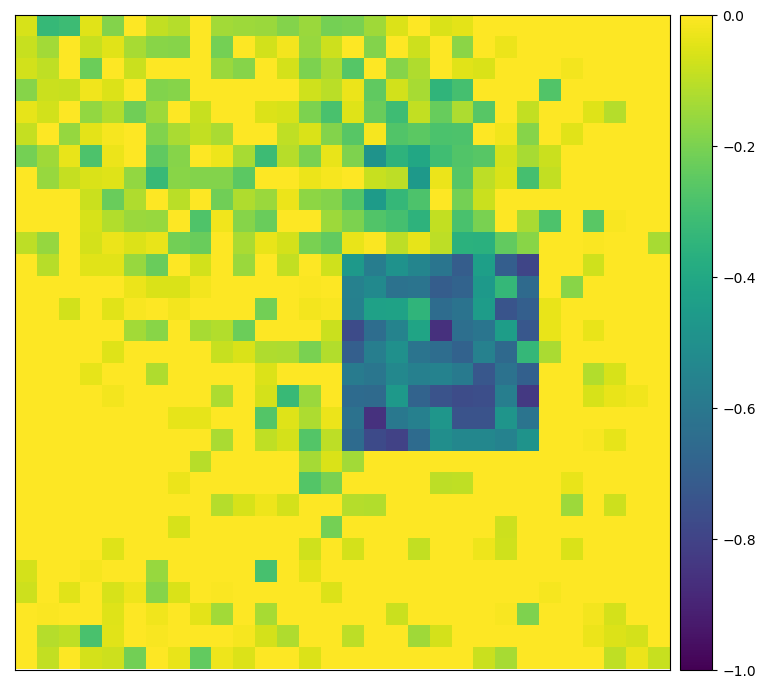}
        \includegraphics[width=0.32\textwidth]{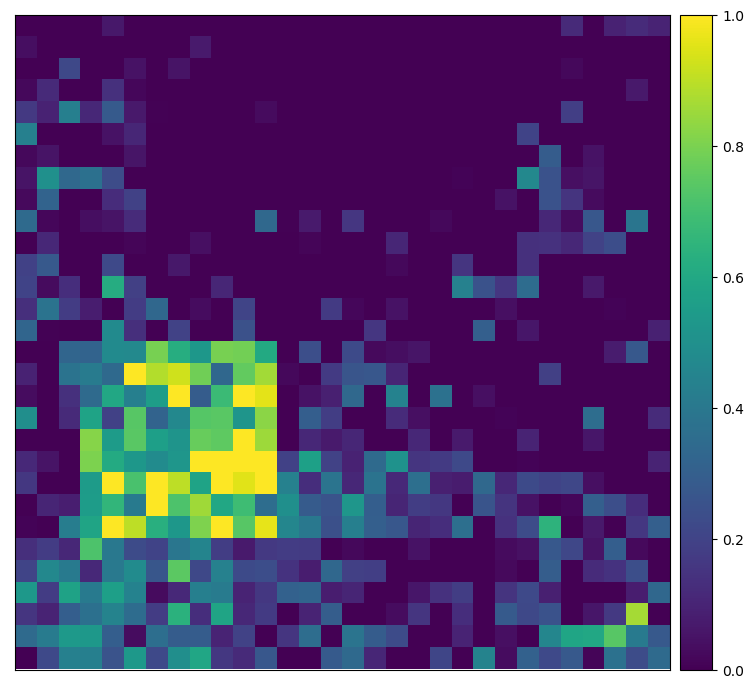}
    \caption*{ $\beta(\vec{\bm{s}})$ }
    \end{subfigure}
     \caption{Second simulated scenario $(\pi = 9\%)$: Example images of $Z(\vec{\bm{s}})$ and $\beta(\vec{\bm{s}})$, where 1st row is generated data; 2nd row is the estimates of BHM; and 3rd row is the estimates of MUA.   }.
    \label{fig:sim.examples.21}
\end{figure}

\begin{figure}[!hbt]
    \centering
    \begin{subfigure}[b]{0.24\textwidth}
        \centering
        \includegraphics[width=1\textwidth]{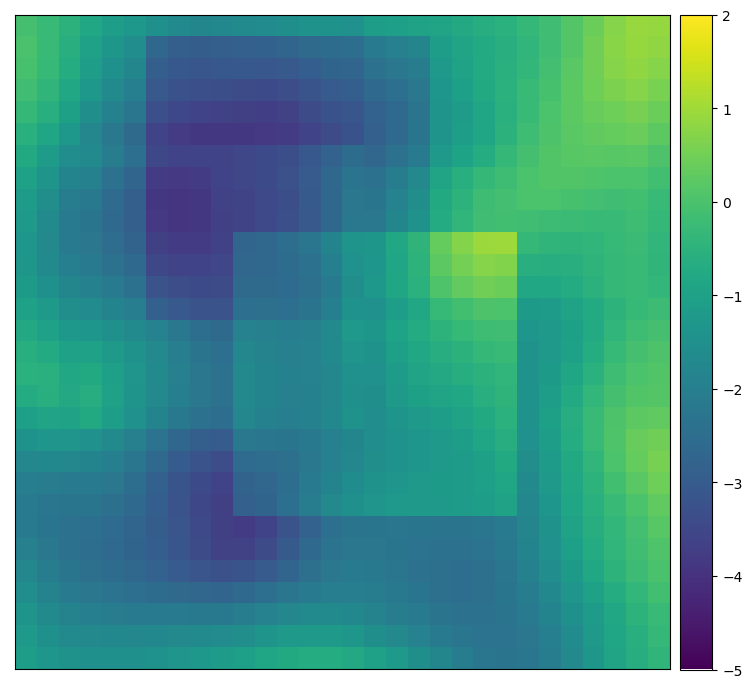}
    \end{subfigure}
    \begin{subfigure}[b]{0.75\textwidth}
        \centering
        \includegraphics[width=0.32\textwidth]{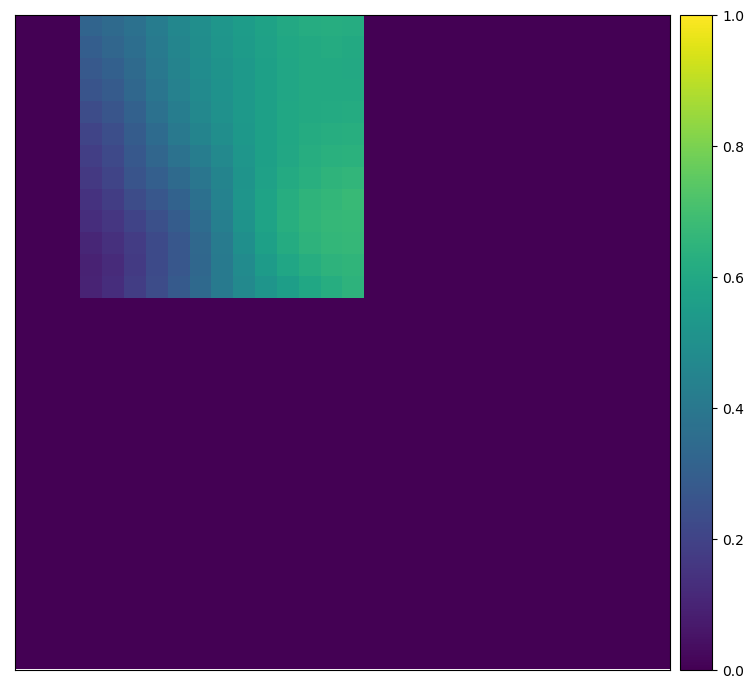}
        \includegraphics[width=0.32\textwidth]{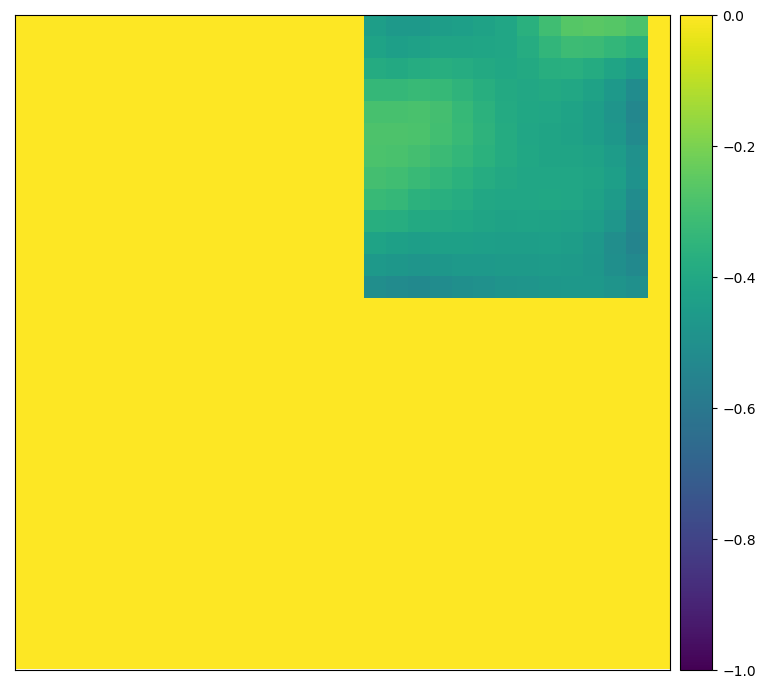}
        \includegraphics[width=0.32\textwidth]{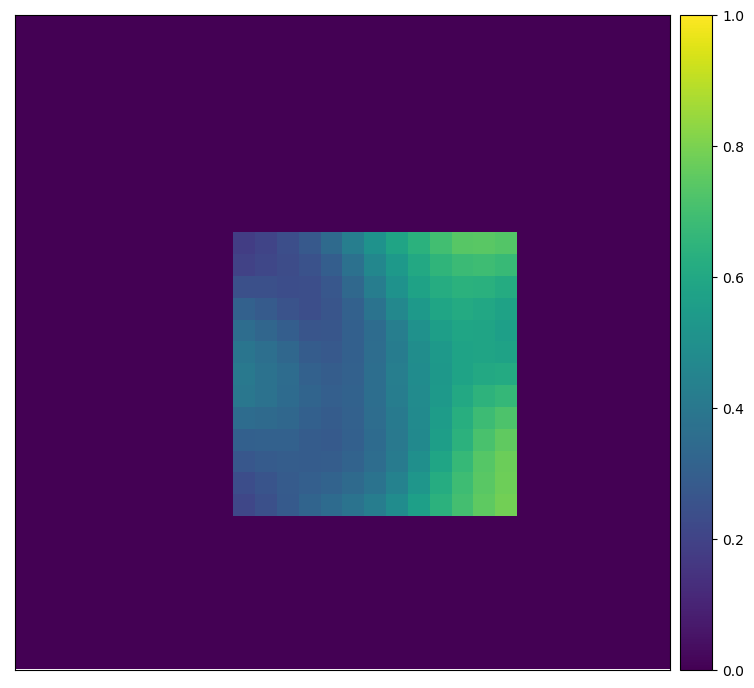}
    \end{subfigure}\\
    \begin{subfigure}[b]{0.24\textwidth}
        \centering
        \includegraphics[width=1\textwidth]{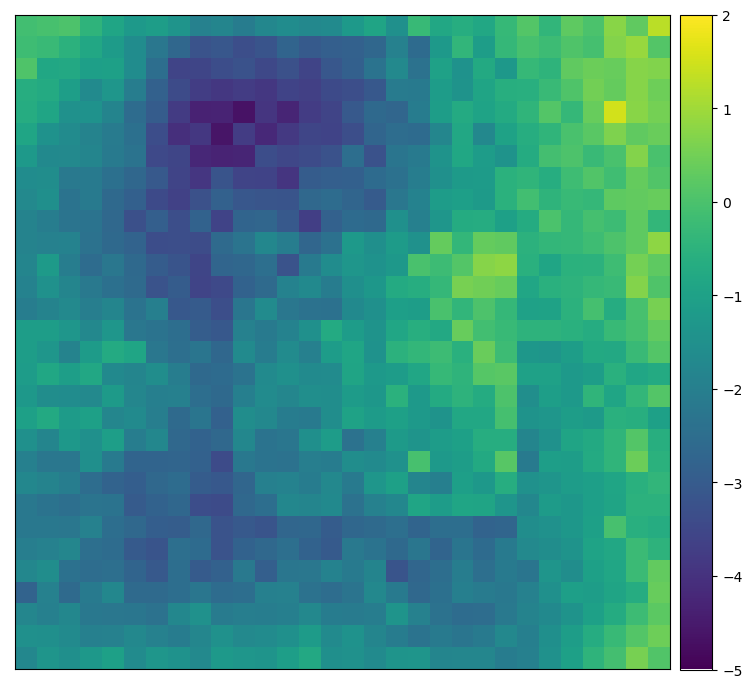}
    \end{subfigure}
    \begin{subfigure}[b]{0.75\textwidth}
        \centering
        \includegraphics[width=0.32\textwidth]{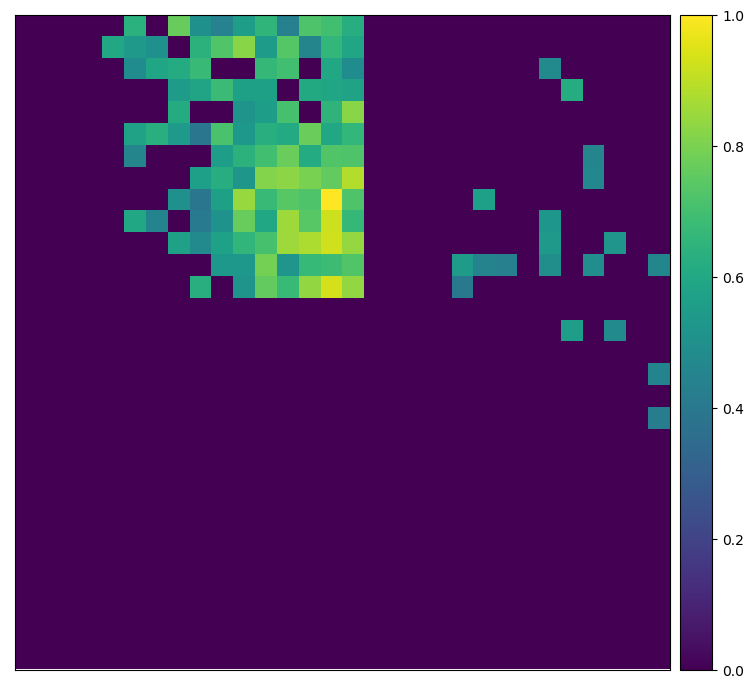}
        \includegraphics[width=0.32\textwidth]{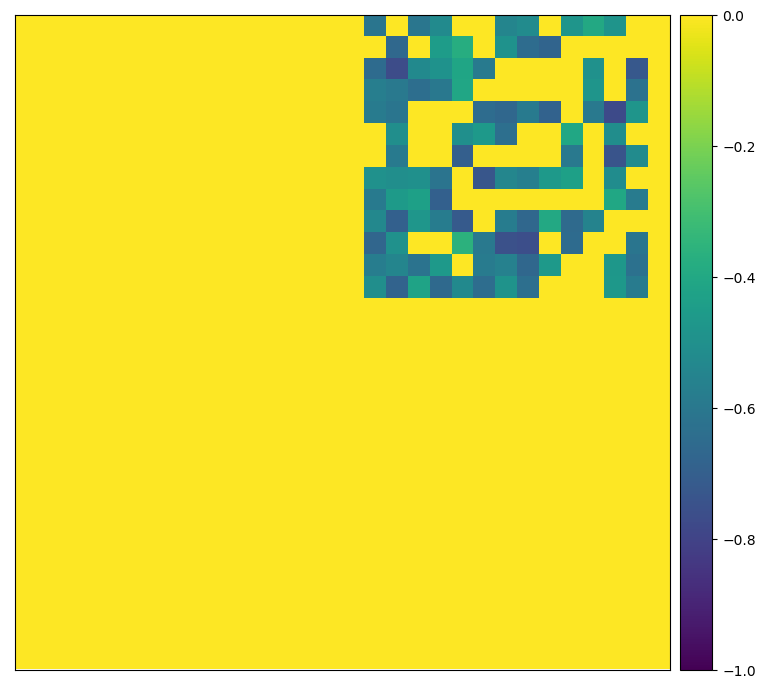}
        \includegraphics[width=0.32\textwidth]{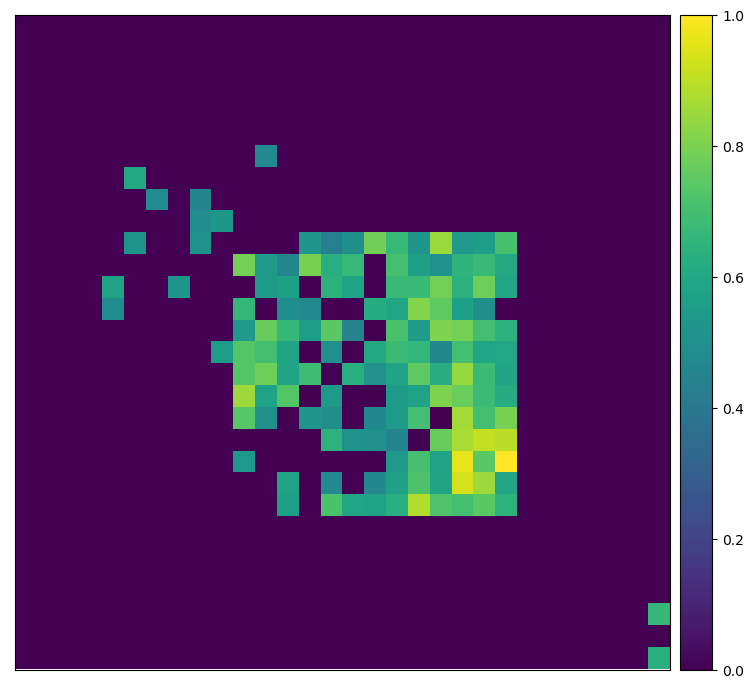}
    \end{subfigure}\\
        \begin{subfigure}[b]{0.24\textwidth}
        \centering
        \includegraphics[width=1\textwidth]{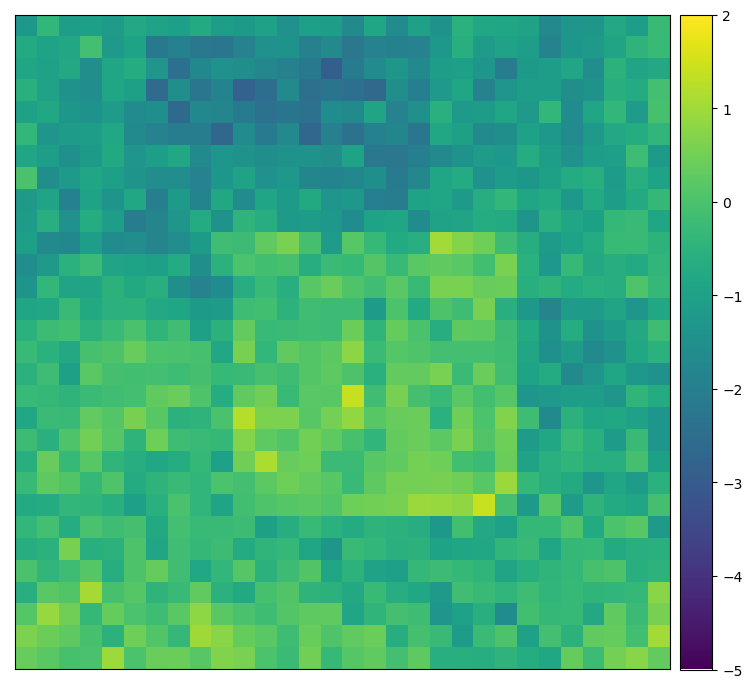}
    \caption*{  $Z(\vec{\bm{s}})$ }
    \end{subfigure}
    \begin{subfigure}[b]{0.75\textwidth}
        \centering
        \includegraphics[width=0.32\textwidth]{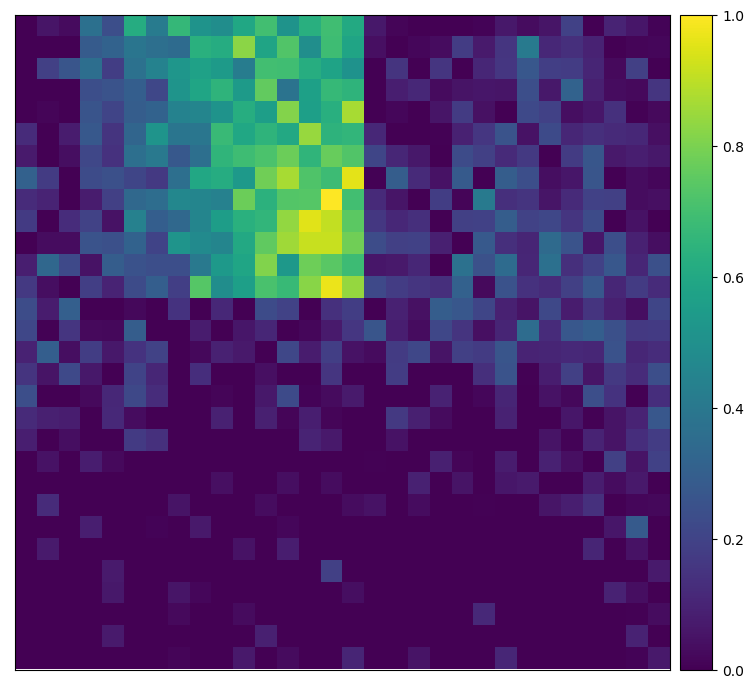}
        \includegraphics[width=0.32\textwidth]{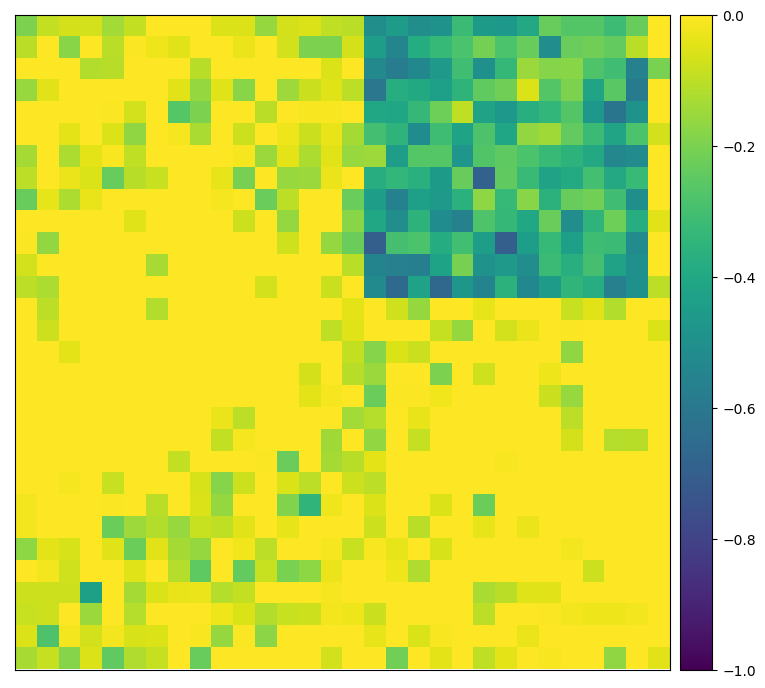}
        \includegraphics[width=0.32\textwidth]{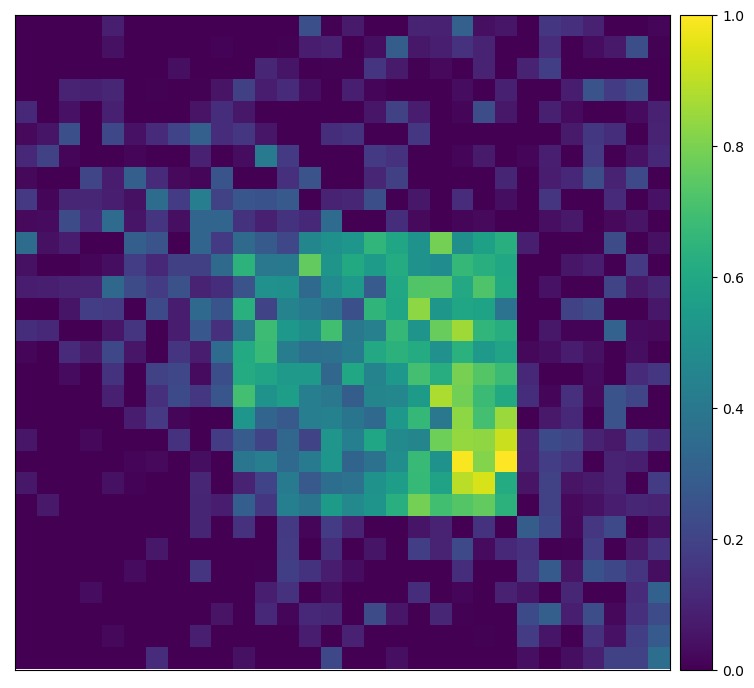}
    \caption*{ $\beta(\vec{\bm{s}})$ }
    \end{subfigure}
     \caption{Second simulated scenario $(\pi \approx 18.8\%)$: Example images of $Z(\vec{\bm{s}})$ and $\beta(\vec{\bm{s}})$, where 1st row is generated data; 2nd row is the estimates of BHM; and 3rd row is the estimates of MUA.   }.
    \label{fig:sim.examples.22}
\end{figure}

We set the number of covariates to $q=15$ and the sample size to $n =100$ and generate image data (i.e., $K=2$) from model \eqref{BHM.eq.1}-\eqref{BHM.eq.2} using a $30$-by-$30$ grid, i.e., $p = 900$. We sample the coefficient images $\{\beta_j(\vec{\bm{s}})\}^{15}_{j=1}$ and the  intercept $\beta_0(\vec{\bm{s}})$ similarly to \cite{Xinyi2020}. At covariate level, we induce sparsity by sampling $\tilde{\beta}_j(\vec{\bm{s}})$ for $j \in \{0,1,2,3,4,5,6,7,8\}$ from a $\mathcal{GP}(\bm{0}, \Sigma_{\beta})$, with $\Sigma_{\beta}$ specified by the Mat\'{e}rn kernel, and setting the remaining $\beta_j(\bm{s}) = \tilde{\beta}_j(\bm{s}) = 0, \forall \bm{s} \in \vec{\bm{s}}, j \in \{9,10,11,12,13,14,15\}$. At location level,  we first rescale  the images as
\begin{equation*}
    \beta_j(\vec{\bm{s}}) = \frac{ \tilde{\beta}_j(\vec{\bm{s}}) + \text{sign}( \tilde{\beta}_j(\bm{s}')) \left| \tilde{\beta}_j(\bm{s}') \right|}{2\left| \tilde{\beta}_j(\bm{s}') \right| }, ~~\bm{s}' = \argmax_{s \in \vec{\bm{s}}} \left| \tilde{\beta}_j(\bm{s}) \right|,
\end{equation*}
which excludes zeros introduced by randomness and then consider two different scenarios to introduce sparsity. In the first scenario we set $10\%$ randomly chosen elements of $\beta_{2}(\vec{\bm{s}})$ and $\beta_{7}(\vec{\bm{s}})$ to zero, $20\%$ randomly chosen elements of $\beta_{3}(\vec{\bm{s}})$ and $\beta_{8}(\vec{\bm{s}})$ to zero, $30\%$ randomly chosen elements of $\beta_{4}(\vec{\bm{s}})$ to zero and $40\%$ randomly chosen elements of $\beta_{5}(\vec{\bm{s}})$. The second scenario addresses a more realistic and challenging case, in which signals are clustered and influential covariates only affect a small portion of the images, hardly distinguishable from the noise covariates. In this case, we randomly select a square with $\pi$ percent pixels/voxels being non-zero. We consider two settings, $\pi_j \approx 10\%$ and $\pi_j \approx 20\%$ $(j = \{1,2,3,4,5,6,7,8\} )$.  Figures~\ref{fig:sim.examples.1},\ref{fig:sim.examples.21} and \ref{fig:sim.examples.22} show some example images of both noise-free images and coefficient images generated from the three scenarios.

Next, we generate the covariates $x_{i,j}$'s, including both continuous and discrete variables. We generate $x_{i,j}$ with $j = 1,2,3,4,5$ from $N(0,1)$, to obtain continuous features, and $x_{i,j}$, with $j = 6,7,8$, from Bernoulli$(0.5)$, for the discrete features. We add noisy features generated from $N(0,1)$, for $j=9,\ldots,15$. Finally, we sample the noise-free mean surface $Z_i(\vec{\bm{s}})$ from Equation~\eqref{BHM.eq.2} using a Mat\'{e}rn kernel for the covariance matrix $\Sigma$, and the image data $Y_i(\vec{\bm{s}})$ from Equation~\eqref{BHM.eq.1} with $\epsilon_{i,\bm{s}}$ sampled from $N(0,1)$ for $i=1,\ldots,n$.
Below we report results using Mat\'{e}rn kernels of the type
$\Sigma = \Sigma_\beta = $Mat\'{e}rn$(1, 1/4, 5/2)$. 

\subsection{Prior Specification}
\label{sec:priorspec}
For the prior specification, we use a weakly-informative prior on the participation rate parameters $\{\pi_j\}_{j=1}^{q}$ in Equation~\eqref{SGLSS.eq.2}, by setting $a_\pi = b_\pi=1$.  For the threshold $d$, we report the results for a conventional sparsity level, $d=0.05$, and then discuss sensitivity in the supplementary material. We center the slab distribution in Equation~\eqref{SGLSS.eq.2} at   $\{\mu_{0j}(\vec{\bm{s}})\}^q_{j=0} = 0$, as commonly done with spike-and-slab priors, and set  $\{\sigma^2_{0j}(\vec{\bm{s}})\}^q_{j=0} = 1$ (see supplementary material for a sensitivity analysis).  As previously discussed, we derive empirical estimates of the Mat\'{e}rn parameters $(\sigma^2_s,\rho)$ in Equation~\eqref{eq:matern.kernel} by minimizing the mean square error between the sample covariance estimate and the Mat\'{e}rn kernel. This provides a prior with the closest kernel to the empirical covariance matrix by Frobenius norm and prevents singularity issues caused by $n < p$.  We set $\delta=5$ for the IWP prior in Equation~\eqref{IWP.eq.1}, following \cite{Cox2016}. Finally, we set a weakly-informative Inverse-Gamma prior on the noise variance $\sigma^2_{\epsilon, \bm{s}}$ by setting $a_{\epsilon} = b_{\epsilon} = 1$.  

\subsection{Results}
All results we report were obtained by running MCMC chains with $2000$ iterations and $500$ burn-in. A single chain took around 8 minutes to run on a 6-core 2.6GHz Intel(R) core i7 CPU. For each chain, convergence was assessed by inspecting the MCMC traces, and more formally using the Geweke test \citep{Geweke1992} to check for signs of non-convergence of the individual parameters. As an example, the z-scores from the Geweke test were 0.9603 for $\left\{\tau_j(\vec{\bm{s}})\right\}^{15}_{j=1}$ and 1.0581 for $\left\{\pi_j \right\}^{15}_{j=1}$, clearly indicating that the MCMC chains were run for a sufficient number of iterations.

We evaluated performance for variable selection and parameter estimation. For variable selection, we calculated
\begin{equation*}
F_1 = 2\cdot \frac{\text{Precision}\cdot \text{Recall}}{\text{Precision} + \text{Recall}}, \quad \text{Precision} = \frac{TP}{TP + FP}, \quad \text{Recall} = \frac{TP}{TP + FN}.
\end{equation*}
For parameter estimation, we evaluated performances by calculating mean squared errors (MSEs) as
\begin{equation*}
    \text{MSE} =\frac{1}{|\bm{A}| } \sum_{\bm{a} \in \bm{A}} (F(\bm{a}) - \hat{F}(\bm{a}))^2,
\end{equation*}
where $F(\cdot)$ and $\hat{F}(\cdot)$ represent the true and estimated parameters, respectively, and $\bm{a}$ represents the vector of the related indices and/or locations, e.g. $i = 1,...,n; j = 1,...,q; \bm{s}, \bm{s}' \in \vec{\bm{s}}$. We report the accumulated MSE of all coefficient images $\left\{ \beta_j(\vec{\bm{s}}) \right\}^{15}_{j=0}$ as a summary measure of performance.

We first showcase inference from our BHM model on one simulated data set and then perform comparisons on 50 replicated data sets. Figures~\ref{fig:sim.examples.1},\ref{fig:sim.examples.21} and \ref{fig:sim.examples.22} show estimates from BHM of the example images for one data set from each of the three simulated scenarios. The MUA estimates are also shown, for comparison. Results show that the coefficient images estimates can capture the pixel-level information relatively well, even with the location-wise independent priors \eqref{SGLSS.eq.1}-\eqref{SGLSS.eq.2} for spatially-dependent coefficient images.

Tables~\ref{tab:sim.ogls} reports precision, recall and $F_1$ scores for both global and local selections and Table~\ref{tab:sim.omse} reports the MSEs for the parameters of interests. The proposed method performs well at the global level selection, leading to precision, recall and $F_1$ scores all relatively high. At the local level selection, some differences are noted among the different simulated scenarios, in particular in the second scenario with the sparser case $\pi = 9\%$, as in this scenario influential covariates are closer to noisy ones. Also, results vary with the covariates' types, with coefficient images for discrete covariates being  challenging for local selection, as shown by the lower recalls and $F_1$ scores. The MSEs of all parameters of interests are relatively small, demonstrating that BHM can estimate those parameters relatively well. Results for BHM with $d = 0.01$ and $ d=0.1$ are reported in the supplementary material. As expected, as $d$ increases, precision tends to increase and recall tends
to decrease. However, when there exists a ‘good separation’ between the influential and noisy covariates, like
in the first scenario and the second scenario with $\pi=$ 18.8\%, good performances overall
can be observed for different choices of $d$.

\begin{table}[!hbt]
\caption{BHM ($d=0.05$): Global-local selection for a representative dataset}
\label{tab:sim.ogls}
\centering
\begin{adjustbox}{max width = \textwidth}
\begin{tabular}{ccccccccccccc}
 \hline
 \hline
 \multicolumn{13}{c}{First simulated scenario} \\
 \hline
 \multicolumn{2}{c}{} && Global && \multicolumn{8}{c}{Local} \\ 
 \hline 
 & &&  && $\tau_1(\vec{\bm{s}})$ & $\tau_2(\vec{\bm{s}})$ & $\tau_3(\vec{\bm{s}})$ & $\tau_4(\vec{\bm{s}})$ & $\tau_5(\vec{\bm{s}})$ & $\tau_6(\vec{\bm{s}})$ & $\tau_7(\vec{\bm{s}})$ & $\tau_8(\vec{\bm{s}})$ \\  \\[-1em]
 \cline{4-4} \cline{6-13}   \\[-1em]
\multicolumn{2}{c}{ Precision } && $1$ &&  $1   $ & $0.901$ & $0.973$ & $0.985$ & $0.981$ & $1   $ & $0.949$ & $0.958$ \\
\multicolumn{2}{c}{ Recall } && $1$ && $0.821$ & $1   $ & $0.983$ & $0.917$ & $0.944$ & $0.742$ & $0.823$ & $0.769$ \\
\multicolumn{2}{c}{ $F_1$ scores } && $1$ && $0.902$ & $0.948$ & $0.978$ & $0.950 $ & $0.962$ & $0.852$ & $0.882$ & $0.854$ \\
\hline
\hline
\multicolumn{13}{c}{Second simulated scenario  ($\pi = 9\%$) } \\
 \hline
 \multicolumn{2}{c}{} && Global && \multicolumn{8}{c}{Local} \\ 
 \hline 
 & &&  && $\tau_1(\vec{\bm{s}})$ & $\tau_2(\vec{\bm{s}})$ & $\tau_3(\vec{\bm{s}})$ & $\tau_4(\vec{\bm{s}})$ & $\tau_5(\vec{\bm{s}})$ & $\tau_6(\vec{\bm{s}})$ & $\tau_7(\vec{\bm{s}})$ & $\tau_8(\vec{\bm{s}})$ \\  \\[-1em]
 \cline{4-4} \cline{6-13}   \\[-1em]
\multicolumn{2}{c}{ Precision } && $1$ &&  $0.743$ & $0.673$ & $0.714$ & $0.551$ & $  -  $ &   $0.505$ & $0.276$ & $  -  $  \\
\multicolumn{2}{c}{ Recall }  && $0.75$ &&  $1    $ & $0.432$ & $0.617$ & $0.728$ & $0    $ & $0.568$ & $0.840 $ & $0    $ \\
\multicolumn{2}{c}{ $F_1$ scores } && $0.857$ && $0.853$ & $0.526$ & $0.662$ & $0.628$ & $  -  $ &   $0.535$ & $0.416$ & $  -  $  \\
\hline
\hline
\multicolumn{13}{c}{Second simulated scenario  ($\pi \approx 18.8\%$) } \\
 \hline
 \multicolumn{2}{c}{} && Global && \multicolumn{8}{c}{Local} \\ 
 \hline 
 & &&  && $\tau_1(\vec{\bm{s}})$ & $\tau_2(\vec{\bm{s}})$ & $\tau_3(\vec{\bm{s}})$ & $\tau_4(\vec{\bm{s}})$ & $\tau_5(\vec{\bm{s}})$ & $\tau_6(\vec{\bm{s}})$ & $\tau_7(\vec{\bm{s}})$ & $\tau_8(\vec{\bm{s}})$ \\  \\[-1em]
 \cline{4-4} \cline{6-13}   \\[-1em]
\multicolumn{2}{c}{ Precision }  && $1$ &&  $0.806$ & $0.923$ & $0.858$ & $0.753$ & $0.890$ & $0.839$ & $0.793$ & $0.480$ \\
\multicolumn{2}{c}{ Recall } && $1$ &&  $0.663$ & $0.355$ & $0.609$ & $0.959$ & $0.769$ & $0.615$ & $0.840$ & $0.852$\\
\multicolumn{2}{c}{ $F_1$ scores } && $1$ && $0.727$ & $0.513$ & $0.713$ & $0.844$ & $0.825$ & $0.710$ & $0.816$ & $0.614$  \\
\hline
\hline
\end{tabular} 
\end{adjustbox}
\end{table} 

\begin{table}[!hbt]
\caption{BHM ($d=0.05$): MSEs for a representative dataset}
\label{tab:sim.omse}
\centering
\begin{adjustbox}{max width = \textwidth}
\begin{tabular}{cccccccccccccc}
 \hline 
 \hline
 \multicolumn{14}{c}{MSE} \\
 \hline \\[-1em]
 \multicolumn{2}{c}{Scenarios} &&
 \multicolumn{2}{c}{$\left\{ Z_i(\vec{\bm{s}})\right\}^{100}_{i=1}$} &&  \multicolumn{2}{c}{$\Sigma(\vec{\bm{s}}, \vec{\bm{s}})$} && \multicolumn{2}{c}{$\left\{ \beta_j(\vec{\bm{s}}) \right\}^{15}_{j=0}$} && \multicolumn{2}{c}{ $\sigma^2_\epsilon$}  \\ \\[-1em]
 \hline \\[-1em]
 \multicolumn{2}{c}{First } &&
 \multicolumn{2}{c}{ $0.1816$ } &&  \multicolumn{2}{c}{ $0.0143$ } && \multicolumn{2}{c}{ $0.0308$ } && \multicolumn{2}{c}{ $0.0015$ } \\
  \multicolumn{2}{c}{Second $(\pi = 9\%)$ } &&
 \multicolumn{2}{c}{ $0.1428$ } &&  \multicolumn{2}{c}{ $0.0113$ } && \multicolumn{2}{c}{ $0.0345$ } && \multicolumn{2}{c}{ $0.0014$ } \\ 
\multicolumn{2}{c}{ Second $(\pi \approx 18.8\%)$ } &&
 \multicolumn{2}{c}{ $0.1639$ } &&  \multicolumn{2}{c}{ $0.0102$ } && \multicolumn{2}{c}{ $0.0363$ } && \multicolumn{2}{c}{ $0.0025$ } \\ 
 \hline
\hline
\end{tabular}
\end{adjustbox}
\end{table}

\subsection{Performance comparisons}
Next, we simulate 50 replicated data sets, according to the same settings described above, and compare the performance of BHM with MUA methods. MUA approaches fit independent linear regressions at each location $\bm{s}$, to estimate coefficient images, and rely on post-inference to do variable selection and smoothing. For global selection, we first use Simes test \citep{Simes1986} to convert multiple p-values at each location to one single p-value for the whole coefficient image, and then control the False Discovery Rate (FDR) at $0.05$ for the $15$ coefficient images. We implement three different FDR control procedures, the Benjamini–Hochberg (BH) procedure \citep{Benjamini1995}, the Benjamini–Yekutieli (BY) procedure \citep{Benjamini2001}, both implemented in R via the function `\textit{p.adjust}', and another Benjamini-Hochberg procedure described in \cite{Strimmer2008}, implemented in the R package `\textit{fdrtool}', which estimates the proportion of null features from data. We denote the third procedure by MUA (SBH). Local level selection is achieved by applying these three procedures to control the FDR at $0.05$ for each coefficient image. 

Table~\ref{tab:sim.rgls.1} reports precision, recall and $F_1$ scores for both global-level selection and local-level selection, averaged over 50 replicates from the first scenario, for MUA (BH), MUA (BY), MUA (SBH) and BHM with SGLSS prior and $d=0.05$. For the global level selection, all the methods achieve similarly high values for all three metrics, indicating that the influential covariates can potentially be well distinguished from the noisy ones. For the local level selection, we observe BHM ($d=0.05$) and MUA (SBH) have similar performance with respect to the continuous covariates, $\tau_{1,2,3,4,5}(\vec{\bm{s}})$. Meanwhile, when it comes to the discrete covariates $\tau_{6,7,8}(\vec{\bm{s}})$, BHM ($d = 0.05$) achieves higher averaged $F_1$ scores than the other methods, due to a relatively better balance between precision and recall. As for the other MUA approaches, MUA (BH) and MUA (BY) have higher precision but much lower recall, leading to lower $F_1$ scores, especially for the discrete covariates.

Table~\ref{tab:sim.rgls.2} reports the three metrics for the second scenario with two settings, $\pi = 9\%$ and $\pi \approx 18.8\%$. As noted above, this scenario is  more challenging since influential covariates are closer to the noisy ones, especially for the sparser case $\pi = 9\%$. However, results are relatively consistent with the previous setting. For the global level selection, all methods achieve comparably high metrics. We notice some precision-recall trade-offs, while high $F_1$ scores result from a relative balance between precision and recall. For the local level selection, BHM ($d=0.05$) obtains similarly high $F_1$ scores as MUA (SBH) and MUA (BH) on the continuous covariates, and higher $F_1$ scores on the discrete covariates. Although the precision of BHM($d=0.05$) is not as high as the other methods, its recall is relatively higher, leading to comparable $F_1$ scores. At the same time, although MUA (BY) has the lowest $F_1$ scores, it has the highest precision in both settings.

Table~\ref{tab:sim.mse} report the MSEs for the parameters of interest and their standard errors (SE). The MUA estimators are best linear unbiased estimators (BLUE) at each location, and indeed lead to relatively accurate estimates for the coefficient images $\left\{ \beta_j(\vec{\bm{s}}) \right\}^{15}_{j=0}$. Meanwhile, the proposed BHM with SGLSS prior can return comparably good estimates since the global level indicators can exclude noisy covariates, leading to zero errors when global-level selection is done correctly. 
In addition, given its hierarchical structure, BHM also produces estimates for noise-free mean surface $\left\{ Z_i(\vec{\bm{s}})\right\}^{100}_{i=1}$ and covariance surface $\Sigma(\vec{\bm{s}}, \vec{\bm{s}})$, which are shown to be relatively accurate.

\begin{table}[!hbt]
\caption{First simulated scenario: Global and local selection for 50 replicates}
\label{tab:sim.rgls.1}
\centering
\begin{adjustbox}{max width = \textwidth}
\begin{tabular}{ccccccccccccc}
 \hline
 \hline
 \multicolumn{13}{c}{Averaged Precision} \\
 \hline
 \multicolumn{2}{c}{Methods} && Global && \multicolumn{8}{c}{Local} \\
 \hline
 & &&  && $\tau_1(\vec{\bm{s}})$ & $\tau_2(\vec{\bm{s}})$ & $\tau_3(\vec{\bm{s}})$ & $\tau_4(\vec{\bm{s}})$ & $\tau_5(\vec{\bm{s}})$ & $\tau_6(\vec{\bm{s}})$ & $\tau_7(\vec{\bm{s}})$ & $\tau_8(\vec{\bm{s}})$ \\
 \cline{4-4} \cline{6-13}
\multicolumn{2}{c}{ BHM ($d = 0.05$) } && $0.944$ &&  $1    $ & $0.937$ & $0.949$ & $0.962$ & $0.952$ & $1    $ & $0.953$ & $0.931$ \\  
\multicolumn{2}{c}{ MUA (SBH) }  && $0.964$  && $1    $ & $0.963$ & $0.952$ & $0.954$ & $0.947$ & $1    $ & $0.982$ & $0.968$ \\  
\multicolumn{2}{c}{ MUA (BH) }  &&  $0.993$ && $1    $ & $0.996$ & $0.988$ & $0.985$ & $0.979$ & $1    $ & $0.993$ & $0.988$ \\ 
\multicolumn{2}{c}{ MUA (BY) }  && $1$ && $1    $ & $0.999$ & $0.998$ & $0.998$ & $0.996$ & $1    $ & $0.999$ & $0.999$ \\  
\hline 
\hline 
\multicolumn{13}{c}{Averaged Recall} \\
 \hline
 \multicolumn{2}{c}{ Methods } && Global && \multicolumn{8}{c}{Local} \\
 \hline
& &&  && $\tau_1(\vec{\bm{s}})$ & $\tau_2(\vec{\bm{s}})$ & $\tau_3(\vec{\bm{s}})$ & $\tau_4(\vec{\bm{s}})$ & $\tau_5(\vec{\bm{s}})$ & $\tau_6(\vec{\bm{s}})$ & $\tau_7(\vec{\bm{s}})$ & $\tau_8(\vec{\bm{s}})$ \\
 \cline{4-4} \cline{6-13}
\multicolumn{2}{c}{ BHM ($d = 0.05$) } &&  $1$ &&  $0.970$ & $0.954$ & $0.920$ & $0.885$ & $0.865$ & $0.764$ & $0.752$ & $0.690$ \\   
\multicolumn{2}{c}{ MUA (SBH) }  && $0.992$ && $0.957$ & $0.936$ & $0.900$ & $0.871$ & $0.846$ & $0.522$ & $0.506$ & $0.422$ \\  
\multicolumn{2}{c}{ MUA (BH) }  && $0.982$ &&  $0.817$ & $0.806$ & $0.811$ & $0.790$ & $0.782$ & $0.257$ & $0.291$ & $0.253$  \\ 
\multicolumn{2}{c}{ MUA (BY) }  && $0.948$ && $0.656$ & $0.628$ & $0.654$ & $0.632$ & $0.622$ & $0.069$ & $0.080$ & $0.070$ \\ 
\hline 
\hline
\multicolumn{13}{c}{Averaged $F_1$ scores } \\
\hline
 \multicolumn{2}{c}{ Methods } && Global && \multicolumn{8}{c}{Local} \\
 \hline
 & &&  && $\tau_1(\vec{\bm{s}})$ & $\tau_2(\vec{\bm{s}})$ & $\tau_3(\vec{\bm{s}})$ & $\tau_4(\vec{\bm{s}})$ & $\tau_5(\vec{\bm{s}})$ & $\tau_6(\vec{\bm{s}})$ & $\tau_7(\vec{\bm{s}})$ & $\tau_8(\vec{\bm{s}})$ \\
 \cline{4-4} \cline{6-13}
\multicolumn{2}{c}{ BHM ($d = 0.05$) } && $0.970$ &&   $\bm{0.983}$ & $0.943$ & $\bm{0.932}$ & $\bm{0.919}$ & $\bm{0.903}$ & $\bm{0.837}$ & $\bm{0.814}$ & $\bm{0.771}$ \\ 
\multicolumn{2}{c}{ MUA (SBH) }  && $0.977$ && $0.975$ & $\bm{0.947}$ & $0.921$ & $0.906$ & $0.888$ & $0.642$ & $0.622$ & $0.555$ \\
\multicolumn{2}{c}{ MUA (BH) }  && $\bm{0.987}$ && $0.892$ & $0.884$ & $0.884$ & $0.868$ & $0.862$ & $0.379$ & $0.440$ & $0.386$ \\ 
\multicolumn{2}{c}{ MUA (BY) }  && $0.972$ &&  $0.775$ & $0.755$ & $0.775$ & $0.751$ & $0.749$ & $0.145$ & $0.154$ & $0.143$ \\
\hline
\hline
\end{tabular} 
\end{adjustbox}
\end{table} 

\begin{table}[!hbt]
\caption{Second simulated scenario: Global and local selection for 50 replicates}
\label{tab:sim.rgls.2}
\centering
\begin{adjustbox}{max width = \textwidth}
\begin{tabular}{ccccccccccccc}
 \hline
 \hline
 \multicolumn{13}{c}{Averaged Precision ($\pi = 9\%$)} \\
 \hline
 \multicolumn{2}{c}{Methods} && Global && \multicolumn{8}{c}{Local} \\
 \hline
 & &&  && $\tau_1(\vec{\bm{s}})$ & $\tau_2(\vec{\bm{s}})$ & $\tau_3(\vec{\bm{s}})$ & $\tau_4(\vec{\bm{s}})$ & $\tau_5(\vec{\bm{s}})$ & $\tau_6(\vec{\bm{s}})$ & $\tau_7(\vec{\bm{s}})$ & $\tau_8(\vec{\bm{s}})$ \\
 \cline{4-4} \cline{6-13}
\multicolumn{2}{c}{ BHM ($d = 0.05$) }  && $0.919$  &&  $0.708$ & $0.754$ & $0.671$ & $0.726$ & $0.762$ & $0.484$ & $0.386$ & $0.454$  \\
\multicolumn{2}{c}{ MUA (SBH) }  && $0.934$  && $0.926$ & $0.948$ & $0.902$ & $0.948$ & $0.964$ & $0.906$ & $0.910$ &  $0.775$  \\
\multicolumn{2}{c}{ MUA (BH) }  &&  $0.970$ && $0.946$ & $0.958$ & $0.930$ &  $0.959$ & $0.970$ &  $0.930$ &  $0.981$ & $0.814$  \\
\multicolumn{2}{c}{ MUA (BY) }  && $0.997$ && $0.990$ &  $0.994$ & $0.993$ & $0.996$ & $0.996$ & $1$ & $1$ &    $0.999$  \\
 \hline
 \multicolumn{13}{c}{Averaged Precision ($\pi \approx 18.8\%$)} \\
 \hline
 \multicolumn{2}{c}{Methods} && Global && \multicolumn{8}{c}{Local} \\
 \hline
 & &&  && $\tau_1(\vec{\bm{s}})$ & $\tau_2(\vec{\bm{s}})$ & $\tau_3(\vec{\bm{s}})$ & $\tau_4(\vec{\bm{s}})$ & $\tau_5(\vec{\bm{s}})$ & $\tau_6(\vec{\bm{s}})$ & $\tau_7(\vec{\bm{s}})$ & $\tau_8(\vec{\bm{s}})$ \\
 \cline{4-4} \cline{6-13}
\multicolumn{2}{c}{ BHM ($d = 0.05$) }  && $0.926$  &&  $0.824$ & $0.830$  & $0.823$ & $0.831$ & $0.829$ & $0.686$ & $0.689$ & $0.652$ \\
\multicolumn{2}{c}{ MUA (SBH) }  && $0.945$  && $0.949$ & $0.952$ & $0.953$ & $0.956$ & $0.959$ & $0.927$ & $0.901$ & $0.918$  \\
\multicolumn{2}{c}{ MUA (BH) }  &&  $0.977$ && $0.960$  & $0.963$ & $0.965$ & $0.966$ & $0.967$ & $0.949$ & $0.913$ & $0.928$  \\
\multicolumn{2}{c}{ MUA (BY) }  && $0.998$ && $0.991$ & $0.996$ & $0.997$ & $0.995$ & $0.997$ & $0.997$ & $0.995$ & $0.993$  \\
\hline 
\hline 
\multicolumn{13}{c}{Averaged Recall ($\pi = 9\%$)} \\
 \hline
 \multicolumn{2}{c}{ Methods } && Global && \multicolumn{8}{c}{Local} \\
 \hline
& &&  && $\tau_1(\vec{\bm{s}})$ & $\tau_2(\vec{\bm{s}})$ & $\tau_3(\vec{\bm{s}})$ & $\tau_4(\vec{\bm{s}})$ & $\tau_5(\vec{\bm{s}})$ & $\tau_6(\vec{\bm{s}})$ & $\tau_7(\vec{\bm{s}})$ & $\tau_8(\vec{\bm{s}})$ \\
 \cline{4-4} \cline{6-13}
\multicolumn{2}{c}{ BHM ($d = 0.05$) }  && $0.742$  &&  $0.681$ &  $0.579$ &  $0.570$  &  $0.592$ &  $0.680$  &  $0.272$ &  $0.232$ &  $0.253$ \\
\multicolumn{2}{c}{ MUA (SBH) }  && $0.755$  && $0.677$ &  $0.569$ &  $0.568$ &  $0.598$ &  $0.699$ &  $0.071$ &  $0.079$ &  $0.107$ \\
\multicolumn{2}{c}{ MUA (BH) }  &&  $0.690$ && $0.672$ &  $0.571$ &  $0.559$ &  $0.589$ &  $0.691$ &  $0.061$ &  $0.074$ &  $0.101$ \\
\multicolumn{2}{c}{ MUA (BY) }  && $0.600$ && $0.514$ &  $0.395$ &  $0.439$ &  $0.424$ &  $0.530$  &  $0.013$ &  $0.023$ &  $0.042$ \\
\hline 
\multicolumn{13}{c}{Averaged Recall ($\pi \approx 18.8\%$)} \\
 \hline
 \multicolumn{2}{c}{ Methods } && Global && \multicolumn{8}{c}{Local} \\
 \hline
& &&  && $\tau_1(\vec{\bm{s}})$ & $\tau_2(\vec{\bm{s}})$ & $\tau_3(\vec{\bm{s}})$ & $\tau_4(\vec{\bm{s}})$ & $\tau_5(\vec{\bm{s}})$ & $\tau_6(\vec{\bm{s}})$ & $\tau_7(\vec{\bm{s}})$ & $\tau_8(\vec{\bm{s}})$ \\
 \cline{4-4} \cline{6-13}
\multicolumn{2}{c}{ BHM ($d = 0.05$) }  && $0.945$  &&  $0.794$ & $0.750$  & $0.797$ & $0.781$ & $0.770$  & $0.397$ & $0.367$ & $0.441$ \\
\multicolumn{2}{c}{ MUA (SBH) }  && $0.870$  && $0.760$  & $0.697$ & $0.762$ & $0.743$ & $0.73 $ & $0.153$ & $0.161$ & $0.175$ \\
\multicolumn{2}{c}{ MUA (BH) }  &&  $0.818$ && $0.738$ & $0.682$ & $0.748$ & $0.730$  & $0.712$ & $0.131$ & $0.148$ & $0.147$ \\
\multicolumn{2}{c}{ MUA (BY) }  && $0.732$ && $0.553$ & $0.532$ & $0.586$ & $0.589$ & $0.530$  & $0.033$ & $0.053$ & $0.039$ \\
\hline 
\hline
\multicolumn{13}{c}{Averaged $F_1$ scores  ($\pi = 9\%$)} \\
\hline
 \multicolumn{2}{c}{ Methods } && Global && \multicolumn{8}{c}{Local} \\
 \hline
 & &&  && $\tau_1(\vec{\bm{s}})$ & $\tau_2(\vec{\bm{s}})$ & $\tau_3(\vec{\bm{s}})$ & $\tau_4(\vec{\bm{s}})$ & $\tau_5(\vec{\bm{s}})$ & $\tau_6(\vec{\bm{s}})$ & $\tau_7(\vec{\bm{s}})$ & $\tau_8(\vec{\bm{s}})$ \\
 \cline{4-4} \cline{6-13}
\multicolumn{2}{c}{ BHM ($d = 0.05$)}  && $0.812$  &&  $0.718$ &  $0.705$ &  $0.670$  &  $\bm{0.716}$ &  $\bm{0.754}$ &  $\bm{0.450}$  &  $\bm{0.387}$ & $\bm{0.412}$ \\
\multicolumn{2}{c}{ MUA (SBH) }  && $\bm{0.823}$  && $0.753$ &  $0.714$ &  $0.680$  &  $0.699$ &  $0.752$ &  $0.227$ &  $0.267$ & $0.345$ \\
\multicolumn{2}{c}{ MUA (BH) }  &&  $0.795$ && $\bm{0.761}$ &  $\bm{0.722}$ &  $\bm{0.689}$ &  $0.693$ &  $0.748$ &  $0.195$ &  $0.266$ & $0.330$ \\
\multicolumn{2}{c}{ MUA (BY) }  && $0.735$ && $0.670$  &  $0.577$ &  $0.611$ &  $0.571$ &  $0.682$ &  $0.124$ &  $0.154$ & $0.241$ \\
\hline
\multicolumn{13}{c}{Averaged $F_1$ scores ($\pi \approx 18.8\%$) } \\
\hline
 \multicolumn{2}{c}{ Methods } && Global && \multicolumn{8}{c}{Local} \\
 \hline
 & &&  && $\tau_1(\vec{\bm{s}})$ & $\tau_2(\vec{\bm{s}})$ & $\tau_3(\vec{\bm{s}})$ & $\tau_4(\vec{\bm{s}})$ & $\tau_5(\vec{\bm{s}})$ & $\tau_6(\vec{\bm{s}})$ & $\tau_7(\vec{\bm{s}})$ & $\tau_8(\vec{\bm{s}})$ \\
 \cline{4-4} \cline{6-13}
\multicolumn{2}{c}{ BHM ($d = 0.05$) }  && $\bm{0.932}$  &&  $0.795$ &  $0.771$ &  $0.796$ &  $0.801$ &  $0.793$ &  $\bm{0.518}$ &  $\bm{0.511}$ &  $\bm{0.527}$ \\
\multicolumn{2}{c}{ MUA (SBH) }  && $0.900$  && $\bm{0.816}$ &  $\bm{0.774}$ &  $\bm{0.813}$ &  $\bm{0.811}$ &  $\bm{0.794}$ &  $0.310$  &  $0.454$ &  $0.308$ \\
\multicolumn{2}{c}{ MUA (BH) }  &&  $0.885$ && $0.803$ &  $0.767$ &  $0.808$ &  $0.805$ &  $0.784$ &  $0.285$ &  $0.455$ &  $0.274$ \\
\multicolumn{2}{c}{ MUA (BY) }  && $0.837
$ && $0.671$ &  $0.636$ &  $0.691$ &  $0.720$  &  $0.666$ &  $0.131$ &  $0.261$ &  $0.185$ \\
\hline
\hline
\end{tabular} 
\end{adjustbox}
\end{table} 

\begin{table}[!htb]
\caption{MSEs for $50$ replicates}
\label{tab:sim.mse}
\centering
\begin{adjustbox}{max width = \textwidth}
\begin{tabular}{cccccccccc}
\hline
\hline
 && \multicolumn{5}{c}{First simulated scenario} \\
 \hline
 \multicolumn{1}{c}{Model} & & \multicolumn{2}{c}{$\left\{ Z_i(\vec{\bm{s}})\right\}^{100}_{i=1}$} &&  \multicolumn{2}{c}{$\left\{ \beta_j(\vec{\bm{s}})\right\}^{15}_{j=0}$} && \multicolumn{2}{c}{$\Sigma(\vec{\bm{s}}, \vec{\bm{s}})$ }  \\
 \hline
 \multicolumn{1}{c}{} && Mean  & SE && Mean  & SE && Mean  & SE \\
 \cline{3-4} \cline{6-7} \cline{9-10}
  \multicolumn{1}{c}{ BHM  ($d = 0.05$)} && $0.180$ & $(1.0 \times 10^{-3})$ &&  $0.472$ & $(1.13\times 10^{-2})$ && $0.014$ & $(0.5 \times 10^{-3})$ \\
   \multicolumn{1}{c}{MUA} && \multicolumn{2}{c}{-} &&  $0.678$ & $(1.15\times 10^{-2})$ && \multicolumn{2}{c}{-}    \\ 
   \hline 
   \hline
   && \multicolumn{5}{c}{Second simulated scenario ($\pi = 9\%$)} \\
 \hline
 \multicolumn{1}{c}{Model} & & \multicolumn{2}{c}{$\left\{ Z_i(\vec{\bm{s}})\right\}^{100}_{i=1}$} &&  \multicolumn{2}{c}{$\left\{ \beta_j(\vec{\bm{s}})\right\}^{15}_{j=0}$} && \multicolumn{2}{c}{$\Sigma(\vec{\bm{s}}, \vec{\bm{s}})$ }  \\
 \hline
 \multicolumn{1}{c}{} && Mean  & SE && Mean  & SE && Mean  & SE \\
 \cline{3-4} \cline{6-7} \cline{9-10}
  \multicolumn{1}{c}{ BHM  ($d = 0.05$)} && $0.140$ & $(1.9 \times 10^{-3})$ &&  $0.540$ & $(2.02\times 10^{-2})$ && $0.013$ & $(0.5 \times 10^{-3})$ \\
   \multicolumn{1}{c}{MUA} && \multicolumn{2}{c}{-} &&  $0.671$ & $(0.88\times 10^{-2})$ && \multicolumn{2}{c}{-}    \\
\hline
\hline
 && \multicolumn{5}{c}{Second simulated scenario ($\pi \approx 18.8\%$)} \\
 \hline
 \multicolumn{1}{c}{Model} & & \multicolumn{2}{c}{$\left\{ Z_i(\vec{\bm{s}})\right\}^{100}_{i=1}$} &&  \multicolumn{2}{c}{$\left\{ \beta_j(\vec{\bm{s}})\right\}^{15}_{j=0}$} && \multicolumn{2}{c}{$\Sigma(\vec{\bm{s}}, \vec{\bm{s}})$ }  \\
 \hline
 \multicolumn{1}{c}{} && Mean  & SE && Mean  & SE && Mean  & SE \\
 \cline{3-4} \cline{6-7} \cline{9-10}
  \multicolumn{1}{c}{ BHM  ($d = 0.05$)} && $0.162$ & $(1.4 \times 10^{-3})$ &&  $0.597$ & $(1.36\times 10^{-2})$ && $0.013$ & $(0.4 \times 10^{-3})$ \\
   \multicolumn{1}{c}{MUA} && \multicolumn{2}{c}{-} &&  $0.664$ & $(0.93\times 10^{-2})$ && \multicolumn{2}{c}{-}    \\
   \hline 
   \hline
\end{tabular}
\end{adjustbox}
\end{table}

\section{Real Data Application}
\label{sec:app}
We demonstrate the proposed model using the Autism Brain Imaging Data Exchange (ABIDE) study of \cite{ABIDE2014}. The study collected resting-state fMRI data from 17 experiment sites including 1112 subjects, with the aim of improving the understanding of neurophysiological mechanisms. For each subject, rs-fMRI data were recorded over time, along with the subject's information such as age, gender, intelligence quotient, etc. To aid computations, we reduced the image size by summarizing the fMRI data to voxel-level imaging statistics and then considered individual brain networks instead of the whole brain image. Following \cite{He2019} and \cite{Zhang2020}, we used the pipeline of \cite{Craddock2013} to preprocess the data and then considerded a parcellation of the brain as defined by the Automated Anatomical Labeling \citep{AAL2002} to select networks. The selected networks are described in Table~\ref{tab:network}, and are known to be associated with cognitive ability based on previous research \citep{Heuvel2009, Wu2013, Hearne2016, Hilger2017, Zhang2020}. As for the covariates, those collected in the 17 experiment sites include diagnostic, age, gender and full-scale intelligence quotient scores (FIQ). The FIQ scores were assessed differently across sites, including DAS-II, WASI, WISC, WAIS, RAVENS and STANFORD scales. After removing missing values, we ended up with 1001 subjects. We standardized the continuous variables, age and FIQ scores, to put them on the same scale with the discrete indicators, diagnostic and gender, which we left unchanged. We included a vector of ones as the intercept to account for those potentially influential covariates which are not available in the study.
We then applied BHM with the SGLSS prior, separately, to the four selected networks. 

\begin{table}[!hbt]
\caption{Networks of interests}
\label{tab:network}
\centering
\begin{adjustbox}{max width = 1\textwidth}
\begin{tabular}{ccccccccc}
 \hline
 \hline
\\[-1em]
 \multicolumn{2}{c}{\multirow{2}{*}{Network}} & & \multirow{2}{*}{\shortstack{Number of\\ voxels}} & & \multicolumn{4}{c}{\multirow{2}{*}{Regions included}}  \\
\\
 \cline{1-2} \cline{4-4} \cline{6-9}  
\\[-1em]
 \multicolumn{2}{c}{Visual} & & 7946 & & Lingual L & Lingual R & Calcarine L & Cuneus R  \\
 \\[-1em]
   \multicolumn{2}{c}{Ventral Attention} & & 9839 & & Temporal Mid L & Temporal Sup L & Temporal Sup R &  -  \\
   \\[-1em]
 \multicolumn{2}{c}{Dorsal Attention} & & 9600 & & Temporal Mid R & Postcentral L & Parietal Sup L &  -  \\
 \\[-1em]
  \multicolumn{2}{c}{Default Mode} & & 7440 & & Temporal Mid R & Frontal Med Orb R & Frontal Med Orb L & Occipital Mid R  \\
 \hline
 \hline
\end{tabular}
\end{adjustbox}
\end{table} 

 We specified the threshold $d$ at the conventional sparsity level $d = 0.05$. As for the slab prior specification, we set $\{\mu_{0j}(\vec{\bm{s}})\}^q_{j=0} = 0$ and specified $\{\sigma^2_{0j}(\vec{\bm{s}})\}^q_{j=0} = 1$. We ran MCMC chains with 2000 iterations and 500 burnin. On average, the z-scores from the Geweke test were 0.9616 for $\left\{\tau_j(\vec{\bm{s}})\right\}^{4}_{j=1}$ and 1.1858 for $\left\{\pi_j \right\}^{4}_{j=1}$, indicating that the MCMC chains were run for a sufficient number of iterations. The MCMC chain took around 35 seconds per iteration on two 20-core 2.4 GHz Intel(R) Xeon CPUs for networks with nearly $10,000$ voxels.

Table~\ref{tab:app.sel_res.1} shows selection results for BHM and the MUA methods. For local level selection, $\pi$ denotes the ratio of selected voxels for MUA-based methods and the posterior mean of the participation rate $\pi_j$ for BHM. The check marks denote whether the covariate is selected at global level. For global level selection, all methods agree on selecting the covariate \textit{age} as influential for human cognitive ability, which makes sense because as people age the brain naturally changes, along with its cognitive functions. The difference is in the selection of the \textit{FIQ} scores. The MUA-based methods tend to include the FIQ in the model, while the proposed BHM $(d=0.05)$ considers FIQ to be related only to Visual network and Dorsal network. At the local level selection, the MUA with Benjamini-Hochberg based procedures tend to select the covariates at almost all voxels. Meanwhile, MUA with Benjamini-Yekutieli procedure tends to select the covariates at fewer voxels than the MUA (BH) especially when it comes to the FIQ scores. The BHM-based methods tend to have similar results as the MUA (BY) on the selection of \textit{age}. As for the selection of FIQ, however, BHM with $d=0.05$ tends to select even fewer voxels. Furthermore, when fitting the BHM model we noticed that a more stringent threshold of $d = 0.1$ would exclude FIQ for all networks entirely, while a less stringent threshold of $d = 0.01$ would include FIQ for all networks but only for very few selected voxels. These results suggest that, although FIQ score may somehow be related to brain signals, this relationship can be hard to recover in this application, possibly because different experimental sites use different standards to measure this covariate. 

\begin{figure}
    \centering
    \begin{subfigure}[b]{1\textwidth}
        \centering
        \includegraphics[width=0.32\textwidth]{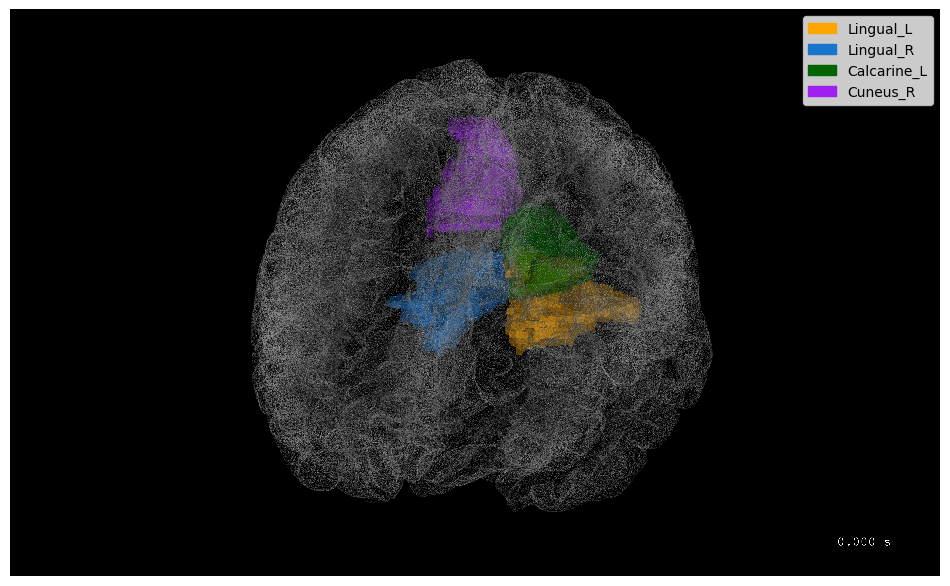}
         \includegraphics[width=0.32\textwidth]{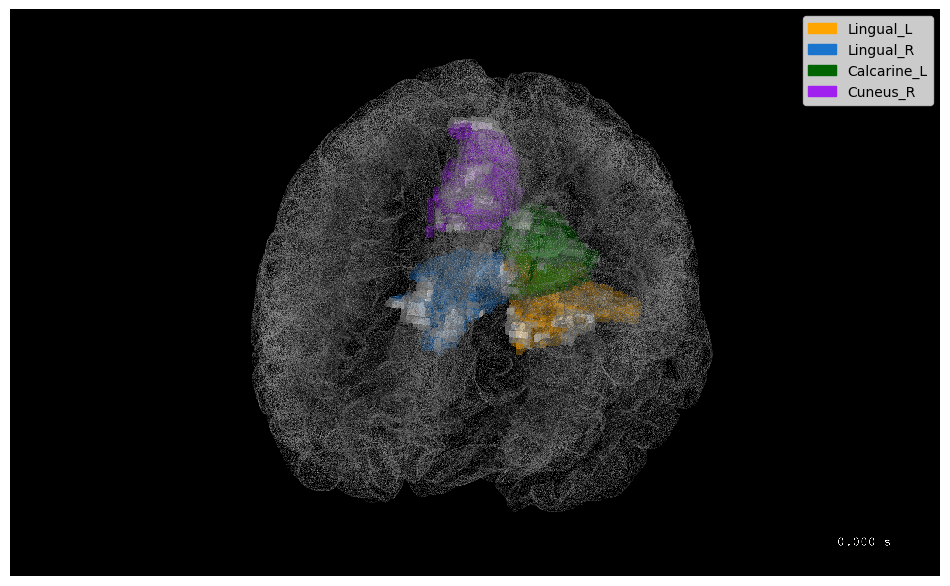}
        \includegraphics[width=0.32\textwidth]{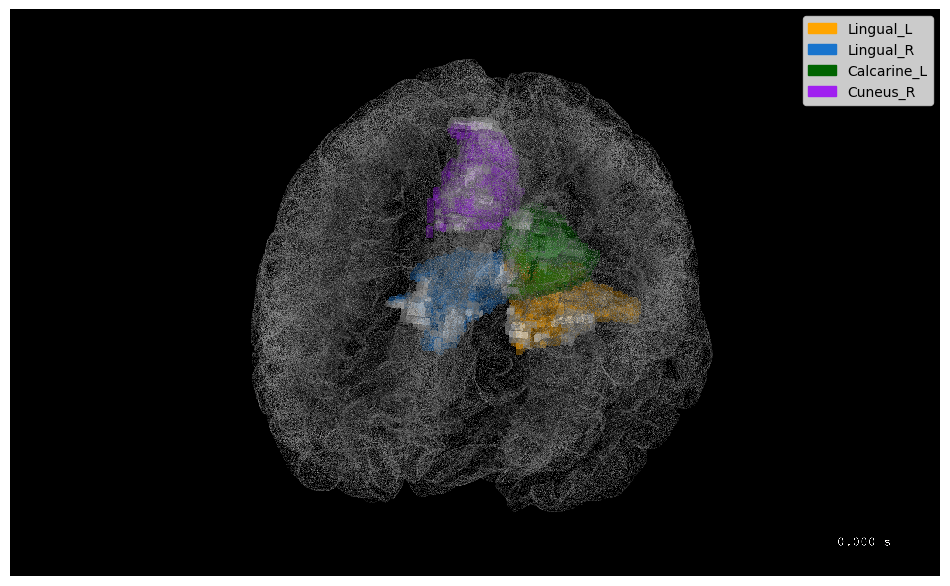} 
        \caption{Visual Network.}
    \end{subfigure}
    \begin{subfigure}[b]{1\textwidth}
        \centering
        \includegraphics[width=0.32\textwidth]{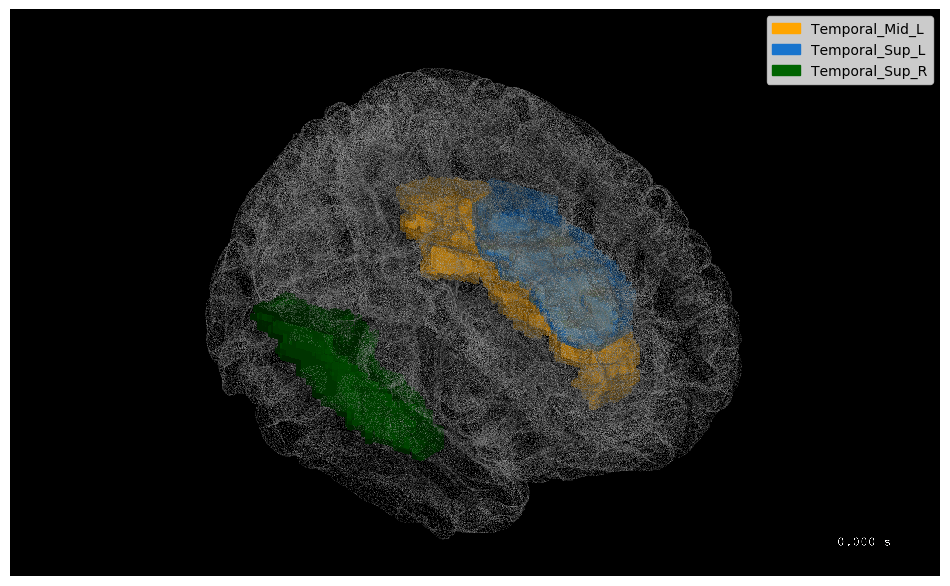}
         \includegraphics[width=0.32\textwidth]{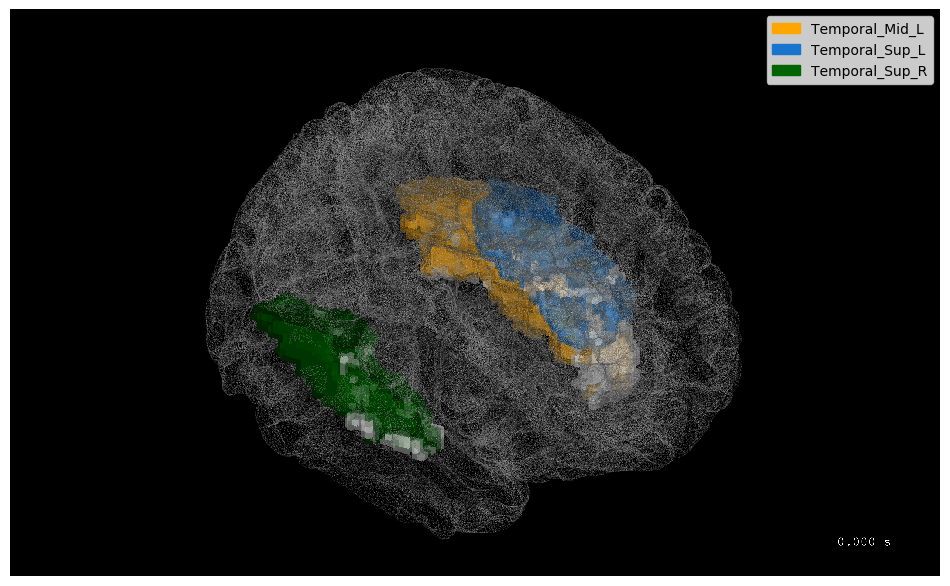}
        \includegraphics[width=0.32\textwidth]{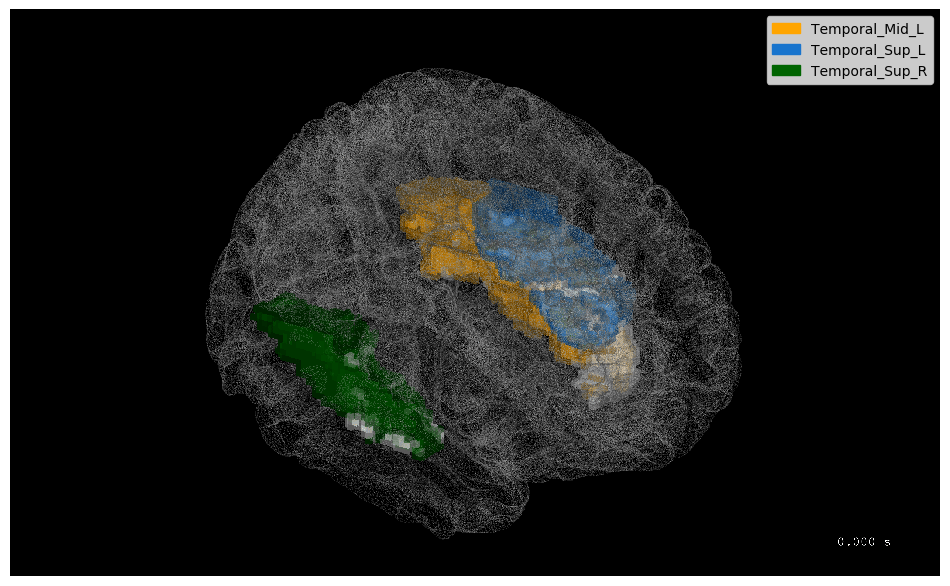} 
        \caption{Ventral Network.}
    \end{subfigure}
    \begin{subfigure}[b]{1\textwidth}
        \centering
        \includegraphics[width=0.32\textwidth]{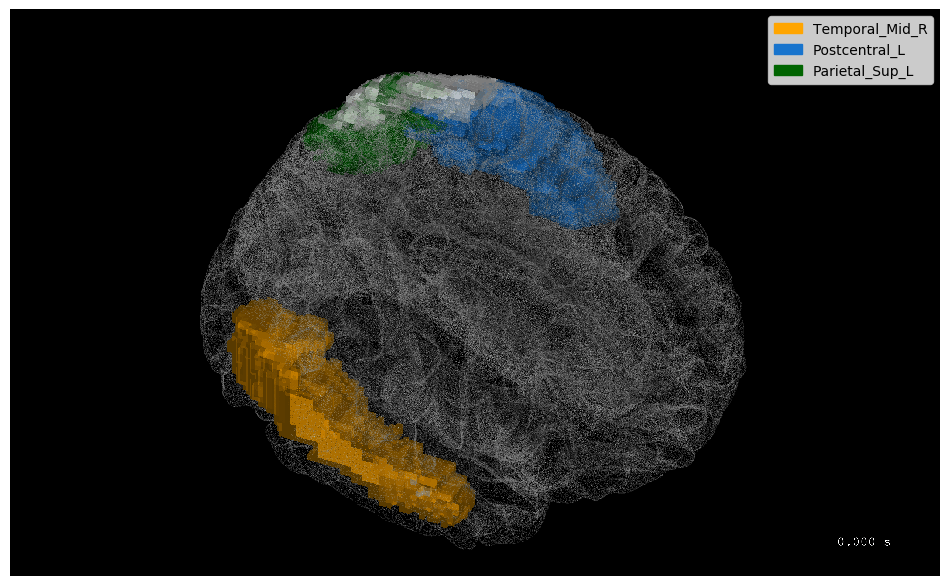}
         \includegraphics[width=0.32\textwidth]{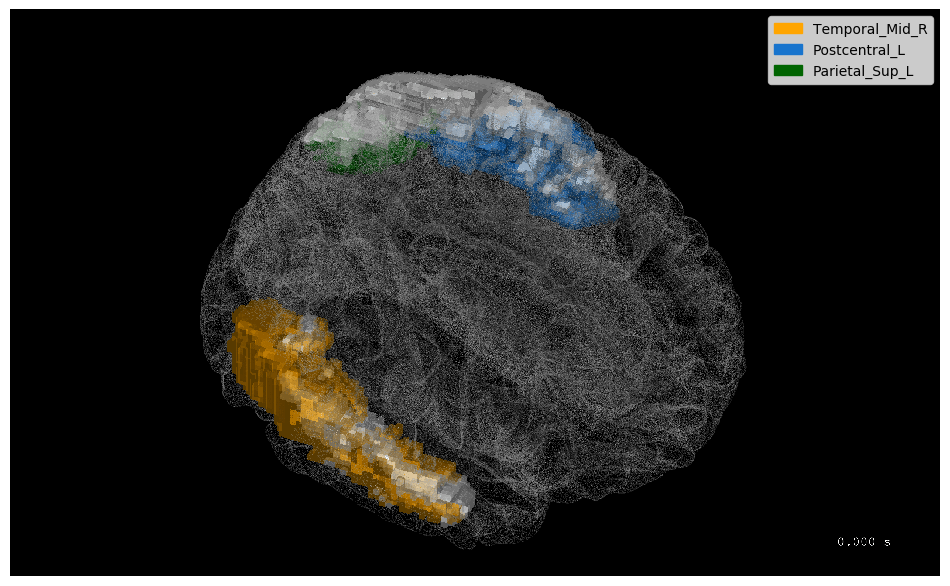}
        \includegraphics[width=0.32\textwidth]{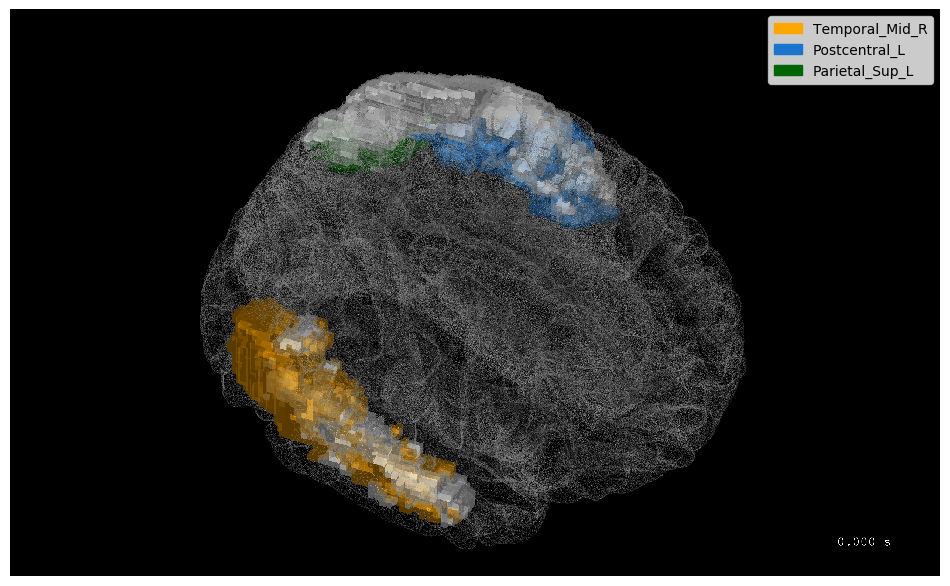} 
        \caption{Dorsal Network.}
    \end{subfigure}
    \begin{subfigure}[b]{1\textwidth}
        \centering
        \includegraphics[width=0.32\textwidth]{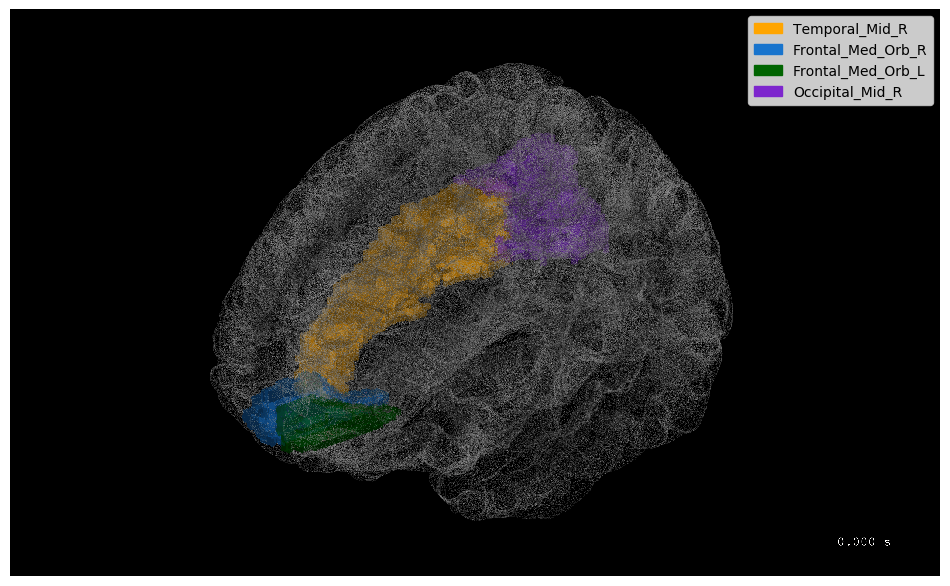}
         \includegraphics[width=0.32\textwidth]{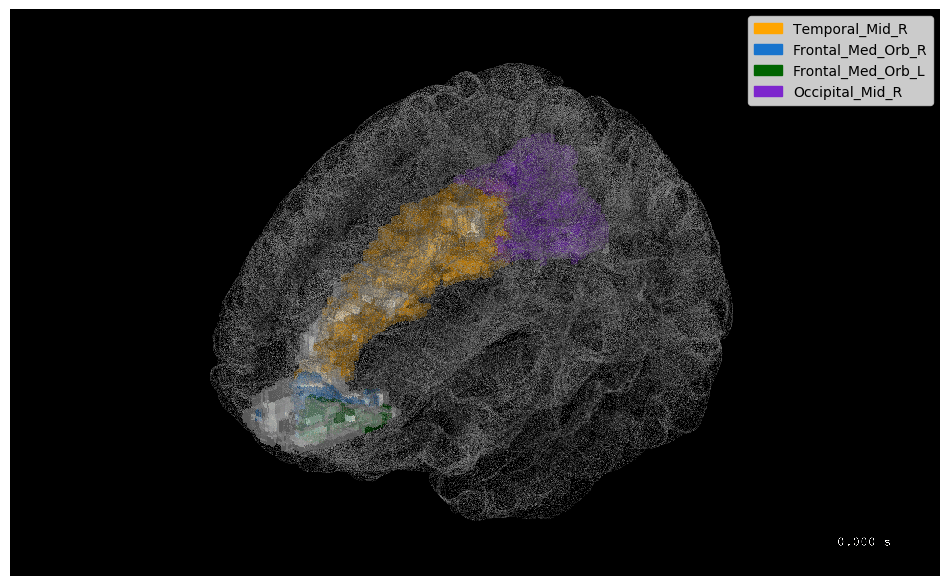}
        \includegraphics[width=0.32\textwidth]{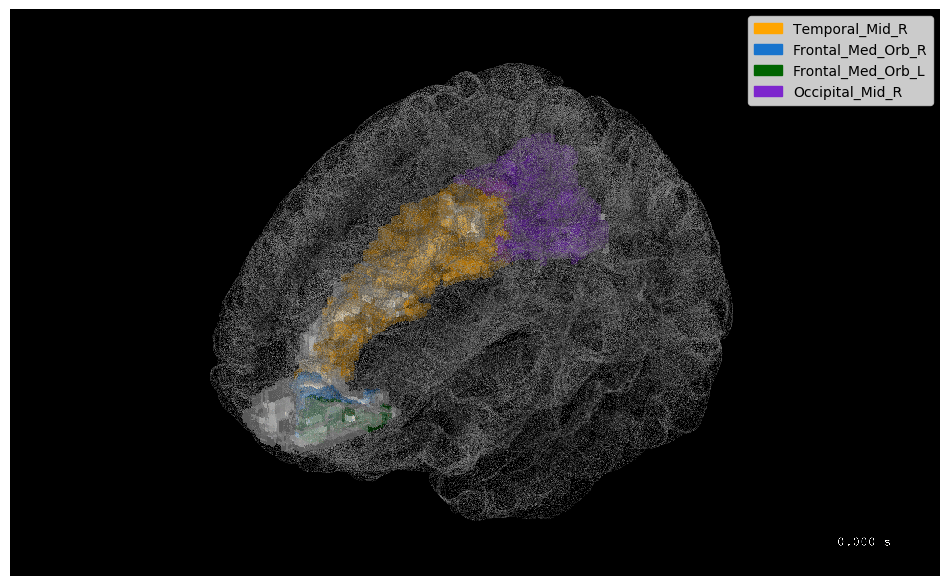} 
        \caption{Default Network.}
    \end{subfigure}
     \caption{Selected voxels for covariate `\textit{age}', by MUA with SBH (Left), MUA with BY (Middle) and BHM with SGLSS ($d = 0.05$) (Right). Figures are plotted using the R package \textit{threeBrain} by \cite{threeBrain}}
    \label{fig:app.sel}
\end{figure}

Figure~\ref{fig:app.sel} shows the selected voxels for the covariate \textit{age} by MUA (SBH) (Left), MUA (BY) (Middle) and BHM ($d=0.05$) (Right). We observe a decreasing number of voxels selected by the methods, and similar local selective results between MUA (BY) and BHM ($0.05$), both of which tend to have a more sparse selection with selected voxels mainly in the central portions of the regions in the functional networks. Table~\ref{tab:app.sel_res.2} reports the ratios of region included in the local selection, showing consistent selection results for BHM. We note that the final selection is determined by the posterior summary, i.e. the median rule, based on the posterior samples, while the threshold $I(\pi \ge d)$ takes effect at each iteration. Hence, although the global indicator can guarantee $\pi \ge d$ at each iteration, the final ratios of selected voxels/pixels are not necessarily greater than $d$, as it is evident from the results. We report results for BHM with $d = 0.01$ and $d=0.1$ in the supplementary material and note here that BHM maintains highly consistent local selection results with different specification of $d$, i.e. when Ceneus R is considered to be affected by FIQ, BHM ($d =0.01$) selects $1.6\%$ of the region and BHM ($d=0.05)$ selects $1.77\%$; when Temporal Mid R is considered to be affected by FIQ, BHM ($d =0.01$) selects $1.7\%$ of the region and BHM ($d=0.05)$ selects $1.5\%$. These results also confirm the previous observation that FIQ scores, converted from different standards, may not show strong relationship to the brain regions.

\begin{table}
\caption{Selection results for the four networks}
\label{tab:app.sel_res.1}
\centering
\begin{adjustbox}{max width = 1\textwidth}
\begin{tabular}{ccccccccccc}
\hline
\hline
& & \multicolumn{4}{c}{MUA (SBH)} &  & \multicolumn{4}{c}{BHM (d = 0.05) }\\
\hline
  &  & diagnostic & age & gender  & FIQ  &  & diagnostic & age & gender  & FIQ   \\
  \cline{3-6} \cline{8-11}
    \multicolumn{1}{c}{\multirow{2}{*}{Visual}} 
    & $\pi (\%)$ & 0.0 & 100.0 & 0.0 & 100.0 & & 0.01 & 77.2 & 0.01 & 9.24 \\
    & if selected   &  & (\checkmark) & & (\checkmark) &   &  &  (\checkmark) & & (\checkmark) \\  
   \multicolumn{1}{c}{\multirow{2}{*}{Ventral}}
    & $\pi (\%)$ & 0.0 & 100.0 & 0.0 & 100.0 & & 0.01 & 83.6 & 0.01 & 1.38 \\
    & if selected   &  & (\checkmark) & & (\checkmark) &   &  &  (\checkmark) & &   \\ 
    \multicolumn{1}{c}{\multirow{2}{*}{Dorsal}} 
    & $\pi (\%)$ & 0.0 & 85.8 & 0.0 & 100.0 & & 0.01 & 42.9 & 0.01 & 7.77\\
    & if selected   &  & (\checkmark) & & (\checkmark) &   &  &  (\checkmark) & & (\checkmark)  \\ 
    \multicolumn{1}{c}{\multirow{2}{*}{Default}} 
    & $\pi (\%)$ & 0.0 & 100.0 & 0.0 & 100.0 & & 0.01 & 71.6 & 0.01 & 0.78 \\
    & if selected   &  & (\checkmark) & & (\checkmark) &   &  &  (\checkmark) & &  \\ 
 \hline
 \hline
 & & \multicolumn{4}{c}{MUA (BH)} &  & \multicolumn{4}{c}{MUA (BY)}\\
\hline
  &  & diagnostic & age & gender  & FIQ  &  & diagnostic & age & gender  & FIQ   \\
  \cline{3-6} \cline{8-11}
    \multicolumn{1}{c}{\multirow{2}{*}{Visual}} 
    & $\pi (\%)$ & 0.0 & 98.8 & 0.0 & 92.3 & & 0.0 & 83.0 & 0.0 & 32.7 \\
    & if selected   &  & (\checkmark) & & (\checkmark) &   &  &  (\checkmark) & & (\checkmark) \\  
   \multicolumn{1}{c}{\multirow{2}{*}{Ventral}}
    & $\pi (\%)$ & 0.0 & 97.1 & 0.0 & 90.6 & & 0.0 & 82.2 & 0.0 & 24.7 \\
    & if selected   &  & (\checkmark) & & (\checkmark) &   &  &  (\checkmark) & &  (\checkmark)  \\ 
    \multicolumn{1}{c}{\multirow{2}{*}{Dorsal}} 
    & $\pi (\%)$ & 0.0 & 74.9 & 0.0 & 88.2 & & 0.0 & 54.3 & 0.0 & 22.7\\
    & if selected   &  & (\checkmark) & & (\checkmark) &   &  &  (\checkmark) & & (\checkmark)  \\ 
    \multicolumn{1}{c}{\multirow{2}{*}{Default}} 
    & $\pi (\%)$ & 0.0 & 90.4 & 0.0 & 89.6 & & 0.0 & 74.5 & 0.0 & 24.4 \\
    & if selected   &  & (\checkmark) & & (\checkmark) &   &  &  (\checkmark) & &  (\checkmark) \\ 
 \hline
 \hline
\end{tabular}
\end{adjustbox}
\end{table}

\begin{table}[!hbt]
\caption{Ratios of Region included within each networks}
\label{tab:app.sel_res.2}
\centering
\begin{adjustbox}{max width = 1\textwidth}
\begin{tabular}{ccccccccccc}
 \hline
 \hline
\\[-1em]
 \multicolumn{2}{c}{\multirow{2}{*}{Network}} & & \multirow{2}{*}{\shortstack{Methods}}& &\multirow{2}{*}{\shortstack{Covariates}} & & \multicolumn{4}{c}{\multirow{2}{*}{Ratio of Region included ($\%$)}}  \\
\\
 \hline
\\[-1em]
 \multicolumn{2}{c}{\multirow{13}{*}{Visual}}  & & &  && & Lingual L & Lingual R & Calcarine L & Cuneus R  \\
 \\[-1em]
\multicolumn{2}{c}{} & & \multirow{3}{*}{ BHM$(d=0.05)$ } && age & & 82.1  &  81.4  &  83.5  &   88.9\\
 \\[-1em]
 \multicolumn{2}{c}{} & && & FIQ & &  5.79  &  7.00   & 3.29 &  1.77 \\
  \\[-1em]
  \cline{6-6} \cline{8-11}
  \\[-1em]
 \multicolumn{2}{c}{} & &  \multirow{3}{*}{MUA (SBH) } & & age & & 100 & 100 & 100 & 100  \\
 \\[-1em]
 \multicolumn{2}{c}{} & && & FIQ & &  100 & 100 & 100 & 100  \\
  \\[-1em]
  \cline{6-6} \cline{8-11}
  \\[-1em]
 \multicolumn{2}{c}{} & &  \multirow{3}{*}{MUA (BH) } & & age & & 98.9 &  97.4 &  99.4 &  99.7  \\
 \\[-1em]
 \multicolumn{2}{c}{} & && & FIQ & &  95.3 & 89.4 &  93.2  & 91.1  \\
  \\[-1em]
\cline{6-6} \cline{8-11}
  \\[-1em]
 \multicolumn{2}{c}{} & &  \multirow{3}{*}{MUA (BY) } & & age & & 81.8 & 81.5 & 82.4 & 88.2  \\
 \\[-1em]
 \multicolumn{2}{c}{} & && & FIQ & &  34.2 & 34.7 & 29.1& 32.9  \\
  \\[-1em]
 \hline
\\[-1em]
  \multicolumn{2}{c}{\multirow{13}{*}{Ventral}} & & & & & & Temporal Mid L & Temporal Sup L & Temporal Sup R &  - \\
  \\[-1em]
\multicolumn{2}{c}{} & & \multirow{3}{*}{ BHM$(d=0.05)$ } && age & &  80.7 & 90.1 & 96.4  &   - \\
 \\[-1em]
 \multicolumn{2}{c}{} & && & FIQ & &  0.0 & 0.0 & 0.0 & - \\
  \\[-1em]
  \cline{6-6} \cline{8-11}
  \\[-1em]
 \multicolumn{2}{c}{} & &  \multirow{3}{*}{MUA (SBH) } & & age & & 100 & 100 & 100 & -  \\
 \\[-1em]
 \multicolumn{2}{c}{} & && & FIQ & &  100 & 100 & 100 & -  \\
  \\[-1em]
  \cline{6-6} \cline{8-11}
  \\[-1em]
 \multicolumn{2}{c}{} & &  \multirow{3}{*}{MUA (BH) } & & age & & 94.5 & 99.7 & 99.8 & -  \\
 \\[-1em]
 \multicolumn{2}{c}{} & && & FIQ & &  84.8 & 96.9 & 95.8 & - \\
  \\[-1em]
\cline{6-6} \cline{8-11}
  \\[-1em]
 \multicolumn{2}{c}{} & &  \multirow{3}{*}{MUA (BY) } & & age & &  76.2 & 81.7 & 93.7 & -  \\
 \\[-1em]
 \multicolumn{2}{c}{} & && & FIQ & &  18.9 & 29.6& 31.2 & - \\
  \\[-1em]
 \hline
\\[-1em]
 \multicolumn{2}{c}{\multirow{13}{*}{Dorsal}}  &&& & & & Temporal Mid R & Postcentral L & Parietal Sup L &   -  \\
 \\[-1em]
 \multicolumn{2}{c}{} & & \multirow{3}{*}{ BHM$(d=0.05)$ } && age & &  62.9 & 38.6 & 13.8  &   - \\
 \\[-1em]
 \multicolumn{2}{c}{} & && & FIQ & &  1.5 & 4.7 & 4.5 & - \\
  \\[-1em]
  \cline{6-6} \cline{8-11}
  \\[-1em]
 \multicolumn{2}{c}{} & &  \multirow{3}{*}{MUA (SBH) } & & age & & 99.9 & 83.8 & 62.7 & -  \\
 \\[-1em]
 \multicolumn{2}{c}{} & && & FIQ & &  100 & 100 & 100 & -  \\
  \\[-1em]
  \cline{6-6} \cline{8-11}
  \\[-1em]
 \multicolumn{2}{c}{} & &  \multirow{3}{*}{MUA (BH) } & & age & & 96.9 & 74.0 & 34.5 & -  \\
 \\[-1em]
 \multicolumn{2}{c}{} & && & FIQ & &  89.6 & 91.7 & 79.1 & - \\
  \\[-1em]
\cline{6-6} \cline{8-11}
  \\[-1em]
 \multicolumn{2}{c}{} & &  \multirow{3}{*}{MUA (BY) } & & age & &  75.8 & 51.6 & 18.0 & -  \\
 \\[-1em]
 \multicolumn{2}{c}{} & && & FIQ & &  21.1 & 26.8 & 18 &  - \\
  \\[-1em]
 \hline
\\[-1em]
  \multicolumn{2}{c}{\multirow{13}{*}{Default}}  && &&  & & Temporal Mid R & Frontal Med Orb R & Frontal Med Orb L & Occipital Mid R  \\
    \\[-1em]
  \multicolumn{2}{c}{} & & \multirow{3}{*}{ BHM$(d=0.05)$ } && age & &   77.7 & 26.2 & 22.7 & 99.5 \\
 \\[-1em]
 \multicolumn{2}{c}{} & && & FIQ & &  0.0 & 0.0 & 0.0 & 0.0\\
  \\[-1em]
  \cline{6-6} \cline{8-11}
  \\[-1em]
 \multicolumn{2}{c}{} & &  \multirow{3}{*}{MUA (SBH) } & & age & & 100 & 100 & 100 & 100 \\
 \\[-1em]
 \multicolumn{2}{c}{} & && & FIQ & &  100 & 100 & 100 & 100  \\
  \\[-1em]
  \cline{6-6} \cline{8-11}
  \\[-1em]
 \multicolumn{2}{c}{} & &  \multirow{3}{*}{MUA (BH) } & & age & & 97.6 & 62.0 & 57.7 & 100.0  \\
 \\[-1em]
 \multicolumn{2}{c}{} & && & FIQ & &  89.7 & 70.1 & 84.1 & 99.2 \\
  \\[-1em]
\cline{6-6} \cline{8-11}
  \\[-1em]
 \multicolumn{2}{c}{} & &  \multirow{3}{*}{MUA (BY) } & & age & &   79.5 & 31.9 & 25.7 & 99.6  \\
 \\[-1em]
 \multicolumn{2}{c}{} & && & FIQ & &  22.7 & 6.2 & 11.1 & 39.6 \\
  \\[-1em]
 \hline
 \hline
\end{tabular}
\end{adjustbox}
\end{table}

\section{Concluding remarks}

In this article, we have extended to image data a Bayesian hierarchical Gaussian process (GP) model that uses a flexible Inverse-Wishart process prior to handle within-image dependency, and have proposed a novel spatial global-local spike-and-slab prior that broadly relates to a rich class of well-studied selection priors. The proposed prior construction achieves simultaneous global (i.e, at covariate-level) and local (i.e., at pixel/voxel-level) selection via participation rate parameters that measure the probability for the individual covariates to affect the observed images. We have used hard-thresholding to decide whether a covariate should be included in the model and have shown on simulated data that parameters are interpretable and lead to efficient selection. The introduced participation rate and threshold parameters establish a bridge between global and local level selection, allowing global selection to be informed
by the selection at the local level. This framework can be applied to more general functional data applications.

There are several interesting future directions to extend our model.  Our rationale for choosing an independent prior on spatial coefficients $\beta_j(s)$ has been largely computational. Our efficient Gibbs sampler takes advantage of this independent prior and only requires to invert the $|\bm{S}|-by-|\bm{S}|$ covariance matrix once for the noise-free mean surface $Z(\cdot)$, roughly $O(|\bm{S}|^3$) at each Gibbs iteration. In the application to real data, our model is able to handle relatively large datasets, with $S$ about 10,000 voxels, and $n$ about 1,000 subject images. 
With a dependent prior, we would not be able to parallelize computations, which would result into having to calculate the inversion of at most $|\bm{S}|-by-|\bm{S}|$ covariance matrices for $q$ covariates at each iteration, with roughly a $O( |\bm{S}|^3q )$ complexity at each iteration. This would have been infeasible for our application. In addition, a construction with a dependent prior would require a more careful interpretation of the participation rate parameters $\pi_j$'s, which measure the probability for the individual covariates to affect the observed image under the assumption of independence.  Given these challenges, we have decided to leave the investigation of dependent priors to future work.  We note, however, that, even though we do not explicitly account for dependency among the coefficients, our model borrows information across voxels via the use of the spatial Gaussian process prior $\mathcal{GP}\left(\mu(\cdot), \Sigma(\cdot, \cdot)\right)$ on $Z_i$.  

In the applications of this paper, when investigating the role of the parameter $d$ and the sensitivity of the results to the specification of this parameter, we found the case $d=0$ interesting. In this degenerate case the model includes all the covariates at each iteration, to explain the observed images, and the traces of the parameters $\pi_j$ inform us on the relative importance of the individual covariates. These trace plots provide an empirical tool that might be helpful in the choice of $d$, particularly in cases where a separation among the traces is observed. In the Supplementary Material we show these plots for one of the simulated scenarios used in this paper, along with comments on how the plots can guide the user in the choice of $d$. We remark, however, that this procedure is ad-hoc and cannot be used as a general method, in particular as the behavior of the trace plots is application-dependent and a clear separation of the traces might not always be observed. We leave further investigation of the role and properties of the parameter $d$ to future work. In the absence of prior information, we recommend to view $d$ as conventional sparsity parameter and use standard values, i.e. $d = 0.05$ or $d = 0.1$. Our sensitivity analyses in the simulations and real data application have shown good performances overall for different choices of $d$, with highly consistent local selection results.

Finally, our proposed global-local selection prior construction can be potentially useful for other modeling settings, such as function-on-scalar and network-on-scalar regressions.

\begin{supplement}
Supplementary Material to “Bayesian Image-on-Scalar Regression with a Spatial Global-Local Spike-and-Slab Prior”; \vspace{0.5em} \\ 
Codes available to “Bayesian Image-on-Scalar Regression with a Spatial Global-Local Spike-and-Slab Prior”, including scripts to recreate the simulated data; \\
Github page: \href{https://github.com/ZijianZeng/BIoS_SGLSS}{https://github.com/ZijianZeng/BIoS\_SGLSS}
\end{supplement}

\bibliographystyle{ba}
\bibliography{BIoS}

\section*{Supplementary Material}
\vspace{0.5cm}
\subsection*{S1. Markov Chain Monte Carlo Sampling (MCMC)}
In this section, we provide the detailed derivations for the Gibbs sampler in Section~\ref{sec.settings}. 
\begin{itemize}
    \item \textbf{Update the BHM parameters $\left\{Z_i(\vec{\bm{s}})\right\}^n_{i=1}$ and $\sigma^2_\epsilon$ conditional on $\{\beta_j(\vec{\bm{s}})\}^q_{j=0} $ and $\Sigma(\vec{\bm{s}},\vec{\bm{s}})$}: \\
    Evaluated on locations $\vec{\bm{s}}$, Equations~\eqref{BHM.eq.1} and~\eqref{BHM.eq.2} yield 
        \begin{equation*}
        \begin{aligned}
        & Y_i(\vec{\bm{s}}) | Z_i(\vec{\bm{s}}), \sigma^2_\epsilon  \sim \text{MVN}( Z_i(\vec{\bm{s}}), \sigma^2_\epsilon I_p), \quad \sigma^2_\epsilon  \sim \text{Inverse-Gamma}(a_\epsilon, b_\epsilon),\\
        & Z_i(\vec{\bm{s}}) | \{\beta_j(\vec{\bm{s}})\}_{j=1}^q, \Sigma( \vec{\bm{s}},\vec{\bm{s}})  \sim \text{MVN}( \mu_i(\vec{\bm{s}}), \Sigma(\vec{\bm{s}},\vec{\bm{s}})), \quad \mu_i(\vec{\bm{s}}) = \beta_0(\vec{\bm{s}}) + \sum^q_{j=1}x_{ij} \beta_j(\vec{\bm{s}}),
        \end{aligned}
    \end{equation*}
    leading to a Normal Inverse-Gamma conjugacy.
    
    With $\mu_i(\vec{\bm{s}}) = \beta_0(\vec{\bm{s}}) + \sum^q_{j=1} x_{ij}\beta_j(\vec{\bm{s}})$, we have
    \begin{equation*}
    \begin{aligned}
    \raggedleft
        &  p(Z_i(\vec{\bm{s}}) | Y_i(\vec{\bm{s}}),\left\{ \beta_j(\vec{\bm{s}})\right\}^q_{j=0}, \Sigma\left(\vec{\bm{s}},\vec{\bm{s}}\right), \sigma^2_\epsilon)\\
         \propto & p(Y_i(\vec{\bm{s}}) | Z_i(\vec{\bm{s}}), \sigma^2_\epsilon) p\left( Z_i(\vec{\bm{s}}) | \left\{\beta_j(\vec{\bm{s}})\right\}^q_{j=0}, \Sigma(\vec{\bm{s}}, \vec{\bm{s}}) \right) \\
         \propto & \exp \left\{ - \frac{1}{2} \left( Y_i(\vec{\bm{s}}) - Z_i(\vec{\bm{s}})\right)^T \sigma^{-2}_\epsilon I_p \left( Y_i(\vec{\bm{s}}) - Z_i(\vec{\bm{s}}) \right)\right\}  \exp \left\{ - \frac{1}{2}\left( Z_i(\vec{\bm{s}}) - \mu_i(\vec{\bm{s}})\right)^T \right. \\
         & \quad \times\left.\Sigma^{-1}(\vec{\bm{s}},\vec{\bm{s}})\left(Z_i(\vec{\bm{s}}) - \mu_i(\vec{\bm{s}}) \right)\right\} \\     
         \propto & \exp \left\{ - \frac{1}{2} \left[ Z^T_{i}(\vec{\bm{s}})\left( \sigma^{-2}_\epsilon I_p + \Sigma^{-1}(\vec{\bm{s}},\vec{\bm{s}})\right)Z_i(\vec{\bm{s}}) - 2Z^T_i(\vec{\bm{s}})\left( \sigma^{-2}_\epsilon  Y_i(\vec{\bm{s}}) \right. \right. \right. \\
         & \quad 
          \left. \left. \left. + \Sigma^{-1}(\vec{\bm{s}},\vec{\bm{s}}) \mu_i(\vec{\bm{s}})\right)\right]\right\}
        \end{aligned}
    \end{equation*} 
    which gives to the posterior distribution,
    \begin{equation*}
        \begin{aligned}
         & Z_i(\vec{\bm{s}}) | Y_i(\vec{\bm{s}}), \mu_i(\vec{\bm{s}}), \Sigma(\vec{\bm{s}},\vec{\bm{s}}), \sigma^2_\epsilon  \sim  MVN\left( \mu_{Z_i}(\vec{\bm{s}}), V_{Z_i}(\vec{\bm{s}})\right), \\
         & V_{Z_i}(\vec{\bm{s}})  =  \left( \sigma^{-2}_\epsilon I_p + \Sigma^{-1}(\vec{\bm{s}},\vec{\bm{s}})\right)^{-1}, \\
          & \mu_{Z_i}(\vec{\bm{s}})  =  V_{Z_i}(\vec{\bm{s}}) \left( \sigma^{-2}_\epsilon Y_i(\vec{\bm{s}}) + \Sigma^{-1}(\vec{\bm{s}},\vec{\bm{s}})\mu_i(\vec{\bm{s}})\right).
        \end{aligned}
    \end{equation*} 
    The posterior distribution of the corresponding variance component is given by
    \begin{equation*}
        \begin{aligned}
        &p(\sigma^2_\epsilon | \{Y_i(\vec{\bm{s}})\}_{i=1}^n, \{Z_i(\vec{\bm{s}})\}_{i=1}^n )  \propto  \prod^n_{i=1}p( Y_i(\vec{\bm{s}}) | Z_i(\vec{\bm{s}}), \sigma^2_\epsilon) p(\sigma^2_{\epsilon}) \\
        \propto & {\hskip1em\relax}   |\sigma^2_{\epsilon} I_p |^{-\frac{n}{2}} \exp\left\{ -\frac{1}{2} \sum^n_{i=1} \left( Y_i(\vec{\bm{s}}) - Z_i(\vec{\bm{s}})\right)^T \sigma^{-2}_\epsilon I_p \left(Y_i(\vec{\bm{s}}) - Z_i(\vec{\bm{s}})\right) \right\} \\
        & {\hskip3em\relax} \times \left( \sigma^{-2}_{\epsilon}\right)^{a_\epsilon + 1} \exp\left\{ - \sigma^{-2}_\epsilon b_{\epsilon}\right\} \\
        \propto & {\hskip1em\relax} \left( \sigma^{-2}_\epsilon \right)^{\frac{np}{2} + a_\epsilon + 1} \exp\left\{ - \sigma^{-2}_\epsilon \left[ b_{\epsilon}+ \frac{1}{2}\sum^n_{i=1} \left( Y_i(\vec{\bm{s}}) - Z_i(\vec{\bm{s}})\right)^T\left(Y_i(\vec{\bm{s}}) - Z_i(\vec{\bm{s}})\right)\right]\right\}, 
        \end{aligned}
    \end{equation*}
    which gives
    \begin{equation*}
        \begin{aligned}
      &\sigma^2_{\epsilon} | \{Y_i(\vec{\bm{s}})\}_{i=1}^n, \{Z_i(\vec{\bm{s}})\}_{i=1}^n  \sim  \\
      & {\hskip4em\relax} \text{InverseGamma}\left( a_\epsilon + \frac{np}{2},  b_{\epsilon}+ \frac{1}{2}\sum^n_{i=1} \left( Y_i(\vec{\bm{s}}) - Z_i(\vec{\bm{s}})\right)^T\left(Y_i(\vec{\bm{s}}) - Z_i(\vec{\bm{s}})\right) \right).
        \end{aligned}
    \end{equation*}
    \item \textbf{Update the SGLSS prior parameters $\left\{\beta_j(\vec{\bm{s}}), \tau_j(\vec{\bm{s}}),\pi_j\right\}^q_{j=0}$ conditional on $\left\{Z_i(\vec{\bm{s}})\right\}^n_{i=1}$ and $\Sigma(\vec{\bm{s}},\vec{\bm{s}})$}:  \\
    In this step, we first sample the indicators $\left\{\tau_j(\vec{\bm{s}})\right\}^q_{j=1}$ and update $\left\{\pi_j\right\}^q_{j=1}$ to obtain selection indicators at both global and local levels. This is achieved via a blocked Gibbs strategy and has the Beta Binomial conjugacy when integrating out $\tilde{\beta}_j(\vec{\bm{s}})$.
    
    Using the blocked Gibbs sampler with respect to each feature $j$, we have $\tilde{Z}_{ij}(\vec{\bm{s}}) = Z_i(\vec{\bm{s}}) - \sum_{j' \ne j} x_{ij'}\bm{\beta}_{j'}(\vec{\bm{s}})$. Furthermore, we denote $\tilde{z}_{ij\bm{s}} = \tilde{Z}_{ij}(\bm{s}), \tilde{\beta}_{j\bm{s}} = \tilde{\beta}_j(\bm{s})$, $\mu_{0j\bm{s}} = \mu_{0j}(\bm{s}), \sigma^2_{0j\bm{s}} = \sigma^2_{0j}(\bm{s}), \sigma^2_{\bm{s}} = \Sigma(\bm{s}, \bm{s}), \bm{s} \in \vec{\bm{s}}$ and $\tilde{z}_{\cdot j\bm{s}} = \{\tilde{z}_{ij\bm{s}}\}^n_{i=1}$ for short, where $\bm{s}$ is one location in the vector of locations $\vec{\bm{s}}$.

    Based on Equation~\eqref{BGBF.eq.2}, we calculate the marginal posterior probability of $\tau_j(\bm{s}) = 1$ by integrating out $\tilde{\beta}_{j\bm{s}}$, 
    \begin{equation*}
    \resizebox{0.9\textwidth}{!}{$
    \begin{aligned}
    && p\left( \tau_j(\bm{s}) = 1 | \tilde{z}_{\cdot j\bm{s}} \right) =  \dfrac{  \int p\left(\tau_j(\bm{s}) = 1, \tilde{\beta}_{j\bm{s}} | \tilde{z}_{\cdot j\bm{s}}\right) d\tilde{\beta}_{j\bm{s}}  }{ p\left(\tau_j(\bm{s}) = 0, \tilde{\beta}_{j\bm{s}} = 0 | \tilde{z}_{\cdot j\bm{s}}\right) +  \int p\left(\tau_j(\bm{s}) = 1, \tilde{\beta}_{j\bm{s}} | \tilde{z}_{\cdot j\bm{s}}\right) d\tilde{\beta}_{j\bm{s}}  }.
    \end{aligned}   
    $}
    \end{equation*}
    Substituting 
    \begin{equation*}
    \resizebox{0.9\textwidth}{!}{$
    \begin{aligned}
      \int p\left(\tau_j(\bm{s}) = 1, \tilde{\beta}_{j\bm{s}} | \tilde{z}_{\cdot j\bm{s}}\right) d\tilde{\beta}_{j\bm{s}} 
        & =   \frac{1}{p\left( \tilde{z}_{\cdot j\bm{s}} \right)} \int p\left(\tilde{z}_{\cdot j\bm{s}} | \tau_j(\bm{s}) = 1, \tilde{\beta}_{j\bm{s}} \right) p\left( \tilde{\beta}_{j\bm{s}}\right) p(\tau_j(\bm{s}) = 1 | \pi_j)d\tilde{\beta}_{j\bm{s}}
    \end{aligned}   
        $}
    \end{equation*}
    and 
    $$
    p\left(\tau_j(\bm{s}) = 0, \tilde{\beta}_{j\bm{s}} = 0 | \tilde{z}_{\cdot j\bm{s}}\right)  
         =   \frac{1}{p\left( \tilde{z}_{\cdot j\bm{s}}  \right)}  p\left(\tilde{z}_{\cdot j\bm{s}} | \tau_j(\bm{s}) = 0, \tilde{\beta}_{j\bm{s}} = 0 \right)  p(\tau_j(\bm{s}) = 0 | \pi_j)
    $$
    yields
    \begin{eqnarray*}
     \resizebox{0.88\textwidth}{!}{$
     \begin{aligned}
        & p\left( \tau_j(\bm{s}) = 1 | \tilde{z}_{\cdot j\bm{s}} \right) \\ 
        & =  \dfrac{ \int p\left(\tilde{z}_{\cdot j\bm{s}} | \tau_j(\bm{s}) = 1, \tilde{\beta}_{j\bm{s}} \right) p\left( \tilde{\beta}_{j\bm{s}}\right)d\tilde{\beta}_{j\bm{s}} \times \pi_j }{ p\left(\tilde{z}_{\cdot j\bm{s}} | \tau_j(\bm{s}) = 0, \tilde{\beta}_{j\bm{s}} = 0 \right)\times (1 - \pi_j) +   \int p\left(\tilde{z}_{\cdot j\bm{s}} | \tau_j(\bm{s}) = 1, \tilde{\beta}_{j\bm{s}} \right) p\left( \tilde{\beta}_{j\bm{s}}\right)d\tilde{\beta}_{j\bm{s}} \times \pi_j   }.
    \end{aligned}
        $}
    \end{eqnarray*}
    For $p\left(\tilde{z}_{\cdot j\bm{s}} | \tau_j(\bm{s}) = 0, \tilde{\beta}_{j\bm{s}} = 0 \right)$, we have
    \begin{equation*}
    p\left(\tilde{z}_{\cdot j\bm{s}} | \tau_j(\bm{s}) = 0, \tilde{\beta}_{j\bm{s}} = 0 \right) = 
         \underbrace{\left(2\pi \sigma^2_{\bm{s}}\right)^{-\frac{n}{2}} \exp\left\{ - \frac{1}{2} \left( \sum^n_{i=1}\tilde{z}^2_{ij\bm{s}}/\sigma^2_{\bm{s}} \right)\right\} }_{\text{common factor (CF)}}.
    \end{equation*}
    For $ \int p\left(\tilde{z}_{\cdot j\bm{s}} | \tau_j(\bm{s}) = 1, \tilde{\beta}_{j\bm{s}} \right) p\left( \tilde{\beta}_{j\bm{s}}\right)d\tilde{\beta}_{j\bm{s}}$, we have
\begin{equation*}
    \resizebox{0.9\textwidth}{!}{$
    \begin{aligned}
        &  \int \left( 2\pi \sigma_{\bm{s}}^2 \right)^{-\frac{n}{2}} \exp\left\{ -\frac{1}{2} \sum^n_{i=1} \left( \tilde{z}_{ij \bm{s}} - x_{ij}\tilde{\beta}_{j\bm{s}}\right)^2/\sigma^{2}_{\bm{s}} \right\}  \left(2\pi \sigma^2_{0j\bm{s}}\right)^{-\frac{1}{2}} \exp\left\{ - \frac{1}{2} \left(\tilde{\beta}_{j\bm{s}} -\mu_{0j\bm{s}}\right)^2/\sigma^{2}_{0j\bm{s}} \right\} d \tilde{\beta}_{j\bm{s}} \\
        & =  \left( 2\pi \sigma^2_{\bm{s}}\right)^{-\frac{n}{2}} \left(2\pi \sigma^2_{0j\bm{s}}\right)^{-\frac{1}{2}}\int \exp\left\{ - \frac{1}{2} \left[ \tilde{\beta}^2_{j\bm{s}}\left( \sum^n_{i=1} x^2_{ij} / \sigma^2_{\bm{s}}\right) - 2\tilde{\beta}_{j\bm{s}}\left( \sum^n_{i=1} x_{ij}\tilde{z}_{ij\bm{s}} / \sigma^{2}_{\bm{s}} \right) \right. \right. \\
        &  \left. \left. + \sum^n_{i=1} \tilde{z}^2_{ij\bm{s}}/ \sigma^{2}_{\bm{s}} \right] -\frac{1}{2} \left[ \tilde{\beta}^2_{j\bm{s}}/\sigma^2_{0j\bm{s}}  - 2\tilde{\beta}_{j\bm{s}}\left( \mu_{0j\bm{s}} /\sigma^{2}_{0j\bm{s}} \right) + \mu^2_{0j\bm{s}}/\sigma^{2}_{0j\bm{s}}\right]\right\}d\tilde{\beta}_{j\bm{s}}  \\
        & =  \left(2\pi \sigma^2_{\bm{s}}\right)^{-\frac{n}{2}} \exp\left\{ - \frac{1}{2} \left( \sum^n_{i=1}\tilde{z}^2_{ij\bm{s}}/\sigma^2_{\bm{s}} \right)\right\} \left(2\pi \sigma^2_{0j\bm{s}}\right)^{-\frac{1}{2}}\exp\left\{ - \frac{1}{2} \left(\mu_{0j\bm{s}}^2 /\sigma^{2}_{0j\bm{s}}\right)\right\}  \\
        & \times \int \exp\left\{ - \frac{1}{2} \left[ \tilde{\beta}^2_{j\bm{s}} \underbrace{\left( \sum^n_{i=1}x^2_{ij}/\sigma^{2}_{\bm{s}} + 1/\sigma^{2}_{0j\bm{s}}\right)}_{\tilde{v}^{-1}_{j\bm{s}}}  - 2\tilde{\beta}_{j\bm{s}} \underbrace{\left( \sum^n_{i=1} x_{ij}\tilde{z}_{ij\bm{s}} /\sigma^{2}_{\bm{s}} + \mu_{0j\bm{s}}/\sigma^{2}_{0j\bm{s}}\right)}_{\tilde{m}_{j\bm{s}}}\right]\right\} d\tilde{\beta}_{j\bm{s}}  \\
        & =  \left(2\pi \sigma^2_{\bm{s}}\right)^{-\frac{n}{2}} \exp\left\{ - \frac{1}{2} \left( \sum^n_{i=1}\tilde{z}^2_{ij\bm{s}}/\sigma^2_{\bm{s}} \right)\right\} \left(2\pi \sigma^2_{0j\bm{s}}\right)^{-\frac{1}{2}}\exp\left\{ - \frac{1}{2} \left(\mu_{0j\bm{s}}^2 /\sigma^{2}_{0j\bm{s}}\right)\right\}  \\
        & \times  \int \left( 2\pi \tilde{\nu}_{j\bm{s}}\right)^{-\frac{1}{2}} \exp\left\{ - \frac{1}{2}\left( \tilde{\beta}^2_{j\bm{s}}  - 2\tilde{\beta}_{j\bm{s}} \tilde{v}_{j\bm{s}} \tilde{m}_{j\bm{s}} + \tilde{\nu}^2_{j\bm{s}}\tilde{m}^2_{j\bm{s}}\right)/\tilde{\nu}_{j\bm{s}} \right\} d\tilde{\beta}_{j\bm{s}} \\
        & \times   \left( 2\pi \tilde{\nu}_{j\bm{s}}\right)^{\frac{1}{2}} \exp\left\{ \frac{1}{2} \tilde{m}^2_{j\bm{s}}\tilde{\nu}_{j\bm{s}} \right\} \\
        & =  \underbrace{\left(2\pi \sigma^2_{\bm{s}}\right)^{-\frac{n}{2}} \exp\left\{ - \frac{1}{2} \left( \sum^n_{i=1}\tilde{z}^2_{ij\bm{s}}/\sigma^2_{\bm{s}} \right)\right\} }_{\text{common factor (CF)}} \times \underbrace{\left( \sigma^2_{0j\bm{s}}\right)^{-\frac{1}{2}}\exp\left\{ - \frac{1}{2} \left(\mu_{0j\bm{s}}^2 /\sigma^{2}_{0j\bm{s}}\right)\right\} }_{\text{prior factor (PF)}} \\
        & \times  \left(  \tilde{\nu}_{j\bm{s}}\right)^{\frac{1}{2}} \exp\left\{ \frac{1}{2} \tilde{m}^2_{j\bm{s}}\tilde{\nu}_{j\bm{s}} \right\} .
    \end{aligned}$}
    \end{equation*}
    Combining both, we obtain that
    \begin{eqnarray*}
        && p\left( \tau_j(\bm{s}) = 1 | \tilde{z}_{\cdot j\bm{s}} \right)  \\
        & = & \frac{ \text{CF} \times \text{PF} \times  \left(  \tilde{\nu}_{j\bm{s}}\right)^{\frac{1}{2}} \exp\left\{ \frac{1}{2} \tilde{m}^2_{j\bm{s}}\tilde{\nu}_{j\bm{s}} \right\}  \pi_j }{
        \text{CF} \times (1-\pi_j) + \text{CF} \times \text{PF} \times \left(  \tilde{\nu}_{j\bm{s}}\right)^{\frac{1}{2}} \exp\left\{ \frac{1}{2} \tilde{m}^2_{j\bm{s}}\tilde{\nu}_{j\bm{s}} \right\} \pi_j} \\
        & = & \frac{ 1 }{ 1 + \theta_{j\bm{s}}},
    \end{eqnarray*}
    with
    $$
        \theta_{j\bm{s}} = \frac{1-\pi_j}{ \pi_j \times \left( \sigma^2_{0j\bm{s}}\right)^{-\frac{1}{2}}\exp\left\{ - \frac{1}{2} \left(\mu_{0j\bm{s}}^2 /\sigma^{2}_{0j\bm{s}}\right)\right\}   \times \left(  \tilde{\nu}_{j\bm{s}}\right)^{\frac{1}{2}} \exp\left\{ \frac{1}{2} \tilde{m}^2_{j\bm{s}}\tilde{\nu}_{j\bm{s}} \right\}  },
    $$
    and
    \begin{eqnarray*}
        \tilde{\nu}_{j\bm{s}}  =  \left[ \sum^n_{i=1}x^2_{ij}/\sigma^{2}_{\bm{s}} + 1/\sigma^{2}_{0j\bm{s}} \right]^{-1}, \quad 
        \tilde{m}_{j\bm{s}} = \sum^n_{i=1} x_{ij}\tilde{z}_{ij\bm{s}} /\sigma^{2}_{\bm{s}} + \mu_{0j\bm{s}}/\sigma^{2}_{0j\bm{s}}.
    \end{eqnarray*}
    This gives the posterior distribution for $\tau_{j}(\bm{s})$ as a Bernoulli distribution, and leads to the Beta distribution for $\pi_j$ by counting the $\tau_{j}(\vec{\bm{s}})$ samples 
    $$
        \pi_j | \tau_j(\vec{\bm{s}}) \sim \text{Beta}\left( a_{\pi_j} + \sum_{\bm{s} \in \vec{\bm{s}}} \tau_{j}(\bm{s}), b_{\pi_j} + p - \sum_{\bm{s} \in \vec{\bm{s}}} \tau_{j}(\bm{s})\right),
    $$
    as written in Equation~\eqref{BGBF.eq.3}.
    
    Conditional on selection indicators at the two levels, we sample the coefficient image $\beta_j(\vec{\bm{s}})$ as summarized in Section~\ref{sec.settings}. 

    \item \textbf{Update the IWP prior parameter $\Sigma(\vec{\bm{s}},\vec{\bm{s}})$ conditional on  $\left\{Z_i(\vec{\bm{s}})\right\}^n_{i=1}$ and $\{\beta_j(\vec{\bm{s}})\}^q_{j=0}$}: \\
    This is a Gibbs step from Multivariate Normal Inverse Wishart conjugacy. We have
    \begin{equation*}
    \begin{aligned}
        & p\left( \Sigma(\vec{\bm{s}}, \vec{\bm{s}}) | \{Z_i(\vec{\bm{s}}\}^n_{i=1}, \{\beta_j(\vec{\bm{s}})\}^q_{j=0}, \Psi(\vec{\bm{s}}, \vec{\bm{s}}) \right) \\
        \propto & {\hskip1em\relax} \prod^n_{i=1} p\left( Z_i(\vec{\bm{s}}) | \left\{\beta_j(\vec{\bm{s}})\right\}^q_{j=0}, \Sigma(\vec{\bm{s}}, \vec{\bm{s}}) \right)  p \left( \Sigma( \vec{\bm{s}}, \vec{\bm{s}}) \right) \\
        \propto & {\hskip1em\relax}  \prod^n_{i=1} \left\{ |\Sigma(\vec{\bm{s}}, \vec{\bm{s}})|^{-\frac{1}{2}} \exp\left\{ -\frac{1}{2} \left( Z_i(\vec{\bm{s}}) - \mu_i(\vec{\bm{s}})\right)^T\Sigma^{-1}(\vec{\bm{s}}, \vec{\bm{s}})\left(Z_i(\vec{\bm{s}}) - \mu_i(\vec{\bm{s}}) \right)\right\}\right\}\\
        &  {\hskip10em\relax} \times |\Sigma(\vec{\bm{s}}, \vec{\bm{s}})|^{-\frac{(\delta +p+1)}{2} }\exp\left\{ -\frac{1}{2} \text{tr}\left(  \Psi(\vec{\bm{s}}, \vec{\bm{s}}) \Sigma^{-1}(\vec{\bm{s}},\vec{\bm{s}})\right)\right\} 
    \end{aligned}
    \end{equation*}
    \begin{equation*}
        \begin{aligned}
        & \propto  |\Sigma(\vec{\bm{s}},\vec{\bm{s}})|^{-\frac{(n+\delta +p + 1)}{2}} \exp\left\{ - \frac{1}{2}\text{tr}\left[\left(  \sum^n_{i=1}\left(Z_i(\vec{\bm{s}}) - \mu_i(\vec{\bm{s}}) \right)\left( Z_i(\vec{\bm{s}}) - \mu_i(\vec{\bm{s}}) \right)^T  \right. \right. \right. \\
        & {\hskip22em\relax} \left. \left. \left. + \Psi(\vec{\bm{s}},\vec{\bm{s}}) \right)\Sigma^{-1}(\vec{\bm{s}}, \vec{\bm{s}}) \right]\right\},
        \end{aligned}  
    \end{equation*}
    which gives the Inverse Wishart distribution
  \begin{equation*}
    \begin{aligned}
     & \Sigma(\vec{\bm{s}},\vec{\bm{s}}) |  \{Z_i(\vec{\bm{s}})\}^n_{i=1}, \{\beta_j(\vec{\bm{s}})\}^q_{j=0},\Psi(\vec{\bm{s}}, \vec{\bm{s}})   \sim \\
     &  {\hskip8em\relax} \text{IW}\left( n +  \delta, \sum^n_{i=1}\left(Z_i(\vec{\bm{s}}) - \mu_i(\vec{\bm{s}}) \right)\left( Z_i(\vec{\bm{s}}) - \mu_i(\vec{\bm{s}}) \right)^T  + \Psi(\vec{\bm{s}},\vec{\bm{s}})  \right).
        \end{aligned} 
    \end{equation*}
    with, again, $\mu_i(\bm{s})  = \beta_0(\vec{\bm{s}}) + \sum^{q}_{j=1} x_{ij}\beta_j(\bm{s})$.
\end{itemize}

\subsection*{S2. Simulation results with $d = 0.01$ and $d = 0.1$}
Tables~\ref{tab:sim.ogls.1} and~\ref{tab:sim.ogls.2} report precision, recall and $F_1$ scores for both global and local selections for our method with $d=0.01, 0.05$ and $d=0.1$. Clearly, a lower $d$ tends to include more covariates, potentially selecting noisy covariates. On the contrary, a higher $d$ tends to exclude more covariates, potentially selecting out influential covariates. Correspondingly, we can observe that as $d$ increases, precision tends to increase and recall tends to decrease especially in the more challenging second simulated scenario. In addition, when there exists a `good separation' between the influential and noisy covariates, like in the first scenario and the second scenario with $\pi \approx 18.8\%$, good performances overall can be observed for different choices of $d$. This is also reflected in the MSE estimates of the model parameters shown in Tables~\ref{tab:sim.omse.1} and~\ref{tab:sim.omse.2}.

\begin{table}[!hbt]
\caption{First simulated scenario: Global-local selection for a representative dataset}
\label{tab:sim.ogls.1}
\centering
\begin{adjustbox}{max width = \textwidth}
\begin{tabular}{ccccccccccccc}
 \hline
 \hline
 \multicolumn{13}{c}{Precision} \\
 \hline
 \multicolumn{2}{c}{Thresholds} && Global && \multicolumn{8}{c}{Local} \\ 
 \hline 
 & &&  && $\tau_1(\vec{\bm{s}})$ & $\tau_2(\vec{\bm{s}})$ & $\tau_3(\vec{\bm{s}})$ & $\tau_4(\vec{\bm{s}})$ & $\tau_5(\vec{\bm{s}})$ & $\tau_6(\vec{\bm{s}})$ & $\tau_7(\vec{\bm{s}})$ & $\tau_8(\vec{\bm{s}})$ \\  \\[-1em]
 \cline{4-4} \cline{6-13}   \\[-1em]
\multicolumn{2}{c}{ BHM ($d = 0.01$) } && $0.727$ &&   $1   $ &  $0.901$ &  $0.973$ &  $0.991$ &  $0.977$ &  $1   $ &  $0.954$ &  $0.946$ \\ 
\multicolumn{2}{c}{ BHM ($d = 0.05$) } && $1$ &&  $1   $ & $0.901$ & $0.973$ & $0.985$ & $0.981$ & $1   $ & $0.949$ & $0.958$ \\
\multicolumn{2}{c}{ BHM ($d = 0.1$) } && $1$ &&   $1    $ & $0.901$ &  $0.978$ &  $0.99 $ &  $0.977$ &  $1    $ & $0.955$ &  $0.961$  \\
\hline 
\hline 
\multicolumn{13}{c}{Recall} \\
 \hline
 \multicolumn{2}{c}{Thresholds} && Global && \multicolumn{8}{c}{Local} \\
 \hline
& &&  && $\tau_1(\vec{\bm{s}})$ & $\tau_2(\vec{\bm{s}})$ & $\tau_3(\vec{\bm{s}})$ & $\tau_4(\vec{\bm{s}})$ & $\tau_5(\vec{\bm{s}})$ & $\tau_6(\vec{\bm{s}})$ & $\tau_7(\vec{\bm{s}})$ & $\tau_8(\vec{\bm{s}})$ \\  \\[-1em]
 \cline{4-4} \cline{6-13}  \\[-1em]
\multicolumn{2}{c}{ BHM ($d = 0.01$) } && $1$ &&  $0.830 $ &  $1   $ &  $0.983$ &  $0.906$ &  $0.943$ &  $0.717$ &  $0.802$ &  $0.772$ \\ 
\multicolumn{2}{c}{ BHM ($d = 0.05$) } && $1$ && $0.821$ & $1   $ & $0.983$ & $0.917$ & $0.944$ & $0.742$ & $0.823$ & $0.769$ \\
\multicolumn{2}{c}{ BHM ($d = 0.1$) } && $1$ && $0.828$ &  $1    $ & $0.983$ &  $0.919$ &  $0.941$ &  $0.748$ &  $0.809$ &  $0.785$  \\
\hline 
\hline
\multicolumn{13}{c}{ $F_1$ scores } \\ 
\hline
 \multicolumn{2}{c}{Thresholds} && Global && \multicolumn{8}{c}{Local} \\
 \hline  
 & &&  && $\tau_1(\vec{\bm{s}})$ & $\tau_2(\vec{\bm{s}})$ & $\tau_3(\vec{\bm{s}})$ & $\tau_4(\vec{\bm{s}})$ & $\tau_5(\vec{\bm{s}})$ & $\tau_6(\vec{\bm{s}})$ & $\tau_7(\vec{\bm{s}})$ & $\tau_8(\vec{\bm{s}})$ \\ \\[-1em]
 \cline{4-4} \cline{6-13} \\[-1em]
\multicolumn{2}{c}{ BHM ($d = 0.01$) } && $0.842$ &&  $0.907$ &  $0.948$ &  $0.978$ &  $0.947$ &  $0.959$ &  $0.835$ &  $0.872$ &  $0.850 $ \\
\multicolumn{2}{c}{ BHM ($d = 0.05$) } && $1$ && $0.902$ & $0.948$ & $0.978$ & $0.950 $ & $0.962$ & $0.852$ & $0.882$ & $0.854$ \\
\multicolumn{2}{c}{ BHM ($d = 0.1$) } && $1$ && $0.906$ &  $0.948$ &  $0.981$ &  $0.953$ &  $0.958$ &  $0.856$ &  $0.876$ &  $0.864$  \\
\hline
\hline
\end{tabular} 
\end{adjustbox}
\end{table} 

\begin{table}[!hbt]
\caption{First simulated scenario: MSEs for a representative dataset}
\label{tab:sim.omse.1}
\centering
\begin{adjustbox}{max width = \textwidth}
\begin{tabular}{cccccccccccccc}
 \hline 
 \hline
 \multicolumn{14}{c}{MSE} \\
 \hline \\[-1em]
 \multicolumn{2}{c}{Thresholds} &&
 \multicolumn{2}{c}{$\left\{ Z_i(\vec{\bm{s}})\right\}^{100}_{i=1}$} &&  \multicolumn{2}{c}{$\Sigma(\vec{\bm{s}}, \vec{\bm{s}})$} && \multicolumn{2}{c}{$\left\{ \beta_j(\vec{\bm{s}}) \right\}^{15}_{j=0}$} && \multicolumn{2}{c}{ $\sigma^2_\epsilon$}  \\ \\[-1em]
 \hline \\[-1em]
  \multicolumn{2}{c}{BHM ($d = 0.01$) } &&
 \multicolumn{2}{c}{ $0.2051$ } &&  \multicolumn{2}{c}{ $0.0154$ } && \multicolumn{2}{c}{ $0.0432$ } && \multicolumn{2}{c}{ $0.0048$ } \\ \multicolumn{2}{c}{BHM ($d = 0.05$) } &&
 \multicolumn{2}{c}{ $0.1816$ } &&  \multicolumn{2}{c}{ $0.0143$ } && \multicolumn{2}{c}{ $0.0308$ } && \multicolumn{2}{c}{ $0.0015$ } \\
 \multicolumn{2}{c}{BHM ($d = 0.1$) } &&
 \multicolumn{2}{c}{ $0.1803$ } &&  \multicolumn{2}{c}{ $0.0141$ } && \multicolumn{2}{c}{ $0.0304$ } && \multicolumn{2}{c}{ $0.0014$ } \\ 
\hline
\hline
\end{tabular}
\end{adjustbox}
\end{table}

\begin{table}[!hbt]
\caption{Second simulated scenario: Global-local selection for a representative dataset}
\label{tab:sim.ogls.2}
\centering
\begin{adjustbox}{max width = \textwidth}
\begin{tabular}{ccccccccccccc}
 \hline
 \hline
 \multicolumn{13}{c}{ Precision ($\pi = 9\%$)} \\
 \hline
 \multicolumn{2}{c}{Thresholds} && Global && \multicolumn{8}{c}{Local} \\
 \hline
 & &&  && $\tau_1(\vec{\bm{s}})$ & $\tau_2(\vec{\bm{s}})$ & $\tau_3(\vec{\bm{s}})$ & $\tau_4(\vec{\bm{s}})$ & $\tau_5(\vec{\bm{s}})$ & $\tau_6(\vec{\bm{s}})$ & $\tau_7(\vec{\bm{s}})$ & $\tau_8(\vec{\bm{s}})$ \\ \\[-1em]
 \cline{4-4} \cline{6-13} \\[-1em]
 \multicolumn{2}{c}{ BHM ($d = 0.01$) } && $0.700$ &&   $0.777$ & $0.618$ & $0.716$ & $0.592$ & $0.793$ & $0.506$ & $0.279$ & $  -  $ \\
 \multicolumn{2}{c}{ BHM ($d = 0.05$) } && $1$ &&  $0.743$ & $0.673$ & $0.714$ & $0.551$ & $  -  $ &   $0.505$ & $0.276$ & $  -  $ \\  
 \multicolumn{2}{c}{ BHM ($d = 0.1$) } && $1$ &&  $0.777$ & $  -  $ &   $  -  $ &   $0.608$ & $  -  $ &   $0.505$ & $0.293$ & $  -  $ \\
 \hline
 \multicolumn{13}{c}{ Precision ($\pi \approx 18.8\%$)} \\
 \hline
 \multicolumn{2}{c}{Thresholds} && Global && \multicolumn{8}{c}{Local} \\
 \hline
 & &&  && $\tau_1(\vec{\bm{s}})$ & $\tau_2(\vec{\bm{s}})$ & $\tau_3(\vec{\bm{s}})$ & $\tau_4(\vec{\bm{s}})$ & $\tau_5(\vec{\bm{s}})$ & $\tau_6(\vec{\bm{s}})$ & $\tau_7(\vec{\bm{s}})$ & $\tau_8(\vec{\bm{s}})$ \\ \\[-1em]
 \cline{4-4} \cline{6-13} \\[-1em]
 \multicolumn{2}{c}{ BHM ($d = 0.01$) } && $0.8$ &&  $0.796$ & $0.873$ & $0.897$ & $0.737$ & $0.876$ & $0.836$ & $0.837$ & $0.483$ \\
 \multicolumn{2}{c}{ BHM ($d = 0.05$) } && $1$ &&  $0.806$ & $0.923$ & $0.858$ & $0.753$ & $0.890$ & $0.839$ & $0.793$ & $0.480$ \\ 
 \multicolumn{2}{c}{ BHM ($d = 0.1$) } && $1$ &&  $0.781$ & $  -  $ &   $0.887$ & $0.76 $ & $0.88 $ & $0.852$ & $0.797$ & $0.497$ \\
\hline 
\hline 
\multicolumn{13}{c}{ Recall ($\pi = 9\%$)} \\
 \hline
 \multicolumn{2}{c}{ Thresholds } && Global && \multicolumn{8}{c}{Local} \\
 \hline
& &&  && $\tau_1(\vec{\bm{s}})$ & $\tau_2(\vec{\bm{s}})$ & $\tau_3(\vec{\bm{s}})$ & $\tau_4(\vec{\bm{s}})$ & $\tau_5(\vec{\bm{s}})$ & $\tau_6(\vec{\bm{s}})$ & $\tau_7(\vec{\bm{s}})$ & $\tau_8(\vec{\bm{s}})$ \\ \\[-1em]
 \cline{4-4} \cline{6-13} \\[-1em]
\multicolumn{2}{c}{ BHM ($d = 0.01$) } && $0.875$ &&  $0.988$ & $0.420$ &  $0.654$ & $0.753$ & $0.284$ & $0.556$ & $0.840$ &  $0    $ \\ 
\multicolumn{2}{c}{ BHM ($d = 0.05$) } && $0.75$ &&  $1    $ & $0.432$ & $0.617$ & $0.728$ & $0    $ & $0.568$ & $0.840 $ & $0    $  \\ 
\multicolumn{2}{c}{ BHM ($d = 0.1$) } && $0.5$ &&  $0.988$ & $0    $ & $0    $ & $0.728$ & $0    $ & $0.605$ & $0.840 $ & $0    $  \\
\hline 
\multicolumn{13}{c}{ Recall ($\pi \approx 18.8\%$)} \\
 \hline
 \multicolumn{2}{c}{ Thresholds } && Global && \multicolumn{8}{c}{Local} \\
 \hline
& &&  && $\tau_1(\vec{\bm{s}})$ & $\tau_2(\vec{\bm{s}})$ & $\tau_3(\vec{\bm{s}})$ & $\tau_4(\vec{\bm{s}})$ & $\tau_5(\vec{\bm{s}})$ & $\tau_6(\vec{\bm{s}})$ & $\tau_7(\vec{\bm{s}})$ & $\tau_8(\vec{\bm{s}})$ \\ \\[-1em]
 \cline{4-4} \cline{6-13} \\[-1em]
 \multicolumn{2}{c}{ BHM ($d = 0.01$) } && $1$ &&  $0.692$ & $0.367$ & $0.568$ & $0.947$ & $0.751$ & $0.544$ & $0.852$ & $0.852$ \\
 \multicolumn{2}{c}{ BHM ($d = 0.05$) } && $1$ &&  $0.663$ & $0.355$ & $0.609$ & $0.959$ & $0.769$ & $0.615$ & $0.840$ & $0.852$ \\
 \multicolumn{2}{c}{ BHM ($d = 0.1$) } && $0.875$ &&  $0.633$ & $0    $ & $0.604$ & $0.935$ & $0.781$ & $0.580$ & $0.834$ & $0.888$ \\
\hline 
\hline
\multicolumn{13}{c}{ $F_1$ scores  ($\pi = 9\%$)} \\
\hline
 \multicolumn{2}{c}{ Thresholds } && Global && \multicolumn{8}{c}{Local} \\
 \hline
 & &&  && $\tau_1(\vec{\bm{s}})$ & $\tau_2(\vec{\bm{s}})$ & $\tau_3(\vec{\bm{s}})$ & $\tau_4(\vec{\bm{s}})$ & $\tau_5(\vec{\bm{s}})$ & $\tau_6(\vec{\bm{s}})$ & $\tau_7(\vec{\bm{s}})$ & $\tau_8(\vec{\bm{s}})$ \\ \\[-1em]
 \cline{4-4} \cline{6-13} \\[-1em]
 \multicolumn{2}{c}{ BHM ($d = 0.01$) } && $0.778$ && $0.870$ &  $0.500$ &   $0.684$ & $0.663$ & $0.418$ & $0.529$ & $0.418$ & $  -  $ \\  
 \multicolumn{2}{c}{ BHM ($d = 0.05$) } && $0.857$ && $0.853$ & $0.526$ & $0.662$ & $0.628$ & $  -  $ &   $0.535$ & $0.416$ & $  -  $ \\   
 \multicolumn{2}{c}{ BHM ($d = 0.1$) } && $0.667$ && $0.870$ &  $  -  $ &   $  -  $ &   $0.663$ & $  -  $ &   $0.551$ & $0.435$ & $  -  $ \\ 
\hline
\multicolumn{13}{c}{ $F_1$ scores ($\pi \approx 18.8\%$) } \\
\hline
 \multicolumn{2}{c}{ Thresholds } && Global && \multicolumn{8}{c}{Local} \\
 \hline
 & &&  && $\tau_1(\vec{\bm{s}})$ & $\tau_2(\vec{\bm{s}})$ & $\tau_3(\vec{\bm{s}})$ & $\tau_4(\vec{\bm{s}})$ & $\tau_5(\vec{\bm{s}})$ & $\tau_6(\vec{\bm{s}})$ & $\tau_7(\vec{\bm{s}})$ & $\tau_8(\vec{\bm{s}})$ \\ \\[-1em]
 \cline{4-4} \cline{6-13} \\[-1em]
 \multicolumn{2}{c}{ BHM ($d = 0.01$) } && $0.889$ && $0.741$ & $0.517$ & $0.696$ & $0.829$ & $0.809$ & $0.659$ & $0.845$ & $0.617$ \\
\multicolumn{2}{c}{ BHM ($d = 0.05$) } && $1$ && $0.727$ & $0.513$ & $0.713$ & $0.844$ & $0.825$ & $0.710$ & $0.816$ & $0.614$ \\
\multicolumn{2}{c}{ BHM ($d = 0.1$) } && $0.933$ && $0.699$ & $  -  $ &   $0.718$ & $0.838$ & $0.828$ & $0.690$ & $0.815$ & $0.637$ \\ 
\hline
\hline
\end{tabular} 
\end{adjustbox}
\end{table}

\begin{table}[!htb]
\caption{Second simulated scenario: MSEs for a representative dataset}
\label{tab:sim.omse.2}
\centering
\begin{adjustbox}{max width = \textwidth}
\begin{tabular}{cccccccccccccc}
 \hline 
 \hline
   \multicolumn{14}{c}{MSE $(\pi = 9\%)$} \\
 \hline  \\[-1em]
 \multicolumn{2}{c}{Thresholds} &&
 \multicolumn{2}{c}{$\left\{ Z_i(\vec{\bm{s}})\right\}^{100}_{i=1}$} &&  \multicolumn{2}{c}{$\Sigma(\vec{\bm{s}}, \vec{\bm{s}})$} && \multicolumn{2}{c}{$\left\{ \beta_j(\vec{\bm{s}}) \right\}^{15}_{j=0}$} && \multicolumn{2}{c}{ $\sigma^2_\epsilon$}  \\ \\[-1em]
 \hline \\[-1em]
\multicolumn{2}{c}{BHM ($d = 0.01$) } &&
 \multicolumn{2}{c}{ $0.1680$ } &&  \multicolumn{2}{c}{ $0.0119$ } && \multicolumn{2}{c}{ $0.0470$ } && \multicolumn{2}{c}{ $0.0045$ } \\ 
 \multicolumn{2}{c}{BHM ($d = 0.05$) } &&
 \multicolumn{2}{c}{ $0.1428$ } &&  \multicolumn{2}{c}{ $0.0113$ } && \multicolumn{2}{c}{ $0.0345$ } && \multicolumn{2}{c}{ $0.0014$ } \\ 
 \multicolumn{2}{c}{BHM ($d = 0.1$) } &&
 \multicolumn{2}{c}{ $0.1294$ } &&  \multicolumn{2}{c}{ $0.0125$ } && \multicolumn{2}{c}{ $0.0334$ } && \multicolumn{2}{c}{ $0.0009$ } \\ 
 \hline
\hline
\multicolumn{14}{c}{MSE $(\pi \approx 18.8\%)$} \\
 \hline  \\[-1em]
 \multicolumn{2}{c}{Thresholds} &&
 \multicolumn{2}{c}{$\left\{ Z_i(\vec{\bm{s}})\right\}^{100}_{i=1}$} &&  \multicolumn{2}{c}{$\Sigma(\vec{\bm{s}}, \vec{\bm{s}})$} && \multicolumn{2}{c}{$\left\{ \beta_j(\vec{\bm{s}}) \right\}^{15}_{j=0}$} && \multicolumn{2}{c}{ $\sigma^2_\epsilon$}  \\ \\[-1em]
 \hline \\[-1em]
\multicolumn{2}{c}{BHM ($d = 0.01$) } &&
 \multicolumn{2}{c}{ $0.1814$ } &&  \multicolumn{2}{c}{ $0.0114$ } && \multicolumn{2}{c}{ $0.0465$ } && \multicolumn{2}{c}{ $0.0049$ } \\ 
\multicolumn{2}{c}{BHM ($d = 0.05$) } &&
 \multicolumn{2}{c}{ $0.1639$ } &&  \multicolumn{2}{c}{ $0.0102$ } && \multicolumn{2}{c}{ $0.0363$ } && \multicolumn{2}{c}{ $0.0025$ } \\ 
 \multicolumn{2}{c}{BHM ($d = 0.1$) } &&
 \multicolumn{2}{c}{ $0.1597$ } &&  \multicolumn{2}{c}{ $0.0099$ } && \multicolumn{2}{c}{ $0.0360$ } && \multicolumn{2}{c}{ $0.0020$ } \\ 
\hline 
\hline
\end{tabular}
\end{adjustbox}
\end{table}

\subsection*{S3. Trace plots for the choice of $d$}

As pointed out in the paper, the sparsity parameter $d$ has the interpretation that covariates affecting more than $d$ percent of the images are included in the model. The challenge in specifying $d$ is to distinguish the low values of the participation parameters $\pi_j$'s corresponding to the noisy covariates from those of the partially influential covariates, i.e., those covariates that affect a small number of voxels/pixels.  To better understand the role of the sparsity parameter $d$, we found helpful to look the MCMC trace plots of $\left\{\pi_j\right\}^{15}_{j=1}$ for the specification $d=0$. In this degenerate case we have $I(\pi_j \ge 0) = 1$ for all $j$ and the model includes all the covariates at each iteration to explain the observed images. 

\begin{figure}[!hbt]
        \centering
        \includegraphics[width = 1\textwidth]{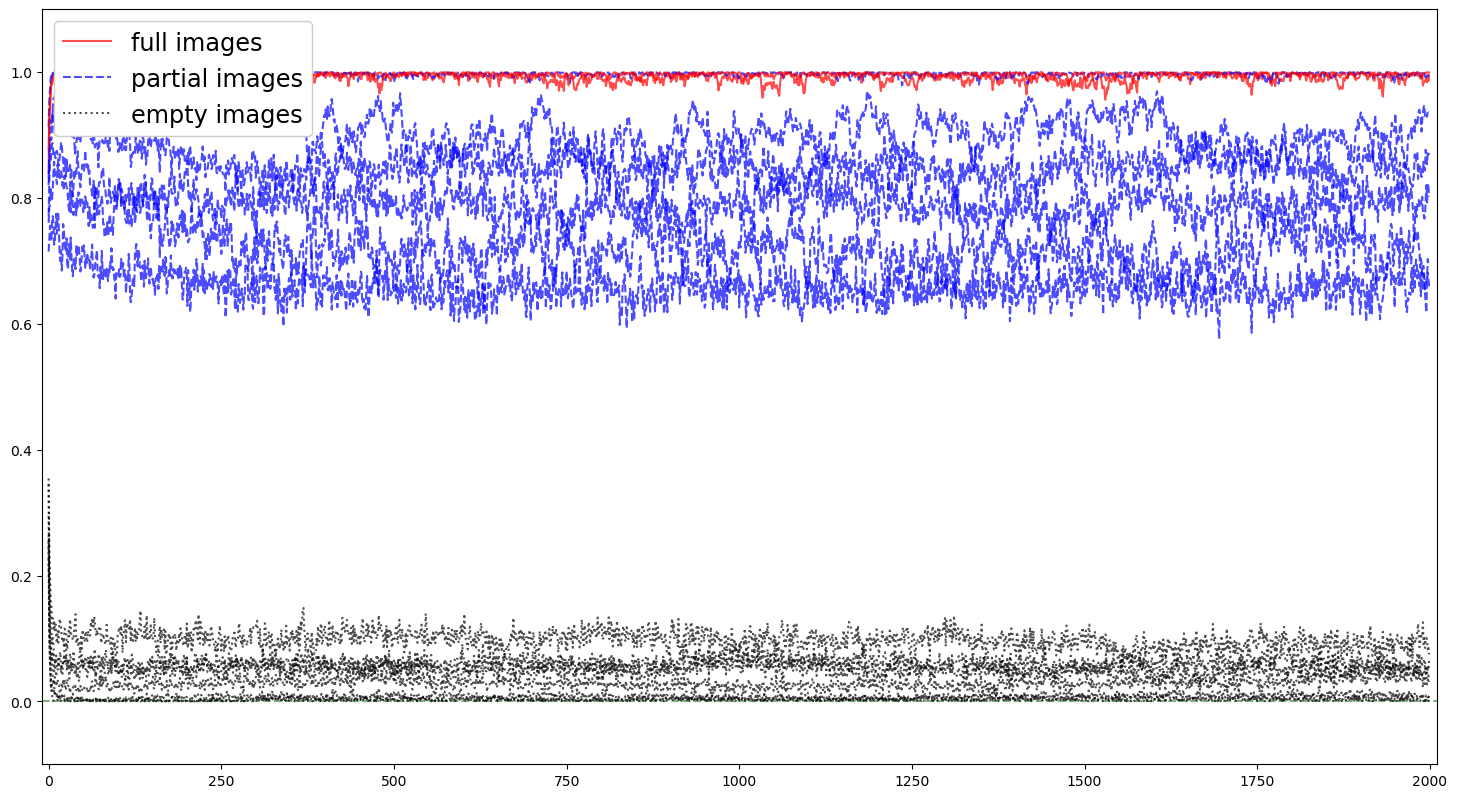}
    \caption{Trace plots of $\left\{ \pi_j \right\}^{15}_{j=1}$ for $d = 0$. Traces corresponding to fully non-zero image coefficients (influential covariates) are in red, those for partially influential covariates are in blue and those for the noisy covariates are in black.}
    \label{fig:sim.pi_spid}
\end{figure}

\begin{figure}
    \centering
        \begin{subfigure}[b]{0.29\textwidth}
        \centering
        \includegraphics[width=\textwidth]{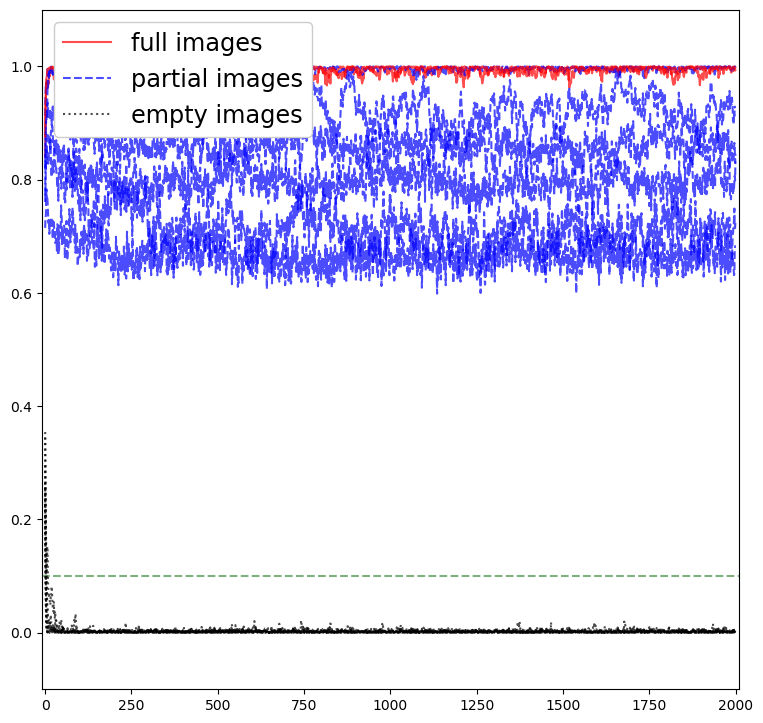}
        \caption{$d = 0.1$}
    \end{subfigure}
    \begin{subfigure}[b]{0.29\textwidth}
        \centering
        \includegraphics[width=\textwidth]{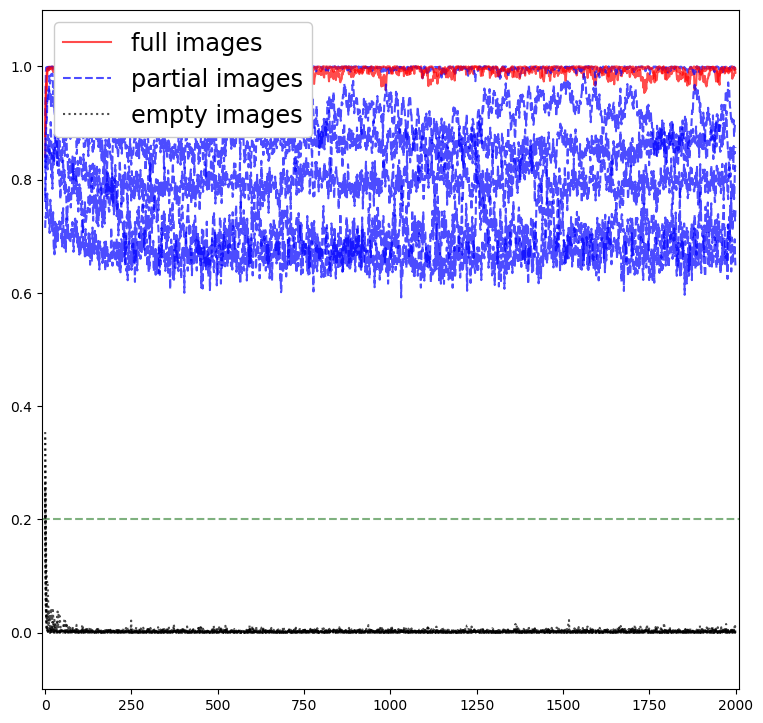}
        \caption{$d = 0.2$}
    \end{subfigure}
        \begin{subfigure}[b]{0.29\textwidth}
        \centering
        \includegraphics[width=\textwidth]{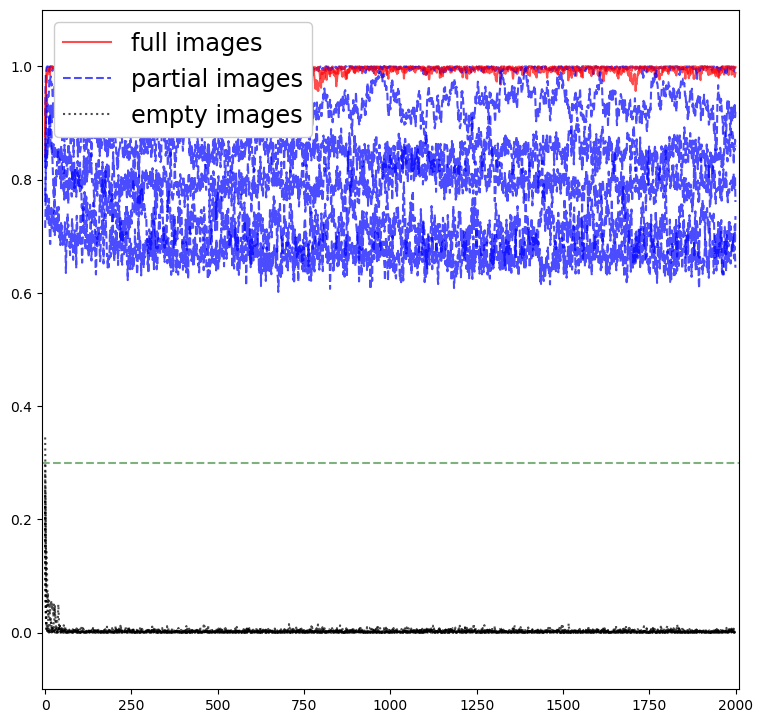}
        \caption{$d = 0.3$}
    \end{subfigure}
        \begin{subfigure}[b]{0.29\textwidth}
        \centering
        \includegraphics[width=\textwidth]{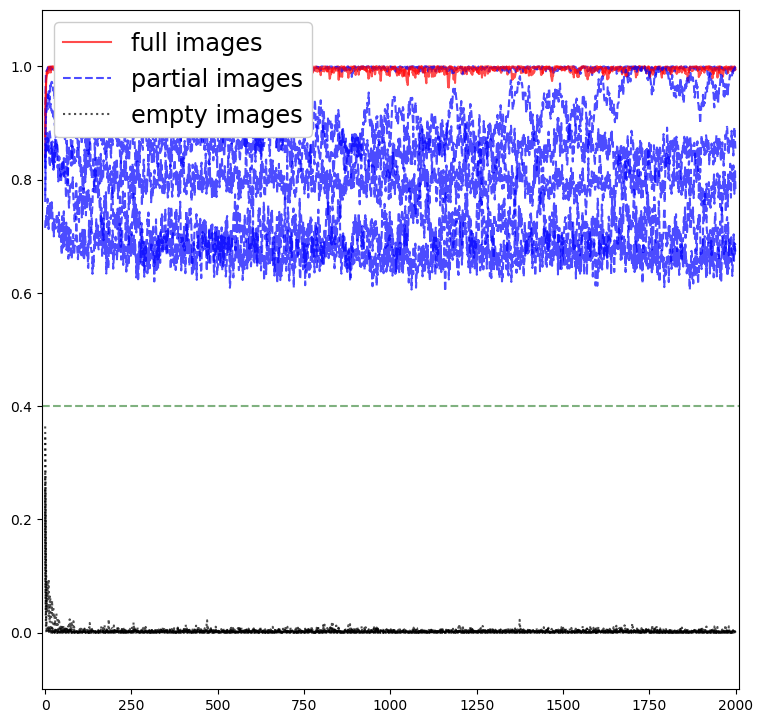}
        \caption{$d = 0.4$}
    \end{subfigure}
        \begin{subfigure}[b]{0.29\textwidth}
        \centering
        \includegraphics[width=\textwidth]{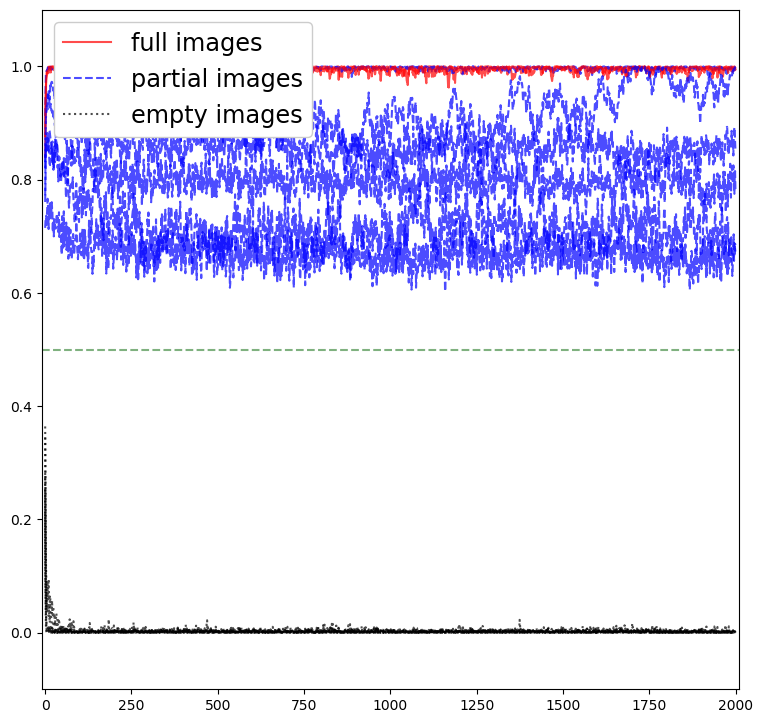}
        \caption{$d = 0.5$}
    \end{subfigure}
    \begin{subfigure}[b]{0.29\textwidth}
        \centering
        \includegraphics[width=\textwidth]{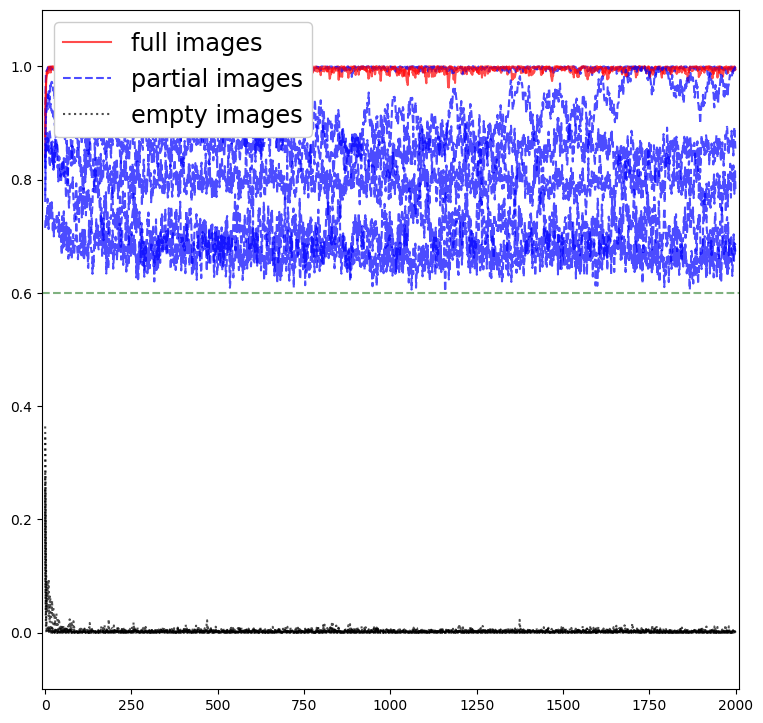}
        \caption{$d = 0.6$}
    \end{subfigure}
    \begin{subfigure}[b]{0.29\textwidth}
        \centering
        \includegraphics[width=\textwidth]{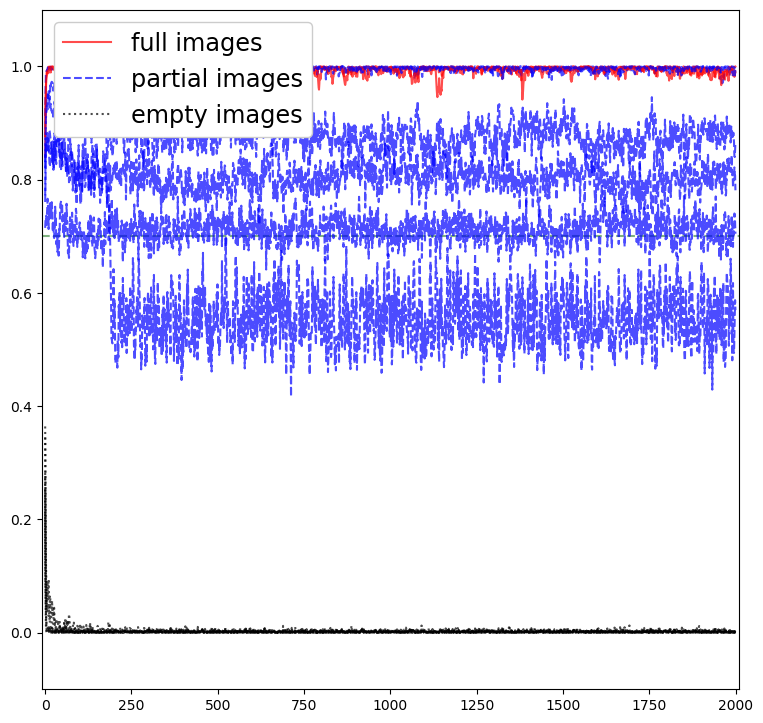}
        \caption{$d = 0.7$}
        \end{subfigure}
        \begin{subfigure}[b]{0.29\textwidth}
        \centering
        \includegraphics[width=\textwidth]{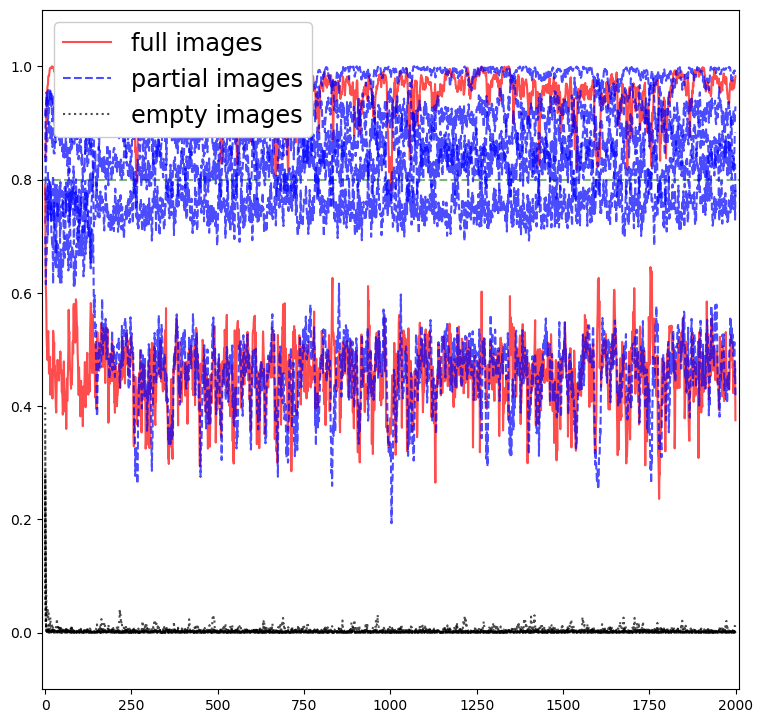}
        \caption{$d = 0.8$}
    \end{subfigure}
    \begin{subfigure}[b]{0.29\textwidth}
        \centering
        \includegraphics[width=\textwidth]{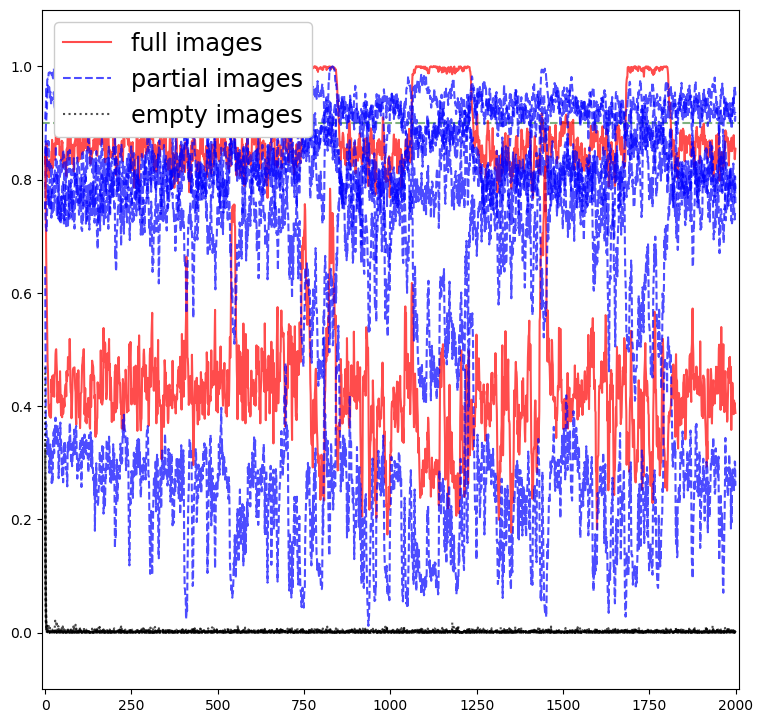}
        \caption{$d = 0.9$}
    \end{subfigure}
    \caption{Trace plots of $\left\{ \pi_j \right\}^{15}_{j=1}$ for a grid of $d \in [0,1]$. Traces corresponding to fully non-zero image coefficients (influential covariates) are in red, those for partially influential covariates are in blue and those for the noisy covariates are in black. }
    \label{fig:sim.pi_send}
\end{figure}

Figure~\ref{fig:sim.pi_spid} (a) shows these traces for a representative simulated case from setting 1. In this figure, traces corresponding to fully non-zero image coefficients (influential covariates) are in red, those for partially influential covariates are in blue and those for the noisy covariates are in black. We observe a clear separation between the three types of traces. In particular, we have $\pi \in [0,0.2]$ for the noisy covariates, $\pi \in [0.6, 0.9]$ for the partially influential covariates  and $\pi \approx 1$ for the covariates affecting the whole images.  This plot suggests that values of $d$ lower than .6 can be good choices, as they separate noisy covariates from the fully and partially influential ones.  Clearly, smaller values of $d$, as those we use in the paper and the sensitivity analysis above, are preferred, as they induce sparsity, excluding covariates effecting very low portions of the images and including those effecting larger portions. As further evidence, in Figure~\ref{fig:sim.pi_send} we show MCMC traces obtained by fitting our model for a grid of values $d \in [0.1,0.9]$. These figures confirm that the model is relatively insensitive to a range of choices $d\in[0.1, 0.6]$ and support the selection of a reasonably small $d$, which separates the traces.  On the contrary, when the choice of $d$ is too large, i.e., $d \geq 0.7$, the covariates get in and out from the model, introducing large fluctuations in several of the traces. 

As a word of caution, we remark that the trace plots we show here are meant to provide an empirical tool that might be helpful in the choice of $d$, particularly in cases where a separation among the traces is observed. However, this procedure is ad-hoc and cannot be used as a general method, in particular as the behavior of the trace plots is application-dependent and a clear separation of the traces might not always be observed.  In such cases, and in the absence of prior information, we recommend to view $d$ as conventional sparsity parameter and use standard values, i.e. $d = 0.05$ or $d = 0.1$.

\subsection*{S4. Sensitivity analysis for slab variance $\sigma^2_0$}

In addition to the hard-threshold $d$, the influential model parameters are the means $\left\{ \mu_{0j}( \vec{\bm{s}})\right\}^{15}_{j=0}$ and variance parameters $\left\{ \sigma^2_{0j} (\vec{\bm{s}}) \right\}^{15}_{j=0}$ of the slab prior distributions. As seen in Equation~\eqref{BGBF.eq.2}, the slab prior is involved in the calculation of the Bayes factors, therefore informing the local selection. Conventional specifications use slab normal distributions with means $0$ and variance parameters in the range $\sigma^2_0 \in [1,100]$. We compare the performances of different $\sigma^2_0$ using the first simulated scenario used in the paper, and fixed $d = 0.05$.

Tables~\ref{tab:sim.sens_sel_50} and~\ref{tab:sim.sens_mse_50} report the averaged precisions, recalls and $F_1$ scores for both global and local selection and MSEs for parameters of interest calculated over the $50$ replicates. We observe that the proposed method shows consistent good performances at the global level selection, achieving similarly high values in  precision, recall and $F_1$ scores, with all prior specifications. At the local level, results show a trade-off between precision and recall, in that larger prior variances tend to achieve higher precisions but lower recalls. As for parameter estimation, the MSEs of the  parameters of interests are relatively similar for different slab prior specifications, showing the estimations are not very sensitive to the slab prior specification. 

\begin{table}[!hbt]
\caption{Global and local selection for the sensitivity analysis - 50 replicates}
\label{tab:sim.sens_sel_50}
\centering
\begin{adjustbox}{max width = \textwidth}
\begin{tabular}{ccccccccccccc}
\hline
\hline
 \multicolumn{13}{c}{Averaged Precision} \\
  \hline
  \multicolumn{2}{c}{Slab} && Global && \multicolumn{8}{c}{Local} \\
 \hline
 \multicolumn{2}{c}{$\left(\mu_{0}(\vec{\bm{s}}), \sigma^2_{0}(\vec{\bm{s}})\right)$}  &&  && $\tau_1(\vec{\bm{s}})$ & $\tau_2(\vec{\bm{s}})$ & $\tau_3(\vec{\bm{s}})$ & $\tau_4(\vec{\bm{s}})$ & $\tau_5(\vec{\bm{s}})$ & $\tau_6(\vec{\bm{s}})$ & $\tau_7(\vec{\bm{s}})$ & $\tau_8(\vec{\bm{s}})$ \\
 \cline{1-2} \cline{4-4} \cline{6-13}
\multicolumn{2}{c}{ $\left( 0, 1 \right)$ } && $0.944$ &&  $1    $ & $0.937$ & $0.949$ & $0.962$ & $0.952$ & $1    $ & $0.953$ & $0.931$ \\  
\multicolumn{2}{c}{ $\left( 0, 10 \right)$ } && $0.998$ &&   $1    $ & $0.986$ & $0.976$ & $0.975$ & $0.969$ & $1    $ & $0.982$ & $0.964$   \\
\multicolumn{2}{c}{ $\left( 0, 100 \right)$ } && $1$ && $1    $ & $0.990$ & $0.984$ & $0.985$ & $0.981$ & $1    $ & $0.989$ & $0.979$ \\
\hline
\hline
\multicolumn{13}{c}{Averaged Recall} \\
  \hline
  \multicolumn{2}{c}{Slab} && Global && \multicolumn{8}{c}{Local} \\
 \hline
 \multicolumn{2}{c}{$\left(\mu_{0}(\vec{\bm{s}}), \sigma^2_{0}(\vec{\bm{s}})\right)$}  &&  && $\tau_1(\vec{\bm{s}})$ & $\tau_2(\vec{\bm{s}})$ & $\tau_3(\vec{\bm{s}})$ & $\tau_4(\vec{\bm{s}})$ & $\tau_5(\vec{\bm{s}})$ & $\tau_6(\vec{\bm{s}})$ & $\tau_7(\vec{\bm{s}})$ & $\tau_8(\vec{\bm{s}})$ \\
 \cline{1-2} \cline{4-4} \cline{6-13}
\multicolumn{2}{c}{ $\left( 0, 1 \right)$ } &&  $1$ &&  $0.970$ & $0.954$ & $0.920$ & $0.885$ & $0.865$ & $0.764$ & $0.752$ & $0.690$ \\ 
\multicolumn{2}{c}{ $\left( 0, 10 \right)$ } && $1$ &&  $0.928$ & $0.904$ & $0.875$ & $0.846$ & $0.822$ & $0.583$ & $0.589$ & $0.529$   \\
\multicolumn{2}{c}{ $\left( 0, 100 \right)$ } && $1$ &&   $0.884$ & $0.857$ & $0.830$ & $0.800$ & $0.777$ & $0.439$ & $0.451$ & $0.393$ \\
\hline
\hline
\multicolumn{13}{c}{Averaged $F_1$ scores } \\
 \hline
  \multicolumn{2}{c}{Slab} && Global && \multicolumn{8}{c}{Local} \\
 \hline
 \multicolumn{2}{c}{$\left(\mu_{0}(\vec{\bm{s}}), \sigma^2_{0}(\vec{\bm{s}})\right)$}  &&  && $\tau_1(\vec{\bm{s}})$ & $\tau_2(\vec{\bm{s}})$ & $\tau_3(\vec{\bm{s}})$ & $\tau_4(\vec{\bm{s}})$ & $\tau_5(\vec{\bm{s}})$ & $\tau_6(\vec{\bm{s}})$ & $\tau_7(\vec{\bm{s}})$ & $\tau_8(\vec{\bm{s}})$ \\
 \cline{1-2} \cline{4-4} \cline{6-13}
\multicolumn{2}{c}{ $\left( 0, 1 \right)$ } && $0.970$ &&   $0.983$ & $0.943$ & $0.932$ & $0.919$ & $0.903$ & $0.837$ & $0.814$ & $0.771$ \\ 
\multicolumn{2}{c}{ $\left( 0, 10 \right)$ } && $0.999$ &&  $0.959$ & $0.940$ & $0.920$ & $0.901$ & $0.885$ & $0.703$ & $0.711$ & $0.662$   \\
\multicolumn{2}{c}{ $\left( 0, 100 \right)$ } && $1$ && $0.933$ & $0.913$ & $0.895$ & $0.875$ & $0.859$ & $0.570$ & $0.592$ & $0.536$ \\
\hline
\hline
\end{tabular} 
\end{adjustbox}
\end{table} 

\begin{table}[!hbt]
\caption{MSEs for sensitivity analysis - $50$ replicates}
\label{tab:sim.sens_mse_50}
\centering
\begin{adjustbox}{max width = 1\textwidth}
\begin{tabular}{cccccccccc}
\hline 
\hline
 && \multicolumn{5}{c}{MSEs} \\
 \hline
 \multicolumn{1}{c}{Slab} & & \multicolumn{2}{c}{$\left\{ Z_i(\vec{\bm{s}})\right\}^{100}_{i=1}$} &&  \multicolumn{2}{c}{$\left\{ \beta_j(\vec{\bm{s}})\right\}^{15}_{j=0}$} && \multicolumn{2}{c}{$\Sigma(\vec{\bm{s}}, \vec{\bm{s}})$ }  \\
 \hline
 \multicolumn{1}{c}{$\left(\mu_{0}(\vec{\bm{s}}), \sigma^2_{0}(\vec{\bm{s}})\right)$} && Mean  & SE && Mean  & SE && Mean  & SE \\
\cline{1-1} \cline{3-4} \cline{6-7} \cline{9-10}
    \multicolumn{1}{c}{$\left(0,1\right)$} && $0.180$ & $(1.0 \times 10^{-3})$ &&  $0.472$ & $(1.13\times 10^{-2})$ && $0.014$ & $(0.5 \times 10^{-3})$ \\
      \multicolumn{1}{c}{$\left(0,10\right)$}&& $0.183$ & $(0.5 \times 10^{-3})$ &&  $0.547$ & $(1.23\times 10^{-2})$ && $0.014$ & $(0.5 \times 10^{-3})$ \\
      \multicolumn{1}{c}{$\left(0, 100\right)$}&& $0.179$ & $(0.4 \times 10^{-3})$ &&  $0.678$ & $(1.15\times 10^{-2})$ && $0.014$ & $(0.5 \times 10^{-3})$ \\ 
\hline
\hline
\end{tabular}
\end{adjustbox}
\end{table}

\subsection*{S5. Read Data Application with $d = 0.01$ and $d = 0.1$}
Table~\ref{tab:app.sel_res.1.d} and~\ref{tab:app.sel_res.2.d} report selection results for $d = 0.01, 0.05$ and $0.1$, along with results from MUA (BY) as a comparison. We observe that, as expected, a more stringent threshold of $d = 0.1$ would exclude FIQ for all networks entirely and a less stringent threshold of $d = 0.01$ would include FIQ for all networks but only very few voxels selected. Without over-interpreting the results, we remark that at the local selection level the ratios of region included are relatively consistent when the covariates are included, i.e. for Cuneus R, for both $d= 0.01$ and $d=0.05$, the ratio is around $1.7\%$.

\begin{table}
\caption{Selection results for the four networks}
\label{tab:app.sel_res.1.d}
\centering
\begin{adjustbox}{max width = 1\textwidth}
\begin{tabular}{ccccccccccc}
\hline
\hline
& & \multicolumn{4}{c}{BHM (d = 0.01)} &  & \multicolumn{4}{c}{BHM (d = 0.05) }\\
\hline
  &  & diagnostic & age & gender  & FIQ  &  & diagnostic & age & gender  & FIQ   \\
  \cline{3-6} \cline{8-11}
    \multicolumn{1}{c}{\multirow{2}{*}{Visual}} 
    & $\pi (\%)$ & 0.01 &  77.1 & 0.01  &  9.22 & & 0.01 & 77.2 & 0.01 & 9.24 \\
    & if selected   &  & (\checkmark) & & (\checkmark) &   &  &  (\checkmark) & & (\checkmark) \\  
   \multicolumn{1}{c}{\multirow{2}{*}{Ventral}}
    & $\pi (\%)$ & 0.01 &  82.8 & 0.01  &  2.68 & & 0.01 & 83.6 & 0.01 & 1.38 \\
    & if selected   &  & (\checkmark) & & (\checkmark) &   &  &  (\checkmark) & &   \\ 
    \multicolumn{1}{c}{\multirow{2}{*}{Dorsal}} 
    & $\pi (\%)$ & 0.01 & 43.1 & 0.01 &  8.02 & & 0.01 & 42.9 & 0.01 & 7.77\\
    & if selected   &  & (\checkmark) & & (\checkmark) &   &  &  (\checkmark) & & (\checkmark)  \\ 
    \multicolumn{1}{c}{\multirow{2}{*}{Default}} 
    & $\pi (\%)$ & 0.01 & 71.4 & 0.01 &  1.67 & & 0.01 & 71.6 & 0.01 & 0.78 \\
    & if selected   &  & (\checkmark) & & (\checkmark) &   &  &  (\checkmark) & &  \\ 
 \hline
 \hline
 & & \multicolumn{4}{c}{BHM (d = 0.1) } &  & \multicolumn{4}{c}{MUA (BY)}\\
\hline
  &  & diagnostic & age & gender  & FIQ  &  & diagnostic & age & gender  & FIQ   \\
  \cline{3-6} \cline{8-11}
    \multicolumn{1}{c}{\multirow{2}{*}{Visual}} 
    & $\pi (\%)$ &  0.01 & 81.6 & 0.01 & 6.33  & & 0.0 & 83.0 & 0.0 & 32.7 \\
    & if selected   &  & (\checkmark) & &  &   &  &  (\checkmark) & & (\checkmark) \\  
   \multicolumn{1}{c}{\multirow{2}{*}{Ventral}}
    & $\pi (\%)$ & 0.01 & 83.5 & 0.01 & 1.29 & & 0.0 & 82.2 & 0.0 & 24.7 \\
    & if selected   &  & (\checkmark) & &  &   &  &  (\checkmark) & &  (\checkmark)  \\ 
    \multicolumn{1}{c}{\multirow{2}{*}{Dorsal}} 
    & $\pi (\%)$ & 0.01 & 44.5 & 0.01 & 5.70 & & 0.0 & 54.3 & 0.0 & 22.7\\
    & if selected   &  & (\checkmark) & & &   &  &  (\checkmark) & & (\checkmark)  \\ 
    \multicolumn{1}{c}{\multirow{2}{*}{Default}} 
    & $\pi (\%)$ & 0.01 & 71.6 & 0.01 & 0.78 & & 0.0 & 74.5 & 0.0 & 24.4 \\
    & if selected   &  & (\checkmark) & &  &   &  &  (\checkmark) & &  (\checkmark) \\ 
 \hline
 \hline
\end{tabular}
\end{adjustbox}
\end{table} 

\begin{table}[!hbt]
\caption{Ratios of Region included within each networks}
\label{tab:app.sel_res.2.d}
\centering
\begin{adjustbox}{max width = 1\textwidth}
\begin{tabular}{ccccccccccc}
 \hline
 \hline
\\[-1em]
 \multicolumn{2}{c}{\multirow{2}{*}{Network}} & & \multirow{2}{*}{\shortstack{Methods}}& &\multirow{2}{*}{\shortstack{Covariates}} & & \multicolumn{4}{c}{\multirow{2}{*}{Ratio of Region included ($\%$)}}  \\
\\
\hline
\\[-1em]
 \multicolumn{2}{c}{\multirow{13}{*}{Visual}}  & & &  && & Lingual L & Lingual R & Calcarine L & Cuneus R  \\
 \\[-1em]
\multicolumn{2}{c}{} & & \multirow{3}{*}{ BHM$(d=0.01)$ } && age & & 82.0 & 81.6 & 83.5 & 88.6\\
 \\[-1em]
 \multicolumn{2}{c}{} & && & FIQ & &  5.6 & 6.8 & 3.4 & 1.6 \\
  \\[-1em]
  \cline{6-6} \cline{8-11}
  \\[-1em]
\multicolumn{2}{c}{} & & \multirow{3}{*}{ BHM$(d=0.05)$ } && age & & 82.1  &  81.4  &  83.5  &   88.9\\
 \\[-1em]
 \multicolumn{2}{c}{} & && & FIQ & &  5.79  &  7.00   & 3.29 &  1.77 \\
  \\[-1em]
  \cline{6-6} \cline{8-11}
  \\[-1em]
\multicolumn{2}{c}{} & & \multirow{3}{*}{ BHM$(d=0.1)$ } && age & & 87.4 & 85.8 & 89.2 & 92.6 \\
 \\[-1em]
 \multicolumn{2}{c}{} & && & FIQ & &  0.0 & 0.0 & 0.0& 0.0 \\
  \\[-1em]
\cline{6-6} \cline{8-11}
  \\[-1em]
 \multicolumn{2}{c}{} & &\multirow{3}{*}{MUA (BY) }  & & age & & 81.8 & 81.5 & 82.4 & 88.2  \\
 \\[-1em]
 \multicolumn{2}{c}{} & && & FIQ & &  34.2 & 34.7 & 29.1& 32.9  \\
  \\[-1em]
 \hline
\\[-1em]
 \multicolumn{2}{c}{\multirow{13}{*}{Ventral}} & & & & & & Temporal Mid L & Temporal Sup L & Temporal Sup R &  - \\
  \\[-1em]
\multicolumn{2}{c}{} & & \multirow{3}{*}{ BHM$(d=0.01)$ } && age & &  79.8 & 88.8 & 95.9 & - \\
 \\[-1em]
 \multicolumn{2}{c}{} & && & FIQ & &  0.7 & 1.2 & 0.2 & - \\
  \\[-1em]
  \cline{6-6} \cline{8-11}
  \\[-1em]
\multicolumn{2}{c}{} & & \multirow{3}{*}{ BHM$(d=0.05)$ } && age & &  80.7 & 90.1 & 96.4  &   - \\
 \\[-1em]
 \multicolumn{2}{c}{} & && & FIQ & &  0.0 & 0.0 & 0.0 & - \\
  \\[-1em]
  \cline{6-6} \cline{8-11}
  \\[-1em]
\multicolumn{2}{c}{} & & \multirow{3}{*}{ BHM$(d=0.1)$ } && age & &  80.7 & 90.0 & 96.5  &   - \\
 \\[-1em]
 \multicolumn{2}{c}{} & && & FIQ & &  0.0 & 0.0 & 0.0 & - \\
  \\[-1em]
\cline{6-6} \cline{8-11}
  \\[-1em]
 \multicolumn{2}{c}{} & &  \multirow{3}{*}{MUA (BY) } & & age & &  76.2 & 81.7 & 93.7 & -  \\
 \\[-1em]
 \multicolumn{2}{c}{} & && & FIQ & &  18.9 & 29.6& 31.2 & - \\
  \\[-1em]
 \hline
\\[-1em]
\multicolumn{2}{c}{\multirow{13}{*}{Dorsal}}  &&& & & & Temporal Mid R & Postcentral L & Parietal Sup L &   -  \\
 \\[-1em]
 \multicolumn{2}{c}{} & & \multirow{3}{*}{ BHM$(d=0.01)$ } && age & &   63.2 & 38.8 & 13.7  &   - \\
 \\[-1em]
 \multicolumn{2}{c}{} & && & FIQ & &   1.7 & 5.3 & 4.7 & - \\
  \\[-1em]
  \cline{6-6} \cline{8-11}
  \\[-1em]
 \multicolumn{2}{c}{} & & \multirow{3}{*}{ BHM$(d=0.05)$ } && age & &  62.9 & 38.6 & 13.8  &   - \\
 \\[-1em]
 \multicolumn{2}{c}{} & && & FIQ & &  1.5 & 4.7 & 4.5 & - \\
  \\[-1em]
  \cline{6-6} \cline{8-11}
  \\[-1em]
 \multicolumn{2}{c}{} & & \multirow{3}{*}{ BHM$(d=0.1)$ } && age & &  65.0 & 41.2 & 14.2  &   - \\
 \\[-1em]
 \multicolumn{2}{c}{} & && & FIQ & &   0.0 & 0.0 & 0.0& - \\
  \\[-1em]
\cline{6-6} \cline{8-11}
  \\[-1em]
 \multicolumn{2}{c}{} & &  \multirow{3}{*}{MUA (BY) } & & age & &  75.8 & 51.6 & 18.0 & -  \\
 \\[-1em]
 \multicolumn{2}{c}{} & && & FIQ & &  21.1 & 26.8 & 18 &  - \\
  \\[-1em]
 \hline
\\[-1em]
    \multicolumn{2}{c}{\multirow{13}{*}{Default}}  && &&  & & Temporal Mid R & Frontal Med Orb R & Frontal Med Orb L & Occipital Mid R  \\
    \\[-1em]
  \multicolumn{2}{c}{} & & \multirow{3}{*}{ BHM$(d=0.01)$ } && age & &   77.7 & 26.0 & 22.8 & 99.5 \\
 \\[-1em]
 \multicolumn{2}{c}{} & && & FIQ & &  0.1 & 0 & 0.0 & 0.4\\
  \\[-1em]
  \cline{6-6} \cline{8-11}
  \\[-1em]
 \multicolumn{2}{c}{} & & \multirow{3}{*}{ BHM$(d=0.05)$ } && age & &   77.7 & 26.2 & 22.7 & 99.5 \\
 \\[-1em]
 \multicolumn{2}{c}{} & && & FIQ & &  0.0 & 0.0 & 0.0 & 0.0\\
  \\[-1em]
  \cline{6-6} \cline{8-11}
  \\[-1em]
 \multicolumn{2}{c}{} & & \multirow{3}{*}{ BHM$(d=0.1)$ } && age & &    77.6 & 26.4 & 22.8 & 99.5\\
 \\[-1em]
 \multicolumn{2}{c}{} & && & FIQ & &  0.0 & 0.0 & 0.0 & 0.0\\
  \\[-1em]
\cline{6-6} \cline{8-11}
  \\[-1em]
 \multicolumn{2}{c}{} & &  \multirow{3}{*}{MUA (BY) } & & age & &   79.5 & 31.9 & 25.7 & 99.6  \\
 \\[-1em]
 \multicolumn{2}{c}{} & && & FIQ & &  22.7 & 6.2 & 11.1 & 39.6 \\
  \\[-1em]
 \hline
 \hline
\end{tabular}
\end{adjustbox}
\end{table}

\end{document}